\documentclass[11pt]{article}
\usepackage[top=50mm, bottom=50mm, left=50mm]{geometry}
\usepackage[labelfont=bf]{caption}
\usepackage{fourier}
\usepackage{lineno}
\usepackage{natbib}
\usepackage{tikz}
\usetikzlibrary{arrows.meta}
\usepackage{amssymb}
\usepackage{algorithm, algpseudocode}
\usepackage{algcompatible}
\usepackage{amsmath}
\usepackage{amsthm}
\usepackage{comment}
\usepackage{epsfig}
\usepackage{booktabs}
\usepackage{graphicx}
\usepackage{graphics}
\usepackage{float}
\usepackage{multirow}
\usepackage{color}
\usepackage{caption}
\usepackage{subcaption}
\usepackage{fullpage}
\usepackage[normalem]{ulem} 
\usepackage{makeidx}
\usepackage{xspace}
\usepackage{wrapfig}
\usepackage{tikz}
\usepackage{hyperref}
\hypersetup{
	colorlinks=true,
	citecolor=blue,
	linkcolor=blue,
	filecolor=blue,      
	urlcolor=blue,
}
\usepackage{setspace}

\newtheorem{proposition}{Proposition}
\newtheorem{assumption}{Assumption}
\newtheorem{remark}{Remark}

\newcommand{\Nik}{\mathcal{N}_{ik}}
\newcommand{\Njk}{\mathcal{N}_{jk}}
\newcommand{\vW}{\mathbf{W}}
\newcommand{\vG}{\mathbf{G}}
\newcommand{\vX}{\mathbf{X}}
\newcommand{\vXik}{\mathbf{X}_{ik}}
\newcommand{\vx}{\mathbf{x}}
\newcommand{\Yik}{{Y}_{ik}}
\newcommand{\Yjk}{{Y}_{jk}}
\newcommand{\Wik}{{W}_{ik}}
\newcommand{\Gik}{{G}_{ik}}
\newcommand{\Wjk}{{W}_{jk}}
\newcommand{\Gjk}{{G}_{jk}}

\newcommand{\E}{\mathbb{E}}
\newcommand{\Var}{\mathbb{V}}

\usepackage{dsfont}
\newcommand{\I}{\mathds{1}}

\usepackage[affil-it]{authblk}

\newcommand*{\estimates}{\mathrel{\widehat=}}
\newcommand\independent{\protect\mathpalette{\protect\independenT}{\perp}}
\def\independenT#1#2{\mathrel{\rlap{$#1#2$}\mkern2mu{#1#2}}}
\renewcommand{\thesubhyp}{\thehyp\alph{subhyp}}

\newcommand{\red}[1]{{\color{red} #1}}
\newcommand{\blue}[1]{{ #1}}

\begin{document}
	\begin{titlepage}
		\title{ Causal Effects with Hidden Treatment Diffusion on\\ Observed or Partially Observed Networks \footnotetext{We are grateful to Patrizia Lattarulo (IRPET – Tuscany’s Regional Institute for Economic Planning,), Marco Mariani (IRPET – Tuscany’s Regional Institute for Economic Planning) and Laura Razzolini (University of Alabama) for having collected, organized and shared with us data about the field experiment on the effect of different types of school incentives on students' museum attendance}}
		
		\author[1]{Costanza Tort\'u}
		\author[2]{Irene Crimaldi}
		\author[3]{Fabrizia Mealli}
		\author[4]{Laura Forastiere}
		
		\affil[1]{\'{E}cole Polytechnique}
		\affil[2]{IMT School for Advanced Studies Lucca}
		\affil[3]{University of Florence}
		\affil[4]{Yale University}

		\date{This version: July 2021.
			\href{https://www.imtlucca.it/it/costanza.tortu}{{\color{blue}Latest version here.}}} 
		\maketitle
		\vspace{-1cm}
		\begin{abstract} 
			In randomized experiments, interactions between units might generate a treatment diffusion process. 
			This is common when the treatment of interest is an actual object or product that can be shared among peers (e.g., flyers, booklets, videos). For instance, if the intervention of interest is an information campaign realized through the distribution of a video  to targeted individuals, some of these treated individuals might share the video they received with their friends. Such a phenomenon is usually unobserved, causing a misallocation of individuals in the two treatment arms: some of the initially untreated units might have actually received the treatment by diffusion. Treatment misclassification can, in turn, introduce a bias in the estimation of the causal effect. Inspired by a recent field experiment on the effect of different types of school incentives aimed at encouraging students to attend cultural events, we present a novel approach to deal with a hidden diffusion process on observed or partially observed networks. 
			Specifically, we develop a simulation-based sensitivity analysis that assesses the robustness of the estimates against the possible presence of a treatment diffusion. We simulate several diffusion scenarios within a plausible range of sensitivity parameters and we compare the treatment effect which is estimated in each scenario with the one that is obtained while ignoring the diffusion process. Results suggest that even a treatment diffusion parameter of small size may lead to a significant bias in the estimation of the treatment effect. 
			\\
			\vspace{0.5cm}\\
			\textbf{Keywords:} causal inference; potential outcomes; treatment diffusion; social networks; school incentives for cultural activities \\
			
			\bigskip
		\end{abstract}
		\setcounter{page}{0}
		\thispagestyle{empty}
	\end{titlepage}
	\pagebreak \newpage


	\maketitle
	
	\linespread{1.5}\selectfont
	\maketitle
	\section{Introduction}
	
	\subsection{Motivation}
	One of the goals of policy evaluation studies is to assess the causal effect of an intervention \citep{rubin1974estimating,imbens2015causal}. However, in a wide variety of real world scenarios, the treatment of interest can be diffused among units \citep{an2018causal,an2019opening}. 
	This phenomenon can occur when individuals interact with one another 
	\citep{liu2017social} and the intervention under evaluation is the distribution of an object or product transferable by nature (e.g., flyers, booklets, videos). Targeted individuals who received the treatment could share it with their social ties who did not receive it.
	
	Treatment diffusion on social networks  
	may arise in the evaluation of many interventions
	: promotional videos or links distributed through social media \citep{rogers2012measuring,la2020diffusion}, advertising fliers or leaflets distributed by hand and passed along \citep{an2018causal}, innovations propagating over networks \citep{steckler1992measuring,valente1993diffusion,nyblom2003statistical,mahajan2010innovation}, and even monetary incentives that can be transferred among economic or social agents \citep{gai2010contagion,leitner2005financial}. 
	The field experiment by \citep{lattarulo2017nudging} provides a motivating example for studying the implications of treatment diffusion. In their study, the intervention of interest is a promotional video to motivate students to visit art museums. However, some of the treated students could have shared the link of the video with their peers.
	
	In the presence of treatment diffusion, some individuals who were originally assigned to the control group, and were not provided with treatment by design, might have actually received the treatment through diffusion thanks to a link with a treated individual. Hence, treatment diffusion alters the original random allocation.
	In most cases the diffusion process is not observed and the researcher is not able to reconstruct the diffusion patterns. Even when there is information about the existing relations among units (\emph{baseline social network}), we rarely know who passed the treatment to whom (\emph{treatment diffusion network}). 
	This circumstance leads to a misspecification of the treatment assignment vector \citep{grandjean2004underestimation,lewbel2007estimation} and, in turn, might introduce a bias in the estimate of the treatment effect, 
	yielding erroneous overestimation or underestimation of the real impact of the intervention.  
	
	Treatment diffusion may be regarded as a specific mechanism of interference 
	\citep{sobel2006, hudgens2008toward,tchetgen2012causal}. Interference is present when the potential outcome of a given unit is affected by the treatment assigned to other units \citep{cox1958planning}. 
	In the treatment diffusion setting, the potential outcome of one unit not only depend on their own treatment assignment, but it also depends on whether  
	their network neighbors were assigned to the intervention, because they could receive the treatment from them through diffusion. 
	
	Treatment diffusion is not the only existing mechanism through which interference can take place. Specifically, 
	we can identify three types of interference mechanisms, that occur at different stages of the causal process: i) the direct effect of one's treatment on their own outcome coupled with the diffusion of the outcome to other individuals \citep{nichol1995effectiveness,bridges2000effectiveness,chin2013impact,chuang1999foreign,del2019causal}; ii) the indirect effect of one's treatment on the outcome of connected units \citep{centola2010spread,zheng2013spreading,paluck2016changing}; iii) the diffusion of the treatment to interacting individuals coupled with the effect of indirectly receiving the treatment on one's own outcom \citep{cohen2002r,gai2010contagion,yang2013credit,lamberson2016diffusion}.

	
	Accounting for interference, when it is likely to arise, not only prevents from facing a bias while estimating the treatment effect \citep{savje2017average,forastiere2020identification}, but most importantly enables policy makers to inspect the global impact of the intervention. 
	In the presence of treatment diffusion, the estimation of the causal effects of being assigned to the treatment would require accounting for interference due to potential treatment diffusion from network ties. This could be done by using recently developed estimators for treatment and spillover effects under partial or network interference \citep[e.g.;][]{hudgens2008toward, aronow2017estimating, forastiere2020identification}. 
	The application of these estimators to a setting affected by interference 
	would result in unbiased estimates of the direct effect of treatment assignment, spillover effect of being exposed to the treatment assignment of others, or the overall effect of the intervention providing treatment in the population under a specif assignment mechanism.   
	However, when interference is due to treatment diffusion, the treatment has been received by individuals beyond those assigned to treatment as part of the intervention. Thus, the use of estimators developed under a general interference setting does not allow us to assess the effect of the actual treatment receipt. 
	When we are primarily concerned with assessing the effect of actually receiving the treatment, directly or indirectly, the estimation strategy should account for the treatment diffusion process in order to recover the missing information about the unobserved treatment receipt. 
	Doing so implies some statistical challenges, 
	since the treatment diffusion process is an unknown and unobserved stochastic process with possibly missing information about network connections among units.

	\subsection{Related Literature}
	The causal inference literature has developed a number of statistical methods to deal with interference \citep{aronow2017estimating,forastiere2020identification,papadogeorgou2019causal,aronow2019design,miles2019causal,loh2020randomization,tortu2020modelling,leung2020treatment}.
	In particular, some recent studies have proposed estimators for treatment and spillover effects under the assumption of on \emph{partial} (or \emph{clustered}) interference \citep{sobel2006,hudgens2008toward, tchetgen2012causal, liu2014large, Liu2016, kang2016peer, forastiere2016clusters, basse2018analyzing, forastiere2019museums}, where units are clustered in exogenous groups and spillover mechanisms are assumed to occur only within groups. However, this assumption is not always reasonable, as often the intervention spills over units according to more complex interactions \citep{arpino2016assessing, del2019causal, tortu2020immigration, tortu2020modelling}. Therefore, some researchers have developed novel methodologies to account for spillovers in a more general setting, where units interfere according to the links observed over a network (\emph{general} or \emph{network} interference) both in experimental settings \citep{aronow2017estimating, AtheyEcklesImbens2018, leung2020,bargagli2020heterogeneous,imai2020causal} and in observational studies \citep{Ogburn2017, Sofrygin2017, forastiere2018estimating, forastiere2016identification, tortu2020modelling}. 
	%
	These estimators tackle the problem of interference  without distinguishing between the specific underlying mechanisms
	\citep{an2019opening}. 
	In the presence of treatment diffusion, these methods would estimate the causal effects of the initial treatment assignments, including the spillover effects from the treatment assigned to other units.
	However, they do not consider
	whether units diffuse the treatment to their neighbors
	preventing the investigation of the effect of the actual treatment receipt. 
	
	Very few works explicitly deal with treatment diffusion \citep{an2018causal,an2019opening}. However, the existing works on treatment diffusion 
	assume to have perfect knowledge on both the baseline social network describing relations among units and the actual treatment diffusion network describing who passes the treatment to whom over time. 
	However, this is a rather rare circumstance, as collecting information on both social ties and diffusion behavior is costly and cumbersome.
	In most real world scenarios the treatment diffusion process is hidden. Although the interest on diffusion processes has rapidly grown in recent years and many researchers have reconstructed and analysed heterogeneous diffusion processes in the field of both theoretical and applied sciences
	\citep{cowan2004network,valente2005network,lopez2008diffusion,mahajan2010innovation,katona2011network,bourigault2014learning,goel2016structural}, existing papers focus on the diffusion process itself, and not on its consequences on the causal evaluation of an intervention. 
	
	The treatment diffusion process gives rise to a \emph{missclassification} in the treatment variable. 
	The statistical literature on misclassification and measurement error \citep{bound2001measurement,carroll2006measurement,fuller2009measurement,grace2017statistical} have recently been extended to 
	the causal inference framework, by exploring the possible bias and developing bias correction methods for settings where the outcome variable \citep{shu2019causal,shu2019weighted,grace2017statistical} or the treatment variable \citep{grandjean2004underestimation,lewbel2007estimation,babanezhad2010comparison,imai2010causal,vanderweele2012inference,mccaffrey2013inverse,braun2014adjustment,braun2016using,yanagi2018inference} are measured with error. These works have all pointed out that a misclassification of the exposure might induce a bias in the estimation of the average treatment effect both when the measurement error is dependent or independent on potential outcomes, given the true value of the treatment and observed covariates (\emph{differential} and \emph{non-differential} misclassification). The direction of this bias 
	depends on several factors, including the treatment effect for those receiving the treatment by diffusion and the relationship between the the misclassifcation process and the outcome (or treatment) variable \citep{lewbel2007estimation}.

	\subsection{Contributions}
	The main purpose of the present work is to characterize the bias in the estimation of the treatment effect due to treatment diffusion and to provide a novel simulation-based method for sensitivity analysis to assess the robustness of results against plausible but hidden treatment diffusion processes. We deal with a setting where the treatment diffusion network is hidden and, hence, it is not possible to explicitly correct for the treatment diffusion bias. 
	
	In particular, this contribution characterizes the diffusion process, formalizes the treatment diffusion bias and clarifies the causal factors, which lead to either an overestimation or an underestimation of the treatment effect. The estimation bias due to having wrongly ignored diffusion cannot be explicitly corrected, as long as the treatment diffusion network is unknown. 
	To address this issue, we propose a sensitivity analysis \citep{rosenbaum1983assessing,robins2000sensitivity,imai2010causal,diaz2013sensitivity,rosenbaum2014sensitivity} to assess the robustness of the estimates with respect to different diffusion scenarios. The main idea is to compare the na\"ive estimate of the treatment effect obtained assuming the absence of any diffusion process, with the estimates under several plausible diffusion scenarios. The estimates are derived using a reworked version of the standard Horvitz-Thompson estimator \citep{horvitz1952generalization}. 
	The performance of the proposed sensitivity analysis is evaluated in some simulated scenarios, where the misclassification due to having wrongly ruled out the presence of a treatment diffusion process introduces either an overestimation or an underestimation of the real effect of the intervention.
	
	The proposed framework is employed in analyzing results related to a recent field experiment, which was designed to assess the effect of different kinds of school-incentives aimed at promoting museums attendance among students \citep{lattarulo2017nudging,forastiere2019exploring}. The intervention of interest is a video presentation on an art exhibit and data provide information about friendship ties linking students enrolled in the same class. In this setting, the treatment diffusion might potentially be plausible because students receiving the video could have shared it with their friends. The empirical setup presents two methodological peculiarities: i) the baseline social network is only partially observed \citep{zhou2009predicting,duong2011modeling,onnela2012spreading,koskinen2013bayesian,pan2016predicting} (the available network data provide information on intra-class links only, while inter-class links are hidden); ii) the experiment is a cluster randomized design \citep{gertler2004conditional,green2008analysis,crepon2013labor} - i.e the intervention is assigned at a class level - and, as a consequence, the treatment assignment of one student also depends on the treatment assignment of other units (the classmates). In analyzing the impact of a possible treatment diffusion process taking place, the proposed sensitivity analysis addresses these two issues by using multiple imputation to predict missing ties, while also accounting for the uncertainty on the complete network structure, and by presenting an appropriate estimator of the real (post-diffusion) treatment effect for clustered randomized designs. 

	The paper is organized as follows. Section \ref{sec: trd} presents the main aspects of the proposed methodology: it characterizes the theoretical framework, formalizes the treatment diffusion process, discusses the direction of the treatment diffusion bias and introduces the key steps of the sensitivity analysis. Section \ref{sec: simul} illustrates how the sensitivity analysis works in some exemplifying simulated scenarios. Section \ref{sec: ea}, focuses on the application: it motivates its empirical relevance, it describes data, it presents the methodological challenges related to the specific empirical context and comments the main findings.  The paper closes with Section \ref{sec: concl}, which recaps the main results of the analysis and outlines potential directions for future research. 
	
	
	\section{Sensitivity Analysis for Treatment Diffusion \label{sec: trd}}
	
	\subsection{Setup and Notation \label{subsec: trd_idea}}
	Let us consider a randomized experiment aimed at evaluating the effect of a given intervention on an outcome, in a given population. We denote by $\mathcal{N}$ the population of interest, where each unit $i$, with $i=1, \dots, card(\mathcal{N})=N$, is randomly assigned to either the active treatment or the control group. Let $Z_{it}\in \{0,1\}$ be the binary variable representing the treatment assigned to unit $i$ at time $t$ and $Y_{it''}$ the individual outcome, observed at the lagged time $t''$, with $t''>t$. For instance, in our motivational study, the treatment of interest represents the kind of incentive to promote museums attendance, while the outcome variable is the number of self-reported museum visits in the eight months between the baseline and the follow-up survey. $\boldsymbol{Z}_{t}$ and $\boldsymbol{Y}_{t''}$ denote the treatment assignment and outcome vectors in the entire population $\mathcal{N}$, while $\boldsymbol{Z}_{-it}$ denotes the vector of the treatments assigned to all the units in $\mathcal{N}$, but $i$. We assume that the randomization design ensures that each unit has a non-zero probability of being initially allocated in each of the two treatment arms. According to the initial randomization, $\boldsymbol{Z}_{t}$, we define the set of treated units at time $t$, $\mathcal{T}_{t}= \{i \in \mathcal{N}: Z_{it}=1 \}$, with cardinality $T_{t}=card(\mathcal{T}_t)=\sum_{i=1}^{N} Z_{it}$. 
	%
	%
	
	We denote by $\mathbf{G}=(\mathcal{N},\mathcal{E})$ the graph describing the relations among units. The graph $\mathbf{G}$ admits an equivalent representation in terms of its adjacency matrix $\boldsymbol{A}=\{a_{ij}: i, j \in \mathcal{N} \}$, where the generic element $a_{ij}$ signals the presence of an edge between unit $i$ and unit $j$. Note that this matrix can be symmetric with $a_{ij}=a_{ji}$ or asymmetric with $a_{ij} \neq a_{ji}$. Therefore, we distinguish between the set of nodes having an in-going link to $i$, $\mathcal{N}^{in}_i = \{j : (j,i) \in \mathcal{E} \} $ with cardinality ${N}_{i}^{in}$, and the set of nodes with an out-going link from $i$, $\mathcal{N}^{out}_i = \{j : (i,j) \in \mathcal{E} \} $ with cardinality ${N}_{i}^{out}$. In a social environment, agents can interact according to friendship ties. In that setting, the graph $\mathbf{G}$ represents the whole friendship network and the social relations are not necessarily reciprocated: for instance, given two agents $i$ and $j$, it may happen that $i$ regards $j$ as a friend -i.e $j$ has an out-going friendship link from $i$ -, but not vice versa - i.e $j$ does not have an in-going link to $i$ -.   
	%
	%
	Network relations are assumed not to change over the time frame from $t$ to $t''$. 
	As we will see, we allow for the possibility that links in $\boldsymbol{A}$ are only partially observed.
	
	Given the existence of connections among individuals, we assume that between the initial randomization at time $t$ and the outcome collection at time $t''$ there could be a treatment diffusion process. If the treatment spreads among individuals, then the initial treatment assignment vector $\boldsymbol{Z}_{t}$ does not truly represent the real allocation of units in the treatment arms. We denote by ${Z}_{it'}$ the actual treatment status of unit $i$ at time $t'$ after the diffusion process, with $\boldsymbol{Z}_{t'}$ being the corresponding treatment vector in the sample. Throughout, we will assume that the $\boldsymbol{Z}_{t'}$ of treatment receipt is not observed. 
	
	
	
	\subsection{The Diffusion Process: Assumptions\label{subsec: trd_fixpr}}
	We model the diffusion process under the following simplifying assumptions.
	\begin{assumption}[Unchangeable Status of the Initially Treated Units]
		\label{ass: initiallytr}
		All the units who receive the intervention by design, remain treated even after the treatment diffusion process, that is $P(Z_{it'}=1|Z_{it}=1)=1$ for all $i\in\mathcal{N}$. 
	\end{assumption}
	
	Assumption \ref{ass: initiallytr} states that initially treated units cannot switch to a control status after the treatment diffusion process.  
	
\begin{figure}[H]
		\centering
		\begin{tikzpicture}
			\begin{scope}[every node/.style={circle,thick,draw}]
				\node (Zit) at (0,4) {Zit};
				\node (Zjt) at (0,0) {Zjt};
				\node (Zit') at (4,4) {Zit'};
				\node (Zjt') at (4,0) {Zjt'};
				\node (Yit'') at (8,4) {Yit''};
				\node (Yjt'') at (8,0) {Yjt''} ;
			\end{scope}
			
			\begin{scope}[>={Stealth[black]},
				every node/.style={fill=white,circle},
				every edge/.style={draw=black,very thick}]
				\path [->] (Zit) edge  (Zit');
				\path [->] (Zit') edge  (Yit'');
				\path [->] (Zit) edge (Zjt');
				\path [->] (Zjt) edge  (Zjt');
				\path [->] (Zjt') edge  (Yjt'');
				\path [->] (Zjt) edge  (Zit');
			\end{scope}
		\end{tikzpicture}    
		\caption[Three-steps diffusion process]
		{Three-steps diffusion process: at the initial time $t$, the treatment is randomly assigned over the population, defining the treatment assignment vector $\boldsymbol{Z}_{t}$; at time $t'$, treatment may spread in the network, that is treated nodes may contaminate untreated neighbors, by sharing the intervention with them, and the new treatment status is represented by the treatment variable $Z_{it'}$: finally, at time $t''$, the outcome $\boldsymbol{Y}_{t''}$ becomes observable.}{\label{fig: timing}}
	\end{figure}
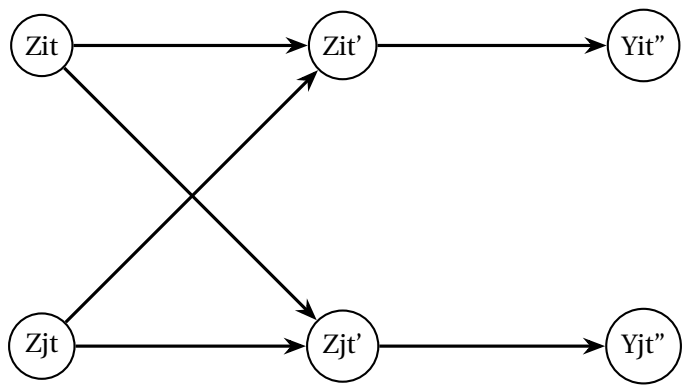

\begin{figure}[H]
		\centering
		\includegraphics[width=100mm]{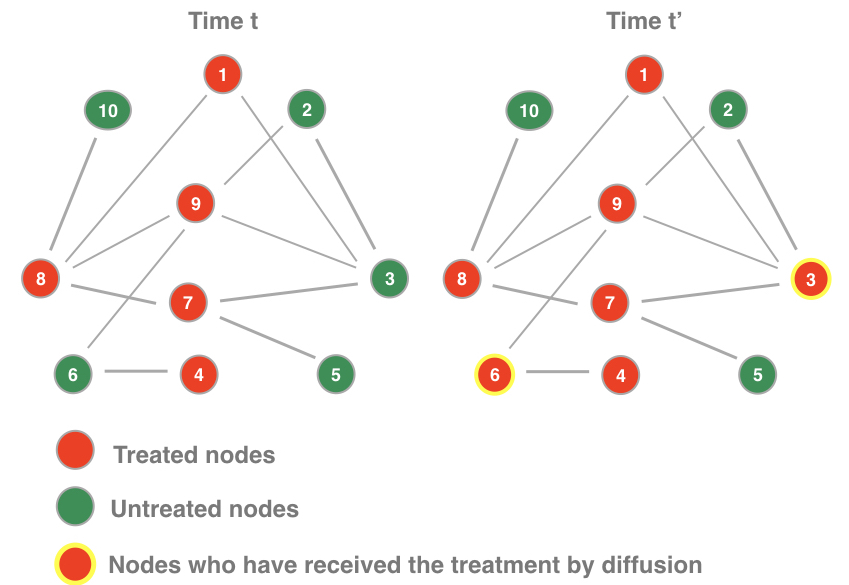}
		\caption[Treatment diffusion process: example]{Treatment diffusion process: example. Colors refer to the treatment status: \emph{red} colored nodes represent treated individuals, while \emph{green} nodes are the untreated ones. Units numbered as 3 and 6 were not initially assigned to the active treatment, but they have received the treatment by diffusion.\label{fig: trdex}}
	\end{figure}

	\begin{assumption}[Diffusion Interval]
		\label{ass: single}
		For all the untreated units, diffusion can occur only at a single point in time  in the time interval $(t,t')$. 
		%
		%
	\end{assumption}
	\noindent The timeline of the diffusion process is graphically represented in Figure \ref{fig: timing} and Figure \ref{fig: trdex} provides a graphical example of how a diffusion process may look like.

	\noindent This assumption avoids the need to consider multiple diffusion steps and to account for possible spreads of the treatment after the fixed time step $t'$.  Note that the simplified scenario considered here, with only three time frames and time-invariant network structure, is reasonable in most settings when the time interval between treatment assignment and follow-up is small.

	\begin{assumption}[Diffusion process with a constant probability]
		\label{ass: diffusion}
		Given the graph $\mathbf{G}$, an untreated node $i$ may receive the treatment only from a treated unit in 
		$\mathcal{N}^{in}_i$ and, each of such units, say $k$, can diffuse the treatment to $i$ with probability $\overline{p}<1$ (diffusion parameter), independently of the other treated units $k'\in \mathcal{N}_i^{in}$. 
		%
		%
	\end{assumption}
	Hence, in this setting, treatment propagates from treated units to their untreated neighbors. Nodes who have initially been assigned to the active treatment cannot subsequently pass to a control status (Assumption \ref{ass: initiallytr}) and cannot receive the treatment by diffusion. Note that Assumption \ref{ass: diffusion} also states that, when the social network $\mathbf{G}$ is directed and ties are not necessarily reciprocated, the agents who are likely to diffuse the intervention to an untreated node $i$ are those who have an in-going link to $i$. This means that we rule out the possibility that treated nodes who have an out-going link from $i$ pass the intervention to $i$. This hypothesis is plausible if we think at the meaning of the direction of a social tie: a given agent $j$ who was initially provided by the intervention by design, is likely to share it with an untreated node $i$ if she regards $i$ as a friend ($a_{ji}=1$), independently on whether the relation is reciprocated.    
	
\begin{figure}[H]
		\centering
		\includegraphics[width=80mm]{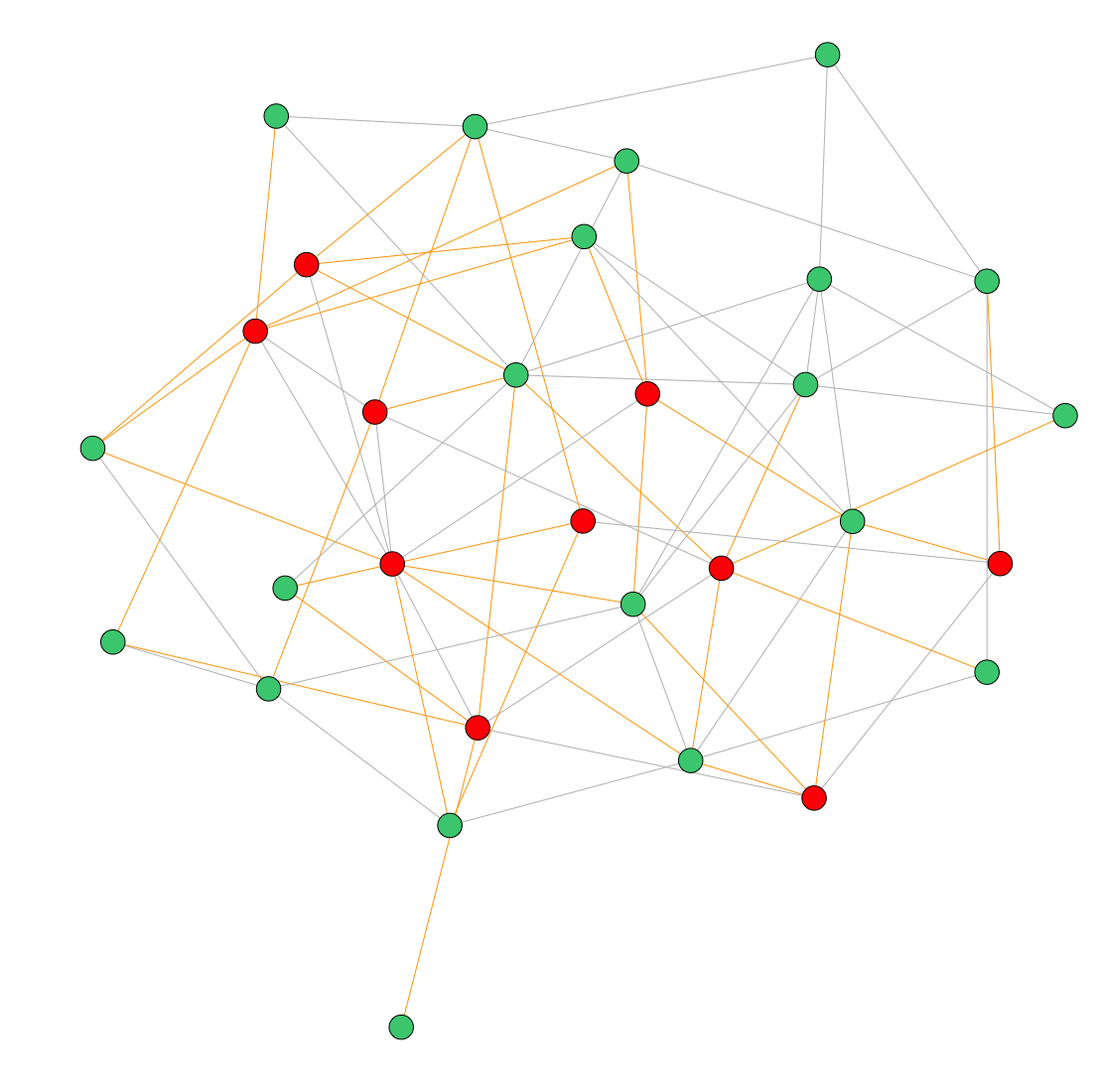}
		\caption[Treatment diffusion: active paths]{Active treatment diffusion paths (\emph{orange} colored edges) in an undirected network}
		\label{fig: diffpaths}
	\end{figure}

	Figure \ref{fig: diffpaths} shows an example of the "active diffusion paths" in an undirected network, that is paths on which the treatment diffusion process might actually occur.  
	The above assumption also states that the event that any untreated nodes switches its status due to the diffusion process occurs independently on the same event affecting any other individual in the network (included neighboring individuals). By assuming independence, we are able to model the treatment diffusion process as a result of independent events. In particular, we model the individual probability of gaining a tangible exposure to the active treatment by adopting a simplified version of the \emph{Independent Cascade Model} \citep{kempe2003maximizing,kempe2005influential,saito2008prediction,wang2012scalable,liu2019influence}, where diffusion may occur in only one time interval by means of a constant probability of infection, treated units are all active spreaders of the treatment and untreated nodes who have treated neighbors are all susceptible to receive the treatment. The possibility of a node to receive the treatment by second-order neighbors is ruled out. 

	Assumptions \ref{ass: single} and \ref{ass: diffusion} characterize the hidden treatment diffusion process, by simplifying the structure of the process and by imposing a plausible statistical model. 
	Note that these assumptions are plausible in a wide ensemble of real-world settings and they can be easily generalized. In addition, alternative and more flexible specifications might be employed if either data provide additional information or the researcher has a deep a priori knowledge on the diffusion phenomenon of interest.   
	%
	%

	\subsection{The Diffusion Process: Preliminaries \label{subsec: dp-prelim}}
	According to the treatment status at time $t$ and the number of treated in-neighbors, we can analyze the possible treatment status, which affects units after the treatment diffusion process. Specifically, we identify the set of \emph{surely treated} nodes at time $t'$,  
	that coincides with the set $\mathcal{T}_{t}$ of units that have been randomly assigned to treatment at time $t$ (see Assumption \ref{ass: initiallytr}), and the set of \emph{surely untreated} nodes at time $t'$, that is the nodes who have been assigned to control at time $t$ and have a zero probability to receive treatment through diffusion because they have not treated nodes in their in-neighborhood.  
	%
	%
	Therefore,  the vector $\mathbf{Z}_{t'}$ of treatment assignments at time $t'$ is partially unknown. Specifically, we know the treatment status of units who are surely treated and surely untreated, with $Z_{it'}=Z_{it}$ , but we do not know whether those who are eligible to gain the treatment by diffusion have actually received it. 
	
	We can express the conditional probability of being treated at time $t'$, given the treatment vector at time $t$ in the rest of the network, i.e. $\mathbf{Z}_{-it}$, and given the graph $\mathbf{G}$, 
	%
	%
	by means of the law of total probabilities, that is
	\begin{align*}
		\pi_{it'}(1; \mathbf{Z}_{-it}, \mathbf{G})&=P(Z_{it'}=1|\mathbf{Z}_{-it}, \mathbf{G})\\ &=P(Z_{it'}=1|Z_{it}=1,\mathbf{Z}_{-it}, \mathbf{G})P(Z_{it}=1|\mathbf{Z}_{-it}, \mathbf{G}) + P(Z_{it'}=1|Z_{it}=0,\mathbf{Z}_{-it},\mathbf{G})P(Z_{it}=0|\mathbf{Z}_{-it},\mathbf{G}))\\
		&=\pi_{it}(1;\mathbf{Z}_{-it},\mathbf{G}) + \rho_i(\mathbf{Z}_{-it},\mathbf{G})(1-\pi_{it}(1;\mathbf{Z}_{-it},\mathbf{G})),
	\end{align*}
	where $\pi_{it}(1;\mathbf{Z}_{-it},\mathbf{G})$ is the conditional probability that unit $i$ is initially  assigned to the treatment, given the treatment vector $\mathbf{Z}_{-it}$ of the other units and the graph $\mathbf{G}$,   
	and $\rho_i(\mathbf{Z}_{-it}, \mathbf{G})=P(Z_{it'}=1|Z_{it}=0,\boldsymbol{Z}_{-it}, \mathbf{G})$ is the conditional probability for unit $i$ of receiving the treatment by diffusion given that unit $i$ was not assigned to the active treatment at time $t$, given the initial treatment assignment vector $\mathbf{Z}_{-it}$ for all units but $i$ and given $\mathbf{G}$. Moreover, the overall individual probability of being treated at time $t'$, given the graph $\mathbf{G}$, is 
	$$
	\begin{aligned}
		\pi_{it'}(1;\mathbf{G})& =P(Z_{it'}=1|\mathbf{G})= P(Z_{it'}=1|Z_{it}=1,\mathbf{G})P(Z_{it}=1|\mathbf{G})+ P(Z_{it'}=1|Z_{it}=0,\mathbf{G})P(Z_{it}=0|\mathbf{G})\\
		& = \pi_{it}(1;\mathbf{G}) + P(Z_{it'}=1|Z_{it}=0,\mathbf{G})(1-\pi_{it}(1;\mathbf{G})) 
		= \pi_{it}(1;\mathbf{G}) + \mathbb{E}[\rho_i(\mathbf{Z}_{-it},\mathbf{G})|Z_{it}=0,\mathbf{G}](1-\pi_{it}(1;\mathbf{G})),
	\end{aligned}
	$$
	where $\pi_{it}(1,\mathbf{G})$ is the conditional probability that unit $i$ is initially  assigned to the treatment, given the graph $\mathbf{G}$, and $\mathbb{E}[\rho_i(\mathbf{Z}_{-it},\mathbf{G})|Z_{it}=0, \mathbf{G}]=\sum_{\mathbf{z}_{-i}}
	\rho_i(\mathbf{z}_{-i},\mathbf{G})P(\mathbf{Z}_{-it}=\mathbf{z}_{-i}|Z_{it}=0,\mathbf{G})$.
	%
	%
	\\
	\indent In our setting, by Assumption \ref{ass: diffusion}, we have  
	$$
	\begin{aligned}
		\rho_{i}&=\rho_i(\mathbf{Z}_{-it},\mathbf{G})=
		P(Z_{it'}=1|Z_{it}=0, \boldsymbol{Z}_{-it}, \mathbf{G}) =P(Z_{it'}=1|Z_{it}=0,T_{it})\\&=1-(1-\overline{p})^{T_{it}}
		\in [0,1)\quad 
		(\mbox{since } \overline{p}<1\;\mbox{by assumption}),
	\end{aligned}
	$$
	where $T_{it}=T_{it}(\mathbf{Z}_{-it},\mathbf{G})=\sum_{j\in \mathcal{N}_i^{in}} Z_{jt}$ is 
	the number of treated in-neighbors of $i$ in $\mathbf{G}$ at time $t$. 
	Therefore, the post-diffusion treatment vector $\mathbf{Z}_t'$ depends on the initial treatment assignment vector $\mathbf{Z}_t$, on the network $\mathbf{G}$ and on the diffusion parameter $\overline{p}$. In particular, the probability $\rho_i$ strongly depends on the number of in-edges unit $i$ has in the underlying network structure $\mathbf{G}$, i.e., $N^{in}$. 
	
	\begin{remark} (Bernoulli Trial) \rm 
		In a Bernoulli randomized experiments units are randomly assigned to treatment independently of each other (and independently of the network $\mathbf{G}$) with probability $0<\pi_{it}(1)<1$.  Hence, $\mathbf{Z}_{t}$ is a vector of independent random variables, independent of the graph  $\mathbf{G}$ and we have  $\pi_{it}(1;\mathbf{Z}_{-it},\mathbf{G})=\pi_{it}(1;\mathbf{G})=\pi_{it}(1)$. 
	\end{remark}
	
	\begin{remark}(Cluster Randomized Experiments) \rm In a cluster randomized experiments clusters are randomly (independently of the network $\mathbf{G}$) assigned to either treatment or control and all units in the same cluster are exposed to the same treatment. In this case, $Z_{it}$ does depend on $\mathbf{Z}_{-it}$, but is independent of the graph $\mathbf{G}$. Therefore, we have
		$\pi_{it}(1;\mathbf{Z}_{-it},\mathbf{G})=\pi_{it}(1;\mathbf{Z}_{-it})$ and $\pi_{it}(1;\mathbf{G})=\pi_{it}(1)\in (0,1)$.
	\end{remark}
	
	In the following we will omit to write explicitly the dependence on $\mathbf{G}$ when it is not necessary.
	
	%
	%
	
	\subsection{Potential Outcomes and Causal Effects}
	Following the \emph{Rubin Causal Model} (RCM) \citep{rubin1974estimating}, we outline the causal effects, under the potential outcomes framework.  
	In principle, we should define the outcome as a function of the whole treatment vector at time $t$ and the whole treatment vector at time $t'$, i.e., $Y_{it''}(\mathbf{Z}_{t}, \mathbf{Z}_{t'})$. To this regard, we make three important assumptions: i) the treatment status of unit $i$ at time $t$ has no effect on their outcome at time $t''$, if not through their treatment status at time $t'$; ii) the treatment status of other units at time $t$ has no effect on the outcome of unit $i$ at time $t''$, if not through the treatment status of unit $i$ at time $t'$; iii) the outcome of unit $i$ at time $t''$ is not affected by the treatment status of other units at time $t'$. 
	These assumptions can be formally expressed as follows: 
	
	\begin{assumption}[Exclusion Restriction and No-interference of other units' treatment at time $t'$]
		\label{ass: interference_diff}
		Given two different assignment treatment vectors $\mathbf{Z}_{t}$ and $\mathbf{Z}'_{t}$, resulting in the same treatment status at time $t'$ for unit $i$, i.e., $Z_{it'}=Z'_{it'}$, but different treatment status for some of the other units, i.e., $Z_{kt'}\neq Z'_{kt'}$ for some $k\neq i$, then $Y_{it''}(\mathbf{Z}_{t}, \mathbf{Z}_{t'})=Y_{it''}(\mathbf{Z}'_{t}, \mathbf{Z}'_{t'}). $
	\end{assumption}
	This means that there is no interference between units if not trough the actual diffusion process. In addition, the treatment assignment at time $t$ does not have any direct effect on the outcome if not through the treatment at time $t'$, that is, the effect of having the treatment at time $t'$ does not depend on whether the treatment was originally assigned at time $t$ or it has been received through diffusion from others.
	Under Assumption \ref{ass: interference_diff}, we can index the outcome for unit $i$ by the treatment status of unit $i$ at post-diffusion time $t'$, that is  $Y_{it''}(\mathbf{Z}_{t}, \mathbf{Z}_{t'})=Y_{it''}(Z_{it'})$. Hence, we postulate the existence of two potential outcomes for each unit, $Y_{it''}(Z_{it'}=0)$ and $Y_{it''}(Z_{it'}=1)$, representing the potential outcome that would be observed for unit $i$ at time $t''$ under control and under active treatment (directly or indirectly  received), respectively. Throughout, we will use the simplified notation $Y_{it''}(z)$ for $Y_{it''}(Z_{it'}=z)$. As a consequence, the observed outcome can be expressed by
	\begin{equation} 
		Y_{it''}= Y_{it''}(Z_{it'}=z)=\begin{cases}  Y_{it''}(0) & \text{if} \;\;\;\;  z=0, \\  Y_{it''}(1) & \text{if} \;\;\;\;  z=1.
		\end{cases}
	\end{equation}  
	
	Moreover, according to the assumed diffusion process, we make the following assumption:
	\begin{assumption}[Unconfoundedness]
		\label{ass:unc}
		$$
		(i)\qquad Y_{it''}(Z_{it'}=z)\independent [\mathbf{Z}_{t},\mathbf{G}] \qquad \forall z\in \{0,1\}
		$$
		$$
		(ii)\qquad Y_{it''}(Z_{it'}=z)\independent{Z_{it'}} \,|\, \mathbf{Z}_{-it},\, Z_{it}=0,\,\mathbf{G}
		\qquad \forall z\in\{0,1\}.
		$$
	\end{assumption}
	The first sub-assumption (i) simply reflects the randomization of the initial treatment assignment, while the second sub-assumption (ii) states that, for any unit $i$ untreated at time $t$, the treatment at time $t'$ is unconfounded given the treatment vector  of the other units at time $t$ and given $\mathbf{G}$. 
	
	Under Assumption \ref{ass: interference_diff}, it is possible to define the Average Treatment Effect (ATE) as follows:
	\begin{equation}
		\tau^{\star}=\mathbb{E}\:\big[Y_{it''}(Z_{it'}=1)\:\big]-\mathbb{E}\:\big[Y_{it''}(Z_{it'}=0)\:\big]=
		\mathbb{E}[Y_{it''}(1)] - \mathbb{E}[Y_{it''}(0)].
	\end{equation}
	where the expectation is taken over the sampling distribution, under the super-population perspective \cite{imbens2015causal}.
	The quantity $  \tau^{\star}$ is a comparison of potential outcomes under the actual (but hidden) treatment status $Z_{it'}$ and it represents the causal effect of receiving the treatment either directly through the initial assignment or indirectly through diffusion. 

	\subsection{Bias Analysis when Diffusion is Neglected \label{subsec: trd_biasdir} }
	Ignoring the diffusion process, when present, may introduce a bias in the estimate of the treatment effect. Here, we show the formula of the bias (proofs are collected in Appendix \ref{app: bias}). We first provide the general formula and then we illustrate how it simplifies under Assumption \ref{ass:unc}. We also discuss the direction of the bias by investigating the settings which that generate either an underestimation or an overestimation of the causal treatment effect.  
	
	Given the hidden nature of the treatment diffusion process, one would be tempted to neglect any diffusion mechanism, even if plausible, and estimate the treatment effect relying on the initial treatment assignment.
	Under the assumption of no-diffusion, and in turn, no-interference, the potential outcomes can be indexed by the individual treatment assignment $Y_{it''}(Z_{it}=z)$ and the causal treatment effect is defined as $\tau^{no-diff}=\mathbb{E}\:\big[Y_{it''}(Z_{it}=1)\:\big]-\mathbb{E}\:\big[Y_{it''}(Z_{it}=0)\:\big]$.
	In a randomized experiment, under the assumption of no-diffusion, the treatment effect $\tau^{no-diff}$ is identified by 
	\begin{equation}
		\label{eq:itt}
		\tau^{no-diff}=\mathbb{E}\:\big[Y_{it''}|Z_{it}=1\:\big]-\mathbb{E}\:\big[Y_{it''}|Z_{it}=0\:\big].
	\end{equation}
	and one one would estimate this quantity to $\tau^{no-diff}$ to assess the effect of receiving the treatment. However, if a treatment diffusion arises, $ \tau^{no-diff}$ does not represent the effect of receiving the treatment, as the initial treatment allocation may have been altered by the  spread of the treatment. 
	
	We define the bias $b$ of the na\"ive approach that neglects diffusion  as the difference between these two quantities, that is $b = \tau^{no-diff}-\tau^{\star}$. This is the difference between the two quantities targeted under either the assumption of no-diffusion and, in turn, no-interference, or taking into account a possible diffusion of the treatment\footnote{Note that this bias is not affected by the potential bias of the estimators used to estimate these quantities.}.

	As proven in Appendix \ref{app: bias}, the bias is expressed by the quantity $b$, where
	\begin{equation}\label{eq:bias-0}
		\begin{split}
			\small
			b&=
			\tau^{no-diff}-\tau^{\star} \\
			&=  \mathbb{E} \; \Big [ Y_{it''}(0) \Big ] -\mathbb{E} \; \Big [ Y_{it''}(0)|Z_{it'}=0, Z_{it}=0 \Big ] \left(1-\mathbb{E}\Big[\rho_i |Z_{it}=0 \Big]\right)- 
			\\ & \;\;\;\;\;\;
			\mathbb{E} \; \Big [ Y_{it''}(1)|Z_{it'}=1, Z_{it}=0 \Big ] \mathbb{E}\Big[\rho_i|Z_{it}=0 \Big]+
			\\ & \;\;\;\;\;\;
			\mathbb{E}\; \Big [Y_{it''}(1)|Z_{it}=1\Big]-\mathbb{E}\;\Big [Y_{it''}(1)\Big]\,,
		\end{split}
	\end{equation}
	where $\mathbb{E}[\rho_i|Z_{it}=0]=P(Z_{it'}=1|Z_{it}=0)$.
	%
	%
	If $b>0$, neglecting the diffusion process implies an overestimation of the real effect of the treatment and the intervention appears to be more effective than it actually is. On the contrary, if $b<0$ our analysis is affected by underestimation and the treatment appears less effective. 
	
	\indent  Under randomization of the initial treatment (Assumption \ref{ass:unc}, Part (i)), we have $\mathbb{E}[Y_{it''}(1)|Z_{it}=1]=\mathbb{E}[Y_{it''}(1)]$ and so the  above quantity $b$ simply becomes \begin{equation}\label{eq:bias}
		\begin{split}
			\small
			b=  \mathbb{E} \; \Big [ Y_{it''}(0) \Big ] -\mathbb{E} \; \Big [ Y_{it''}(0)|Z_{it'}=0, Z_{it}=0 \Big ] \left(1-\mathbb{E}[\rho_i |Z_{it}=0]\right)- 
			\mathbb{E} \; \Big [ Y_{it''}(1)|Z_{it'}=1, Z_{it}=0 \Big ] \mathbb{E}[\rho_i|Z_{it}=0 ]\,. 
		\end{split}
	\end{equation}
	Clearly, in the absence of the diffusion process, i.e.,  $\rho_i=0$ for all $i$, 
	we have $b=0$. This is because we have 
	$Z_{it'}=0\Leftrightarrow Z_{it}=0$ and so $\mathbb{E}[Y_{it''}(0)|Z_{it'}=0,Z_{it}=0]=\mathbb{E}[Y_{it''}(0)|Z_{it}=0]=\mathbb{E}[Y_{it''}(0)]$
	. Moreover, we note that, if $\mathbb{E} \; \Big [ Y_{it''}(0)\Big ]=0$ and 
	$\mathbb{E} \; \Big [ Y_{it''}(0)|Z_{it}=0, Z_{it'}=0 \Big ]=0$, then: 
	\begin{itemize}
		\item  $\mathbb{E} \; \Big [ Y_{it''}(1)|Z_{it'}=1, Z_{it}=0 \Big ]<0$ leads to $b>0$, that is, an overestimation of the real effect;
		\item $\mathbb{E} \; \Big [ Y_{it''}(1)|Z_{it'}=1, Z_{it}=0 \Big ]>0$ leads to $b<0$, that is, an underestimation of the real effect.
	\end{itemize}
	This means that, in this case, the direction of the treatment diffusion bias is driven by the sign of the average potential outcome under the active intervention on those units who actually receive the treatment by diffusion: if this quantity is lower than zero, then units who receive the treatment by diffusion do not really benefit from the intervention and ignoring the treatment diffusion process lead to overestimate the real impact of the treatment; if, vice versa, this quantity is greater than zero, then agents who were not initially provided by the intervention but who gain the treatment from their treated peers positively respond to the intervention and ignoring diffusion leads to underestimate the overall impact of the treatment. 
	
	Finally, we observe that, if the entire Assumption \ref{ass:unc} (first and second part) is satisfied, then we have  $\mathbb{E} \; \Big [ Y_{it''}(0)|Z_{it}=0, Z_{it'}=0 \Big ]=\mathbb{E}[Y_{it''}(0)]$ and  
	$\mathbb{E} \; \Big [ Y_{it''}(1)|Z_{it'}=1, Z_{it}=0 \Big ]=\mathbb{E}[Y_{it''}(1)]$ (for the proof, see Appendix \ref{app: bias}) and so we get 
	\begin{equation}\label{eq:bias-unc}
		b=\left(\mathbb{E}[Y_{it''}(0)]-\mathbb{E}[Y_{it''}(1)]\right)\mathbb{E}[\rho_i|Z_{it}=0 ]
		= - \tau^\star\mathbb{E}[\rho_i|Z_{it}=0 ] .
	\end{equation} 
	Hence,  the sign of the bias $b$ mostly depends on the sign of the average treatment effect of those units who have a greater probability of receiving the treatment by diffusion: specifically, if those units have a positive treatment effect, then, neglecting the possible diffusion may lead to an underestimation of the overall effect; while, if those units have a negative mean treatment effect, then, neglecting the possible diffusion may lead to an overestimation of the overall effect. This is the finding that has inspired the simulations in Section \ref{sec: simul}.

	\subsection{Horvitz-Thomson Estimators 
	}
	\label{subsec: trd_estim}
	In this section, we present the proposed estimators for the average treatment effect under no-diffusion, i.e. $\tau^{no-diff}$ (identified by $\tau^{b}_{obs}$), and the average effect under diffusion, i.e.  $\tau^{\star}$.
	We propose estimators for both Bernoulli randomized designs, where each unit receives the treatment or not, independently of the other units, according to an individual treatment assignment probability, and cluster randomized experiments, where clusters are randomly assigned to treatment and all units of the same cluster receive the same treatment. 
	
	The quantity $\tau_{obs}^b$ can be estimated using a difference-in-means estimator or the standard Horvitz-Thomson estimator \citep{horvitz1952generalization} for randomized experiments, based on the initial treatment assignments $\mathbf{Z}_t$ (see Section \ref{subsec: trd_sens} below).
	For instance, the Horvitz-Thomson estimator for unit-level randomized experiments is given by:
	\begin{equation}
		\label{eq: HTnodiff_Ber}
		\widehat{\tau}_{obs}^{b}=
		\frac{1}{N}\left[\sum_{i=1}^N Z_{it} \frac{Y_{it''}}{\pi_{it}(1)} - \sum_{i=1}^N (1-Z_{it}) \frac{Y_{it''}}{1-\pi_{it}(1)}\right]\,,
	\end{equation}
	where $\pi_{it}(1)\in (0,1)$ denotes the probability of unit $i$ to be initially assigned to the treatment group. A similar Horvitz-Thomson estimator has been developed for cluster randomized experiments \cite{aronow2013class}.
	However, $\widehat{\tau}_{obs}^b$ is a biased estimate of the real treatment effect $\tau^{\star}$. 
	\footnote{
		$\widehat{\tau}_{obs}^b$ cannot be employed even for estimating an intent-to-treat effect, that is, the effect of the treatment assignment. Instead, in order to estimate the causal effect of being assigned to treatment, one would need to take into account interference and consider the treatment assigned to other individuals \cite{hudgens2008toward, aronow2017estimating}. In a general network setting, this could be hard, because treatment diffusion does not allow us to restrict interference to the neighboring units. 
		Furthermore, under interference due to treatment diffusion, the average direct effect of being assigned to treatment is defined by keeping treatment assignment vector of others fixed or marginalizing over its distribution. Therefore, this does not represent the effect of receiving the treatment, regardless of others. 
	}
	
	
	The only way to explicitly account for treatment diffusion process is to consider the real (but unknown) treatment vector $Z_{it'}$, evaluated after the diffusion process. If the post-diffusion treatment assignment $Z_{it'}$ were observed for all units the causal effect $\tau^{\star}$ would be estimated from data.
	However, one would need to control for the fact that the actual treatment receipt observed after diffusion is no longer randomized. In a Bernoulli randomized setting, we propose 
	%
	%
	the following "Horvitz-Thomson" estimator  
	\citep{horvitz1952generalization}
	\begin{equation}
		\begin{aligned}
			\label{eq: HTdiff_Ber}
			\widehat{\tau}^{\star}_1 &= \frac{1}{N} \:\Big[ \sum_{i=1}^{N}Z_{it'}\frac{Y_{it''}}{\pi_{it'}(1; \mathbf{Z}_{-it},\mathbf{G})}-\sum_{i=1}^{N}(1-Z_{it'})\frac{Y_{it''}}{1-\pi_{it'}(1; \mathbf{Z}_{-it},\mathbf{G})}\:\Big]\\
			&=\frac{1}{N} \:\Big[ \sum_{i=1}^{N}Z_{it'}\frac{Y_{it''}}{\pi_{it}(1)+\rho_i(\mathbf{Z}_{-it},\mathbf{G}) (1- \pi_{it}(1))}-\sum_{i=1}^{N}(1-Z_{it'})\frac{Y_{it''}}{(1-\pi_{it}(1))(1-\rho_i(\mathbf{Z}_{-it},\mathbf{G}))}\:\Big].
		\end{aligned}
	\end{equation}
	where $\pi_{it'}(1; \mathbf{Z}_{-it},\mathbf{G})$ is the conditional probability for $i$ of receiving the treatment at time $t'$ conditional on the initial treatment vector $\mathbf{Z}_{-it}$ of the other units and the graph $\mathbf{G}$. Indeed, in a Bernoulli randomized setting, when all the random variables $Z_{it}$, with $i\in\mathcal{N}$, are independent and independent of $\mathbf{G}$, the conditional probability for $i$ of receiving the treatment at time $t'$,  conditional on the initial treatment vector $\mathbf{Z}_{-it}$ of the other units and $\mathbf{G}$, is (see Subsec. \ref{subsec: dp-prelim}) 
	\begin{align*}
		\pi_{it'}(1; \mathbf{Z}_{-it}, \mathbf{G})&=
		\pi_{it}(1; \mathbf{Z}_{-it},\mathbf{G})+
		\rho_i(\mathbf{Z}_{-it},\mathbf{G}) \left(1- \pi_{it}(1; \mathbf{Z}_{-it},\mathbf{G})\right)\\
		&(\mbox{when } Z_{it} \;\mbox{is independent of } \mathbf{Z}_{-it} \;\mbox{and } \mathbf{G})\\
		&=
		\pi_{it}(1)+\rho_i(\mathbf{Z}_{-it},\mathbf{G}) \left(1- \pi_{it}(1)\right)\in (0,1)
	\end{align*}
	(since $\pi_{it}(1)\in (0,1)$ and $\rho_i<1$).  
	In Appendix \ref{app:estimators_indep} we prove that the above estimator is unbiased under Assumption \ref{ass:unc}. 
	
	
	In a cluster randomized experiment, the treatment assignment of one unit depends on the treatment assignment of other units. 
	In particular, $\pi_{it}(1; \mathbf{Z}_{-it},\mathbf{G})$ will always be 0 or 1 and the denominator of the Horvitz-Thomson estimator, $\pi_{it'}(1; \mathbf{Z}_{-it}, \mathbf{G})$, will always be 1 for units in the treated clusters and  0 for isolated units in the control clusters, resulting in a biased estimator of ${\tau}^{\star}$.
	However, we can replace it with an alternative estimator, $\widehat{\tau}^{\star}_2$, which 
	can be used for cluster randomized designs. The alternative estimator can be written as  
	\begin{align}
		\label{eq: HT_diff_cr}
		\widehat{\tau}^{\star}_2 = 
		\frac{1}{N} \:\Big[ \sum_{i=1}^{N}Z_{it'}\frac{Y_{it''}}{\pi_{it'}(1;\mathbf{G})}-
		\sum_{i=1}^{N}(1-Z_{it'})\frac{Y_{it''}}{1-\pi_{it'}(1;\mathbf{G})}\:\Big]\,
	\end{align}
	where, under the independence between $Z_{it}$ and the graph $\mathbf{G}$, we have (see Subsec. \ref{subsec: dp-prelim}) $\pi_{it'}(1;\mathbf{G})=\pi_{it}(1)+\mathbb{E}[\rho_i|Z_{it}=0,\mathbf{G}](1-\pi_{it}(1))$, that (since $\pi_{it}(1)\in (0,1)$ and $\rho_i<1$) belongs to $(0,1)$. In Appendix \ref{app:estimators_clust} we prove that the estimator $\widehat{\tau}^{\star}_2$ is unbiased under Assumption \ref{ass:unc}.
	%
	%
	The computation of $\mathbb{E}[\rho_i|Z_{it}=0,\mathbf{G}]$
	can be challenging, but
	we can replace it by
	its empirical mean (simulating a set of possible scenarios of initial treatments, see Appendix \ref{app: techstarstar} for the computation of $\widehat{\tau}^{\star}_2$ specifically in the proposed empirical setting).
	The proposed estimator performs well in reducing the estimation bias due to treatment diffusion in cluster randomized settings (see Appendix \ref{app: simul_clust} for more details).

	\subsection{Sensitivity Analysis for Treatment Effect in the presence of an Unknown Diffusion Process \label{subsec: trd_sens}}
	We propose here a sensitivity analysis  for the unobserved treatment diffusion process, with the aim of assessing the degree of sensitivity of the na\"ive estimates $\widehat{\tau}_{obs}^b$ of the treatment effect.
	In most settings, the treatment spread is completely unobserved and the diffusion process is unknown. However, sometimes we do have some full or partial information on the relationships between units. In fact, some studies might collect social interactions among participants. Alternatively, oftentimes we have geographic information on participants or some information on the social structure (e.g. schools and classes, social groups, ...). Knowledge of the social network $\mathbf G$ could be used to perform a sensitivity analysis for treatment diffusion. The sensitivity analysis developed in the present work relies on the information of the network to predict treatment diffusion scenarios that might have plausibly occurred.
	The key idea is to simulate a set of diffusion scenarios and compare estimates of the treatment effect accounting for diffusion with the na\"ive estimates under the assumption of no-diffusion. This sensitivity analysis will allow us to assess whether ignoring treatment diffusion would lead to an overestimate or underestimate of the treatment effect or whether any hidden diffusion process would not have a significant impact on results. 
	
	We distinguish between two settings depending on whether the social network is fully or partially observed.
	%
	%
	
	\subsubsection{Fully Observed Network}
	
	The strategy that we propose consists in the following steps: 
	\begin{enumerate}
		\item \emph{Na\"ive estimates under the assumption of no-diffusion}.  Estimate the treatment effect under the assumption of no-diffusion $\widehat{\tau}_{obs}^b$
		by means of a suitable estimator, such as 
		\ref{eq: HTnodiff_Ber} for unit-level randomized experiments (or the one proposed by \citet{aronow2013class} for cluster randomized experiments),
		and compute also the corresponding estimated standard error $\widehat{\sigma}(\widehat{\tau}_{obs}^{b})$ \footnote{Estimated standard errors are computed by accounting for on asymptotic expansions, as motivated in \cite{tan2006distributional,tan2010bounded,tan2013variance}}.
		\item \emph{Dealing with the unknown diffusion process}. Let $\overline{p}$ be the diffusion parameter, that is the probability that a single treated unit passes the treatment to an untreated out-neighbor. If no a-priori knowledge about this parameter is available, we will let $\overline{p}$ vary over a grid of $P$ values, thus $\overline{p} \in \overline{\mathcal{P}}=\{\overline{p}_{1},\dots,\overline{p}_{P}\}$. For each fixed $\overline{p} \in \overline{\mathcal{P}}$:
		\begin{enumerate}
			\item Consider given the network $\mathbf G$ (hence, in the following, we omit to write explicitly the dependence on $\mathbf G$) and:
			\begin{itemize}
				\item Compute for each node the number of treated neighbors $T_{it}$.
				\item Compute the elements of the $N$-dimensional vector $\boldsymbol{\rho}^{\overline{p}}= (\rho^{\overline{p}}_1, \dots, \rho^{\overline{p}}_N)$, where each element $\rho_{i}^{\overline{p}}=\rho_i(\mathbf{Z}_{-it},\overline{p})$ represents the unit-level conditional probability to switch status due to the diffusion process, according to the fixed probability $\overline{p}$.
				\item Using the vector $\boldsymbol{\rho}^{\overline{p}}$, compute the unit level conditional probability of being exposed to the active treatment at time $t'$, say $\pi^{\overline{p}}_{it'}$,  needed for the computation of the estimator  
				$\widehat{\tau}_1^\star$ (Equation \ref{eq: HTdiff_Ber}) or $\widehat{\tau}^{\star}_2$ (Equation \ref{eq: HT_diff_cr}).
				\item The individual treatment status at time $t'$, for those units who have been initially assigned to the control group, say $Z^{\overline{p}}_{1t'|Z_{it}=0}$, is obtained by sampling from a Bernoulli distribution with parameter $\rho_i^{\overline{p}}$, that is 
				$
				Z^{\overline{p}}_{it'|Z_{it}=0}\sim Bernoulli(\rho_i^{\overline{p}}).
				$
				Initially treated units will remain treated, that is $Z^{\overline{p}}_{it'|Z_{it}=1}=1$.
				\item Sample from $\mathbf{Z}^{\overline{p}}_{t'}$ several times, say $R$, where $r= 1, \dots, R$,  and obtain a certain sample space 
				$\mathcal{Z}^{\overline{p}} = \{\mathbf{Z}^{\overline{p},1}_{t'}, \dots, \mathbf{Z}^{\overline{p},R}_{t'}\}$.  
				In other words, $\mathcal{Z}^{\overline{p}}$ collects $R$ different configurations of treatment assignment vectors at time $t'$, given the fixed diffusion probability $\overline{p}$. For each configuration $\mathbf{Z}^{\overline{p},r}_{t'} \in \mathcal{Z}^{\overline{p}}$: 
				\begin{itemize}
					\item Compute an estimate of the treatment effect $\widehat{\tau}^{\overline{p},r}$  using the "Horvitz-Thompson type" estimator $\widehat{\tau}_1^\star$ (Equation \ref{eq: HTdiff_Ber}) or $\widehat{\tau}^{\star}_2$ (Equation \ref{eq: HT_diff_cr}).
					\item Compute the standard error of the estimate $\widehat{\sigma}(\widehat{\tau}^{\overline{p},r})$.
				\end{itemize}
				\item Obtain an entire set of estimates of the overall effect of the treatment, under the network $\mathbf{G}$ and given a fixed diffusion probability $\overline{p}$:  $\Psi^{\overline{p}} = \{\widehat{\tau}^{\overline{p},r}: r = 1, \dots, R\}$.
			\end{itemize} 
			\item The set $\Psi^{\overline{p}}$ contains all the treatment effect estimates, computed under a fixed probability $\overline{p}$. These estimates evaluate distinct random realizations of the unknown after-diffusion treatment assignment vector. In other terms, the values in $\Psi^{\overline{p}}$ represent an empirical distribution of the estimated effects, under diffusion probability $\overline{p}$. Hence,  we can compute the key quantities that allow us to characterize the distribution of the effects, under the diffusion parameter $\overline{p}$: 
			\begin{itemize}
				\item The average value of estimated effects, under $\overline{p}$, $\overline{\widehat{\boldsymbol{\tau}}^{\overline{p}}}$, where
				$$\overline{\widehat{\boldsymbol{\tau}}^{\overline{p}}}= \overline{\Psi^{\overline{p}}} = \frac{1}{R } \sum_{r=1}^R  \widehat{\tau}^{\overline{p},r}.$$
				\item The total variance in the estimated effects, that results from the sum of two components,  a within variance $s^{2^{(W)}}_{\widehat{\boldsymbol{\tau}}^{\overline{p}}}$ and a between variance $s^{2^{(B)}}_{\widehat{\boldsymbol{\tau}}^{\overline{p}}}$, that is
				$s^{2}_{\widehat{\boldsymbol{\tau}}^{\overline{p}}} = s^{2^{(T)}}_{\widehat{\boldsymbol{\tau}}^{\overline{p}}}= s^{2^{(B)}}_{\widehat{\boldsymbol{\tau}}^{\overline{p}}}+s^{2^{(W)}}_{\widehat{\boldsymbol{\tau}}^{\overline{p}}}$, where
				$$s^{2^{(B)}}_{\widehat{\boldsymbol{\tau}}^{\overline{p}}}  = \frac{1}{R }\sum_{r=1}^R ( \widehat{\tau}^{\overline{p},r} - \overline{\widehat{\boldsymbol{\tau}}^{\overline{p}}})^2\quad\mbox{and}\quad
				s^{2^{(W)}}_{\widehat{\boldsymbol{\tau}}^{\overline{p}}}=\frac{1}{R}\sum_{r=1}^R (\widehat{\sigma}(\widehat{\tau}^{\overline{p},r}))^{2}.$$
				The between variance $s^{2^{(B)}}_{\widehat{\boldsymbol{\tau}}^{\overline{p}}}$ captures the estimates' variance in the $ R$ realization of  the treatment assignment vector at time $t'$. Conversely, the within variance component  $s^{2^{(W)}}_{\widehat{\boldsymbol{\tau}}^{\overline{p}}}$ averages on the estimated standard errors which have been computed within each realization $r$. The composite the variance is introduced to account for both the variability that comes from the sensitivity analysis and the intrinsic variability of the effect.
			\end{itemize}
		\end{enumerate}
		\item The key point of the sensitivity analysis stands in the comparison between the average value of the estimated effects under $\overline{p}$, i.e.  $\overline{\widehat{\boldsymbol{\tau}}^{\overline{p}}}$, together with its corresponding total standard error, i.e. $s_{\widehat{\boldsymbol{\tau}}^{\overline{p}}}$, and the estimated treatment effect obtained while ignoring the possibility of treatment diffusion, i.e. $\widehat{\tau}_{obs}^b$, together with its estimated standard error. For assessing the significance of the estimation bias, due to the diffusion process, one strategy could be to assume the estimated effects to be normally-distributed and to compare the confidence intervals 
		$[\widehat{\tau}_{obs}^{b}- z_{\alpha/2} \widehat{\sigma}(\widehat{\tau}_{obs}^{b}),\widehat{\tau}_{obs}^{b} + z_{\alpha/2} \widehat{\sigma}(\widehat{\tau}_{obs}^{b})]$ and  $[\overline{\widehat{\boldsymbol{\tau}}^{\overline{p}}}- z_{\alpha/2} s_{\widehat{\boldsymbol{\tau}}^{\overline{p}}},\overline{\widehat{\boldsymbol{\tau}}^{\overline{p}}} + z_{\alpha/2}s_{\widehat{\boldsymbol{\tau}}^{\overline{p}}}]$, where $z_{\alpha/2}$ represents the critical value of the Normal distribution, associated to the significance level $(1-\alpha)$.
	\end{enumerate}
	
	Note that the sensitivity analysis that we propose implicitly assumes that
	the hidden diffusion process satisfies the assumptions that we have advanced in Subsection \ref{subsec: trd_fixpr}. Relying on specific parametric assumptions is unusual in a pure sensitivity analysis. However, we call this approach a \emph{sensitivity analysis} as it still allows to assess the robustness of results, with respect to plausible realizations of the treatment diffusion process (as pointed out by recent reviews of sensitivity analysis methodologies  \cite{hamby1994review, iooss2015review}, there exist sensitivity approaches which implicitly rely on specific models).
	
	Moreover, it is worthwhile to note that the above approach can be used also to identify a "critical" threshold for the diffusion parameter: that is, testing various values of $\overline{p}$, it is possible to detect the value above which diffusion has a relevant impact on the results. Therefore, if this threshold is plausible in the specific empirical scenario, then researchers cannot really trust in the no-diffusion estimates and must also keep into account the simulated results in summarizing their findings. If instead the threshold appears to be greater than the reasonable diffusion parameters in the considered empirical framework, then no-diffusion hypothesis can be fairly assumed to be valid.
	
	\subsubsection{Partially Observed Network}
	\label{subsec: ea_sensanal}
	In some real world scenarios, the baseline social network $\mathbf{G}$ is only partially observed. For instance, in our motivating application, it is possible to fully observe friendship ties connecting students enrolled in the same class, but data do not provide information on the inter-class ties. When the information on the network $\mathbf{G}$ is not complete, the sensitivity analysis procedure described above cannot be fairly employed, without any modifications: the whole procedure requires to be slightly reworked, to account for the uncertainty on the network structure. To address the issue of a partially observed network, we use multiple imputation \citep{rubin1996multiple,rubin2004multiple}. The idea is to use observed ties to multiply impute missing links, so to generate an ensemble of $M$ completed networks, which include both observed and predicted ties. We denote by $\mathrm{G}^{m}=(\mathcal{N},\mathcal{E}^{m})$, where $m=\{1,\dots,M \}$, the $m$ completed network. Completed networks are collected in the ensemble $\mathcal{G}$. In this setting, the impact of a possible treatment diffusion process is evaluated on \emph{each} of the $M$ completed networks. Therefore, the sensitivity analysis uses the $M$ completed networks obtained through multiple imputation, but explicitly accounts for the uncertainty on the real network structure. 
	
	Hence, the procedure computes an estimate of the post-diffusion treatment effect under each joint realization of i) the fixed diffusion parameter $\overline{p}$; ii) the reconstructed network $m$ and iii) the given realization of the Bernoulli process $r$. Specifically, the estimate $\widehat{\tau}^{\overline{p},m,r}$ is identified by the triple $(\overline{p},m,r)$ and is obtained by using the "Horvitz-Thompson type" estimator $\widehat{\tau}^{\star}$, together with its estimated variance $\widehat{\sigma}(\widehat{\tau}^{\overline{p},m,r})$.
	
	As a consequence, the average value of estimated effects, under a given diffusion parameter $\overline{p}$, $\overline{\widehat{\boldsymbol{\tau}}^{\overline{p}}}$, results from averaging the $R \times M$ estimated effects under $\overline{p}$. Formally,
	$$\overline{\widehat{\boldsymbol{\tau}}^{\overline{p}}}= \overline{\Psi^{\overline{p}}} = \frac{1}{R \times M} \sum_{m=1}^M \sum_{r=1}^R  \widehat{\tau}^{\overline{p},m,r}.$$
	Similarly, the two components which constitute the total variance computation $s^{2}_{\widehat{\boldsymbol{\tau}}^{\overline{p}}}$, the within variance  $s^{2^{(W)}}_{\widehat{\boldsymbol{\tau}}^{\overline{p}}}$ and a between variance $s^{2^{(B)}}_{\widehat{\boldsymbol{\tau}}^{\overline{p}}}$ such that
	$s^{2}_{\widehat{\boldsymbol{\tau}}^{\overline{p}}} = s^{2^{(T)}}_{\widehat{\boldsymbol{\tau}}^{\overline{p}}}= s^{2^{(B)}}_{\widehat{\boldsymbol{\tau}}^{\overline{p}}}+s^{2^{(W)}}_{\widehat{\boldsymbol{\tau}}^{\overline{p}}}$ can be written as
	
	$$s^{2^{(B)}}_{\widehat{\boldsymbol{\tau}}^{\overline{p}}}  = \frac{1}{R \times M-1}\sum_{m=1}^{M}\sum_{r=1}^R ( \widehat{\tau}^{\overline{p},m,r} - \overline{\widehat{\boldsymbol{\tau}}^{\overline{p}}})^2\quad\mbox{and}\quad
	s^{2^{(W)}}_{\widehat{\boldsymbol{\tau}}^{\overline{p}}}=\frac{1}{R \times M-1}\sum_{m=1}^{M}\sum_{r=1}^R (\widehat{\sigma}(\widehat{\tau}^{\overline{p},m,r}))^{2}.$$
	The between variance $s^{2^{(B)}}_{\widehat{\boldsymbol{\tau}}^{\overline{p}}}$ captures the estimates' variance in the $M \times R$ joint realization of different (imputed) networks and different realization of the treatment assignment vector at time $t'$. Conversely, the within variance component  $s^{2^{(W)}}_{\widehat{\boldsymbol{\tau}}^{\overline{p}}}$ averages on the estimated standard errors which have been computed within each joint realization $m,r$. 
	
	Note that the sensitivity analysis that we propose here implicitly assumes that i) the model for imputing missing links accurately predict hidden relationships and ii) the hidden diffusion process satisfies the assumptions that we have advanced in Subsection \ref{subsec: trd_fixpr}. 
	
	
	\section{Illustrative Simulations}
	\label{sec: simul}
	In this section, we illustrate how the proposed procedure performs in some simulated scenarios. Here we focus on the situation, where the baseline social network $\mathbf{G}$ is fully observed. As extensively motivated in the last section, the sensitivity analysis reconstructs an ensemble of plausible diffusion scenarios, by relying on the assumptions that characterize the diffusion process. Then, it inspects the robustness of results with respect to these reasonable realizations of the process. If the treatment diffusion process actually occurs, the researcher must front a miss-classification in the treatment variable, so that the treatment assignment vector that she actually observes does not truly represent the individual allocation in the two treatment arms. In these illustrative scenarios, we simulate a real diffusion process which introduces a bias in the estimates and leads to either an overestimation or to an underestimation of the treatment effect. As we explain in detail in the upcoming subsection, in these illustrative examples the real effect of intervention is set to be heterogeneous with respect to an individual characteristic -i.e the degree-, which directly affects the unit-level probability of receiving the treatment by diffusion: in such a setting, the sensitivity analysis might be employed not only for assessing the robustness of estimates under several diffusion scenarios, but also for actually reducing the treatment diffusion bias. Hence, once introduced the estimation bias, we observe how the sensitivity analysis performs in shrinking this bias and in towing the estimates towards the right treatment effect. 
	
	\subsection{Data Generating Process (DGP)}
	\label{subsec: simul_dgp}
	In this subsection we detail the data generating process. We consider a sample made up by $N=1000$ units. We generate the structure of interactions by simulating an Erd\H{o}s-R\'{e}nyi random graph \citep{erdos1959random} $\mathbf{G}$ with $N$ nodes and a fixed probability $(0.01)$ to have a link. The initial individual treatment assignment at time $t$, i.e. $Z_{it}$is sampled from independent Bernoulli distributions with probability $0.5$: $Z_{it} \sim Bernoulli(0.5)$. Thus, the individual probability to be treated at time $t$ is $\pi_{it}=\pi_{it}(1)=0.5$. According to the graph $\mathbf{G}$ and to the initial treatment assignment vector $\boldsymbol{Z}_{t}$, we compute the number of treated neighbors $T_{it}$ at time $t$. Then, we randomly generate a treatment diffusion process, which we assume to be the one who truly realizes. The real treatment diffusion process is ruled by the fixed contagion parameter $\overline{p}^{*}$.
	Then, we set (omitting the symbol $\mathbf G$): 
	\begin{equation*}
		\rho_i^*=\rho^{p_i^*}_i(\mathbf{Z}_{-it})= 1-(1-\overline{p}^{*})^{T_{it}}  \qquad\mbox{and}\qquad \pi^{*}_{it'}=\pi^*_{it'}(1;\mathbf{Z}_{-it})=\pi_{it}+ \rho^{*}_{i}(1-\pi_{it}).
	\end{equation*}
	(Note that $\pi_{it}(1;\mathbf{Z}_{-it})=\pi_{it}$ by the initial independence of all the random variables $Z_{it}$, with $i\in\mathcal{N}$.) 
	
	The real individual treatment status at time $t'$, for those units who have been initially assigned to the control group, say $Z^{p^{*}_{i}}_{1t'|Z_{it}=0}$, is obtained by uniquely sampling from a Bernoulli distribution with parameter $ \rho^{*}_{i}$, that is $Z^{p^{*}_{i}}_{it'|Z_{it}=0}\sim Bernoulli( \rho^{*}_{i})$. For those untreated units who have not treated individuals in their neighborhood and those units who have been treated at time $t$ the treatment status remains unchanged (to the control and to the active status, respectively).
	The vector $\boldsymbol{Z}^{p^{*}_{i}}_{it'}$, that, for ease of notation, we also denote by $\boldsymbol{Z}^{*}_{it'}$, represents the real individual treatment status after the treatment diffusion process. 
	
	Once we have generated the treatment diffusion process, we introduce the estimation bias. In the Subsection \ref{subsec: trd_biasdir}, we have discussed the causal mechanisms, which may lead to overestimate or underestimate the real causal effect of the intervention. Following those considerations, we have set the individual potential outcome of the unit $i$ under no exposure to the intervention to be sampled from a Standard Normal Distribution,
	$Y_{it''}(Z^{*}_{it'}=0)=Y_{it''}(0) \sim \mathcal{N}(0,1)$, while the potential outcome under an active exposure to the intervention after the diffusion process includes the individual response to the intervention, say $\tau_{i}^{*}$, that is, $Y_{it''}(Z^{*}_{it'}=1)=Y_{it''}(1)=Y_{it''}(0) + \tau^{*}_{i}$.
	Hence, the observed outcome is
	$$Y_{it''}= Y_{it''}(0)(1-Z^{*}_{it'})+ Y_{it''}(1)Z^{*}_{it'}.$$
	
	The quantity $\tau_{i}^{*}$ depends on a fixed parameter $k$ and it is a function of the indicator $N_i^{high}$, which identifies those units who have an high in-degree: that is, $\tau_{i}^{*}=\tau_i^*(N_i^{high})$. The dummy  $N_i^{high}$ equals 1 if the observed in-degree of unit $i$ is above a given cutoff $\nu$,  ($N^{in}_{i} > \nu$). This cutoff is identified from the in-degree distribution at time $t$: specifically, $\nu$ is the median of the in-degree distribution. (Note that this dummy variable allows us to detect those units who are more likely to experience the diffusion process, that is not affected by individual characteristics and only depends on the degrees of the units.)  
	More precisely, the effect $\tau_{i}^{*}$ is defined as follows. Given that $q_{n}$ denotes the proportion of individuals such that $N_i^{high}=n$, we introduce an \emph{overestimation bias} by setting the real individual treatment effect $\tau_{i}^{*}$ as 
	\begin{equation*}
		\tau_{i}^{*}=\tau_i^*(n)=
		\begin{cases}
			-k ; \qquad \text{if} \qquad N_i^{high}=n=1   \nonumber  \\
			\frac{+k*q_{1}}{q_{0}}; \qquad \text{if} \qquad N_i^{high}=n=0 \,.   \nonumber  \\
		\end{cases}
	\end{equation*}
	Conversely, we introduce an \emph{underestimation bias} setting 
	\begin{equation*}
		\tau_{i}^{*}=\tau_i^*(n)=
		\begin{cases}
			+k ; \qquad \text{if} \qquad N_i^{high}=n=1    \nonumber  \\
			\frac{-k*q_{1}}{q_{0}}; \qquad \text{if} \qquad N_i^{high}=n=0 \,.   \nonumber  \\
		\end{cases}
	\end{equation*}
	Note that in both scenarios the effect is normalized, so that the average effect in the whole population is equal to zero, that is we have $\sum_{i=1}^N \tau_i^*/N=0$ by construction. Also note that the effect is defined so to be heterogeneous with respect to the individual probability of switching the treatment status due to the diffusion process: indeed, this probability is higher for those controls who have an high number of neighbors. If, for instance, we move in the underestimation setting, we are assigning a positive treatment effect at those controls who are more likely to receive the treatment by diffusion. It follows, that we can refer to what said in Section \ref{subsec: trd_biasdir}, formula \eqref{eq:bias-unc}, for the direction of the bias. Consequently, for those units who have a higher probability of receiving the treatment by diffusion, we have:
	\begin{itemize}
		\item  $Y_{it''}(1)-Y_{it''}(0)=\tau_i^*<0$  in the overestimation scenario (that is, $b>0$); while
		\item  $Y_{it''}(1)-Y_{it''}(0)=\tau_i^*<0$ in the underestimation scenario (that is, $b<0$).
	\end{itemize}
	%
	%
	
	At this stage, we can estimate the real treatment effect in the whole population through $ \widehat{\tau}^{\star}_1$. This estimate is primarily compared with the  estimated treatment effect under the assumption of no treatment diffusion, $ \widehat{\tau}_{obs}^{b}$. The difference between these two estimates is the estimated bias $\widehat{b}$. The goal of the simulation design is to see whether the sensitivity analysis is able to reduce the bias and to produce estimates which are closer to the real value of the treatment effect (i.e. 0, recall that the real treatment effect is designed to be $0$ from the DGP).

	\subsection{Results}
	\label{subsec: simul_res}
	Here we present the main simulations' results in the Bernoulli randomized setting. The simulated scenarios differ in terms of i) the size of the effect $k$, ii) the real fixed diffusion probability $\overline{p}^{*}$ and the iii) direction of the bias (overestimation or underestimation). 
	
	We start from the underestimation scenario. Here those units who are more likely to have received the treatment by diffusion exhibit a positive response to the intervention. In each figure, the matrix of plot needs to be read as follows: from the left to the right the plots show the results under increasing values of the real fixed diffusion probability $\overline{p}^{*}$ (specified in the title of the plot). In all the plots are depicted: i) the line corresponding to the estimated treatment effect under no diffusion (blue dotted line), with a related band depicting the $95 \% $ confidence interval; ii) the line corresponding to the estimated treatment effect under the real diffusion process (orange dotted line), together with its corresponding confidence band; iii) the zero line, which signals the real treatment effect generated by the DGP (black line); iv) the box-plots representing the distribution of the estimated treatment effect obtained under the $R$ simulated diffusion processes, which are all ruled by the fixed contagion probability $\overline{p}$ represented on the x-axis (colored box plots); v) the mean estimated treatment effect, under a fixed contagion probability $\overline{p}$ (red dots); vi) the corresponding confidence intervals of these estimates (colored triangles). 
	
\begin{figure}[H]
		\centering
		\includegraphics[width=150mm]{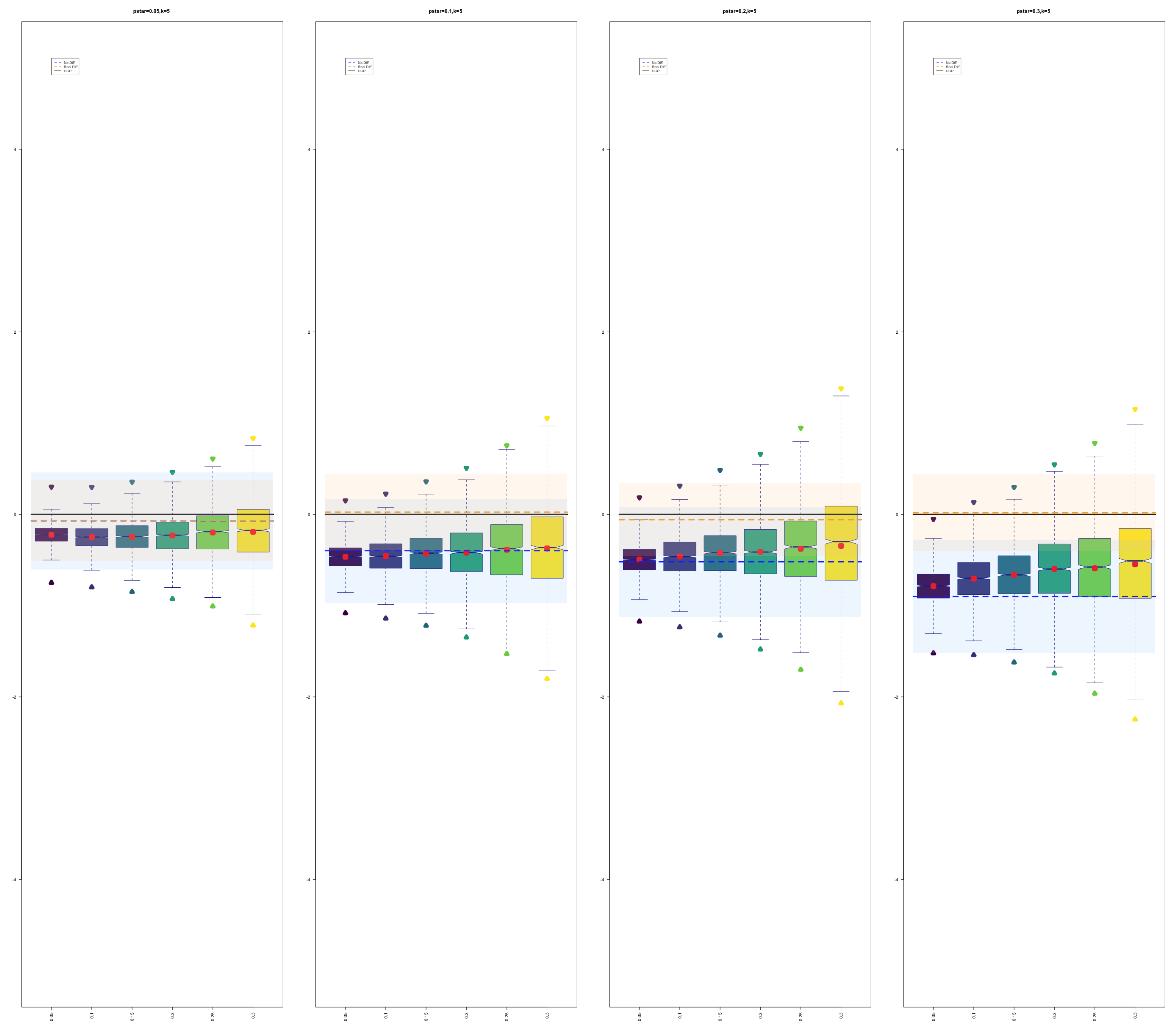}
		\caption[Underestimation, $k=5$]{Underestimation, $k=5$}
		\label{fig: underest_low}
	\end{figure}

\begin{figure}[H]
		\centering
		\includegraphics[width=150mm]{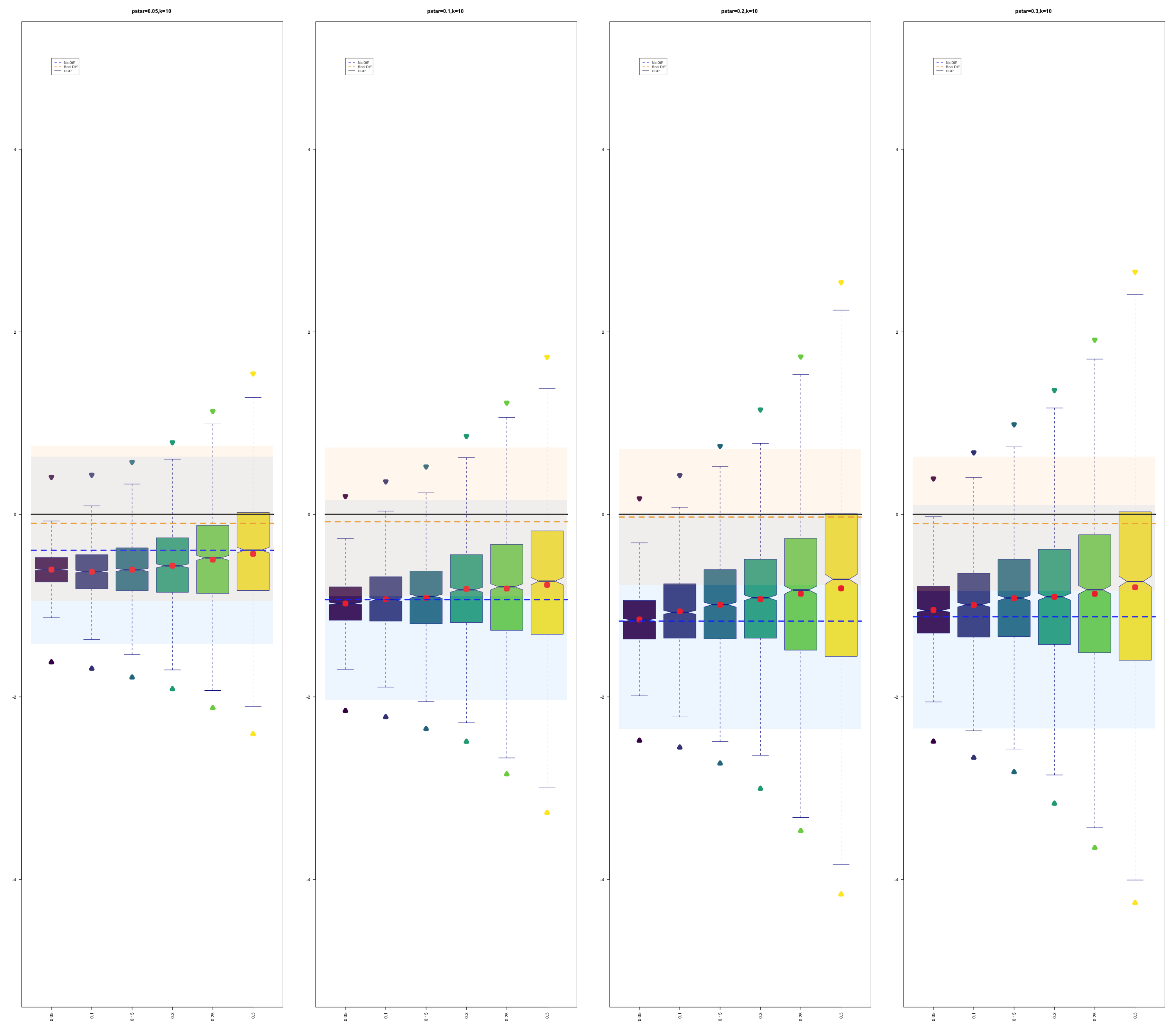}
		\caption[Underestimation, $k=10$]{Underestimation, $k=10$}
		\label{fig: underest_high}
	\end{figure}

	Figure \ref{fig: underest_low} presents the simulations' results in the underestimation setting, where $k=5$. As we may observe in the figure, ignoring the treatment diffusion process leads to an underestimation of the real treatment effect. When the analysis accounts for the possibility of a treatment spreading, the estimation bias gets reduced and the sensitivity analysis leads to estimates, which are closer to the true value 0. Globally, we can state that in all scenarios the sensitivity analysis, when the treatment diffusion process actually happens, performs well in reducing the estimation bias and moves the estimates closer to the real effect. 
	Figure \ref{fig: underest_high} shows the same underestimation scenario, but with the global effect $k$ set to 10. The general conclusions are similar to the ones that we have advanced to the previous scenario. So, we can similarly observe that i) the estimated effect under no diffusion really underestimates the effect of the intervention and that ii) the sensitivity analysis contributes in reducing the estimation bias, moving the estimates towards 0. Moreover, in the presence of an higher overall effect, the estimation bias increases and the sensitivity analysis remains accurate in catching the real treatment effect. Finally, note that the more the general effect $k$ gets higher, the higher is the variability of the estimates. 
	We pass now to the overestimation scenario. In this setting, those units who are more likely of having actually received the treatment through the diffusion process, have a negative treatment effect. 
	
\begin{figure}[H]
		\centering
		\includegraphics[width=0.48\textwidth,width=150mm]{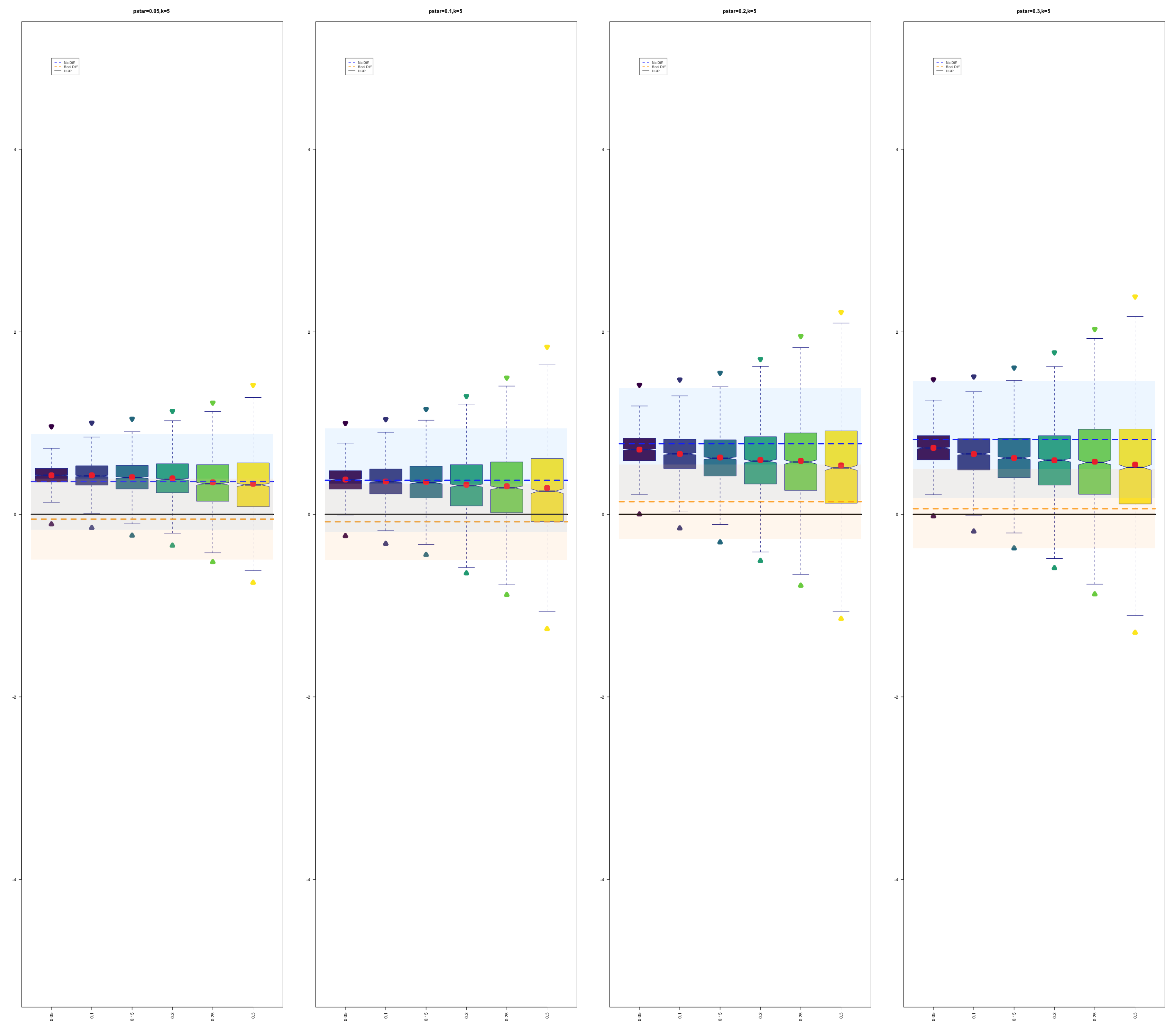}
		\caption[Overestimation, $k=1$]{Overestimation, $k=1$}
		\label{fig: overest_low}
	\end{figure}

\begin{figure}[H]
		\centering
		\includegraphics[width=150mm]{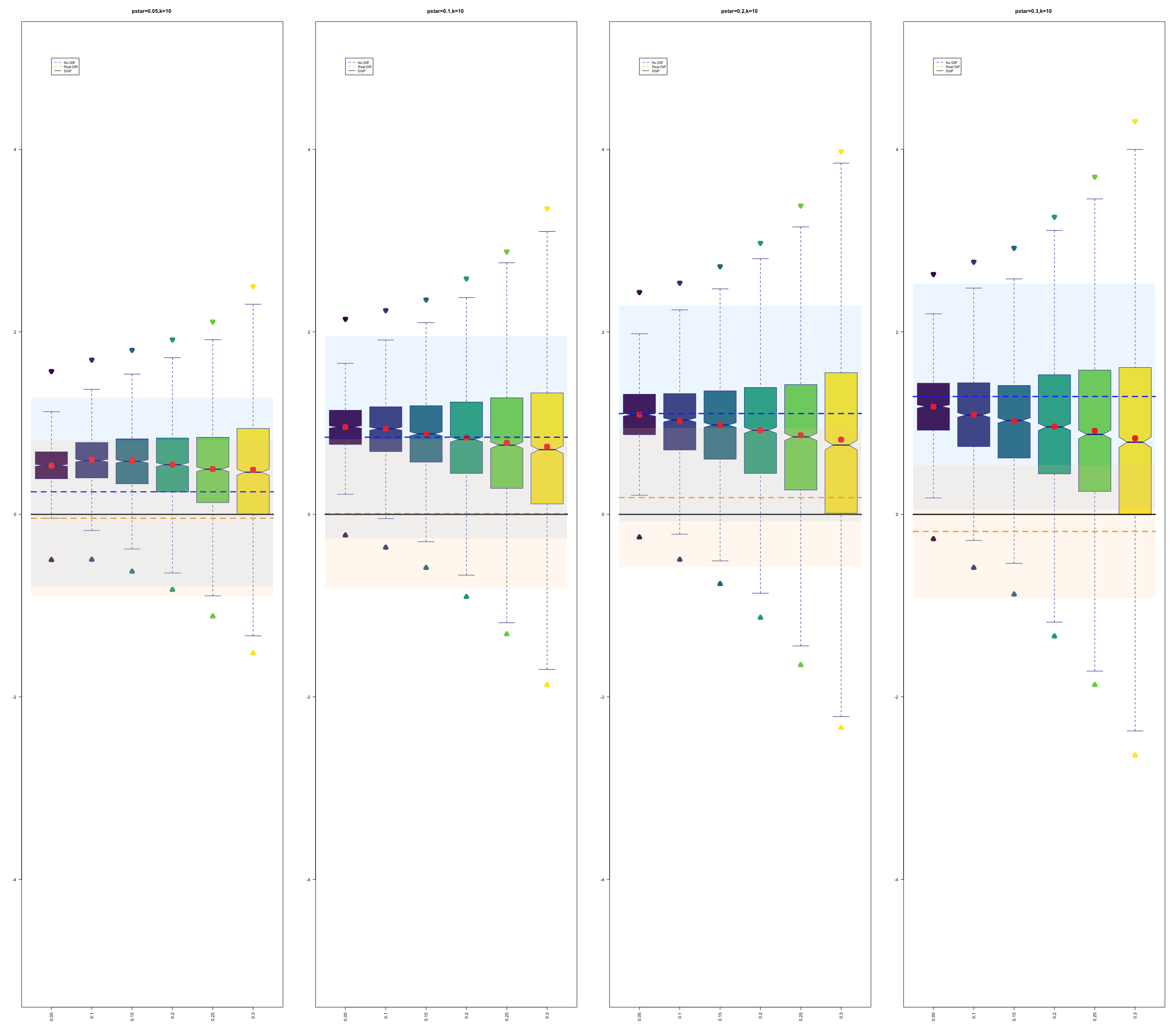}
		\caption[Overestimation, $k=2$]{Overestimation, $k=2$}
		\label{fig: overest_high}
	\end{figure}

	Figure \ref{fig: overest_low} shows the main results of the overestimation setting, under $k=5$. As we notice from the figure, the estimated treatment effect under no diffusion overestimates the real effect of the intervention. The sensitivity analysis causes a downward shifting in the estimates, by moving them towards 0.  As expected, the estimation bias increases as the true diffusion probability increases.  However, we can globally state that the procedure shows a good capability of towing the estimates towards the true value, by reducing the estimation bias due to having wrongly ignored the diffusion process.
	Finally, Figure \ref{fig: overest_high} depicts the simulations' results for the overestimation scenario, where $k=2$. As in the underestimation setting, an increasing in the size of the overall effect leads to an higher initial estimation bias. However, this estimation bias is effectively reduced by the sensitivity analysis, which rapidly tows the estimates towards the true value. 
	We can definitely state that the sensitivity analysis, when the treatment diffusion process occurs, helps the researcher in reducing the estimation bias, by towing the estimates towards the real value.  
	
	
	
	
	
	
	\section{Encouraging students to visit museums: issue and data \label{sec: ea}}
	\subsection{Empirical Motivation \label{subsec: ea_mot}}
	The empirical application is taken from the field of education economics. The school system plays a relevant role in 
	cultural heritage education.
	This can be achieved through the active involvement and participation to theatrical performances, museum visits and art exhibitions. Although several studies \citep{bourdieu2011forms} have pointed out that the family represents the primary focal entity in transmitting the cultural capital to children, other contributions \citep{dimaggio1982cultural,dimaggio1978origins} have highlighted the centrality of school as an institution, which may heavily contribute to provide a cultural exposure to those children who have not adequately benefited from it in their domestic environment. Recently, \cite{kisida2014creating} has underlined that cultural exposure of scholars sparks a real virtuous circle, according to which students become active cultural consumers, who are always more motivated to acquire extra cultural capital.
	
	Although this issue is relevant for  social development, the literature about the impact of school-promoted incentives to students
	is still scarce
	(\cite{lattarulo2017nudging,forastiere2019exploring}). The randomized experiment motivating this work contributes to filling this gap. The field experiment was a \emph{Cluster Randomized Encouragement Designs (CEDs)} implemented in Florence, Italy in 2014, with the aim of assessing the effect of different kinds of school-promoted incentives on encouraging students to visit art museums \footnote{The experiment was conducted by Patrizia Lattarulo (IRPET – Tuscany’s Regional Institute for Economic Planning,), Marco Mariani (IRPET – Tuscany’s Regional Institute for Economic Planning) and Laura Razzolini (University of Alabama). An extensive description of the data can be found in \cite{lattarulo2017nudging}}. The general goal of the experiment was to detect the most effective strategy to increase the teens' museum attendance and change their attitude towards art. In the study, classes of a school in Florence were randomly assigned to experience two different incentives: some of the classes received a flier about the importance of museum attendance, while the remaining classes received, in addition to the flier, a video presentation about an art exhibition \footnote{The original experiment by \cite{lattarulo2017nudging} includes a third type of encouragement: extra-credit points towards their final school grade. For the sake of simplicity, we omit here this third arm from the analysis.}. 
	
	Given the transferable nature of the link of the video presentation,the diffusion of the video link among students could have altered the experiment and the assessment of the 
	effect of providing the flier only as opposed to the video presentation together with the flier. Therefore, we will apply the proposed sensitivity analysis to investigate the robustness of the estimated effect against a possible diffusion process among students.
	
	\subsection{Data \label{subsec: ea_data}}
	Data involve $N=176$ students, enrolled in $C=10$ different classes, with $\mathcal{C}=\{1,\dots, C\}$, in a high school in the city of Florence. A time $t$ (Spring 2014) a set ${\mathcal C}_T\subset \mathcal{C}$ of $C_T=5$ classes were randomly assigned to each of the two types of cultural encouragement. Given the cluster randomized design, all the students enrolled in the same class were exposed to the treatment assigned to their class. Therefore, denoting by $C(i)$ the class of student $i$, the initial probability for each student $i$ of being assigned to the video presentation is $\pi_{it}(1)=P(Z_{it}=1)=P(C(i)\in \mathcal{C}_T) =C_T/C=0.5$. At time $t''$, $8$ months after  the initial assignment, students were asked to report the number of museum visits they had attended during those 8 months: this variable represents our outcome variable $Y_{it''}$. 
	
\begin{figure}[H]
		\centering
		\includegraphics[width=80mm]{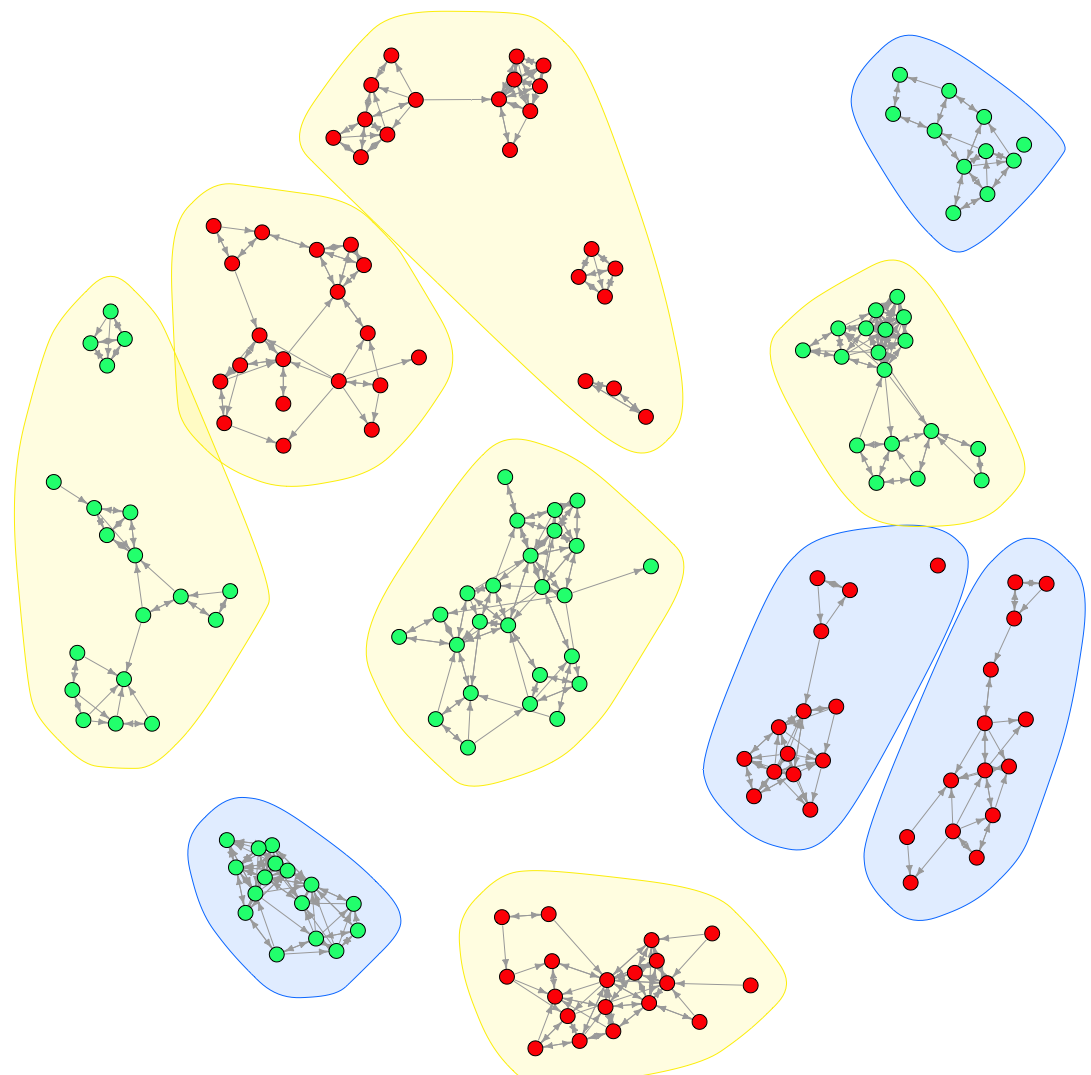}
		\caption[Observed network]{Observed network: nodes are colored with respect to their treatment status (red identifies treated nodes, while green untreated units; polygons encircle classes (note that students belonging to the same classes are characterized by an identical treatment status), while polygons' color represents the school membership. }
		\label{fig: obsnet}
	\end{figure}

	At baseline, students were also asked to report their friendship ties. Specifically, they were asked to declare who among their classmates they consider as friends.
	\footnote{
		also ranking the existing ties based on the strength of the relationships. Here, we only consider the presence or the absence of the friendship tie, regardless of the friendship strength.
	}
	The whole network structure is described by the graph $G=(\mathcal{N}, \mathcal{E})$, which consists of $C$ subgraphs, $\mathrm{G}_{c}=(\mathcal{N}_c, \mathcal{E}_c), \; \; c=1, \dots, C$. 
	Students were only asked to nominate friends in their own class, neglecting relationship between students of different classes.
	In fact, in the observed data there are no links between units belonging to different clusters, that is $A_{ij}=0 \; \forall i,j \; \;  \text{s.t} \; \; C(i) \neq C(j)$. 
	As a consequence, the adjacency matrix $\boldsymbol{A}$ corresponding to the graph $\mathbf{G}$, is a block-diagonal matrix with $C$ blocks, $\boldsymbol{A}_c, \; \; c=1 \dots C$.
	Figure \ref{fig: obsnet} provides a graphical representation of the overall network structure. 
	
\begin{figure}[H]
		\centering
		\begin{subfigure}{.5\textwidth}
			\centering		\includegraphics[width=75mm]{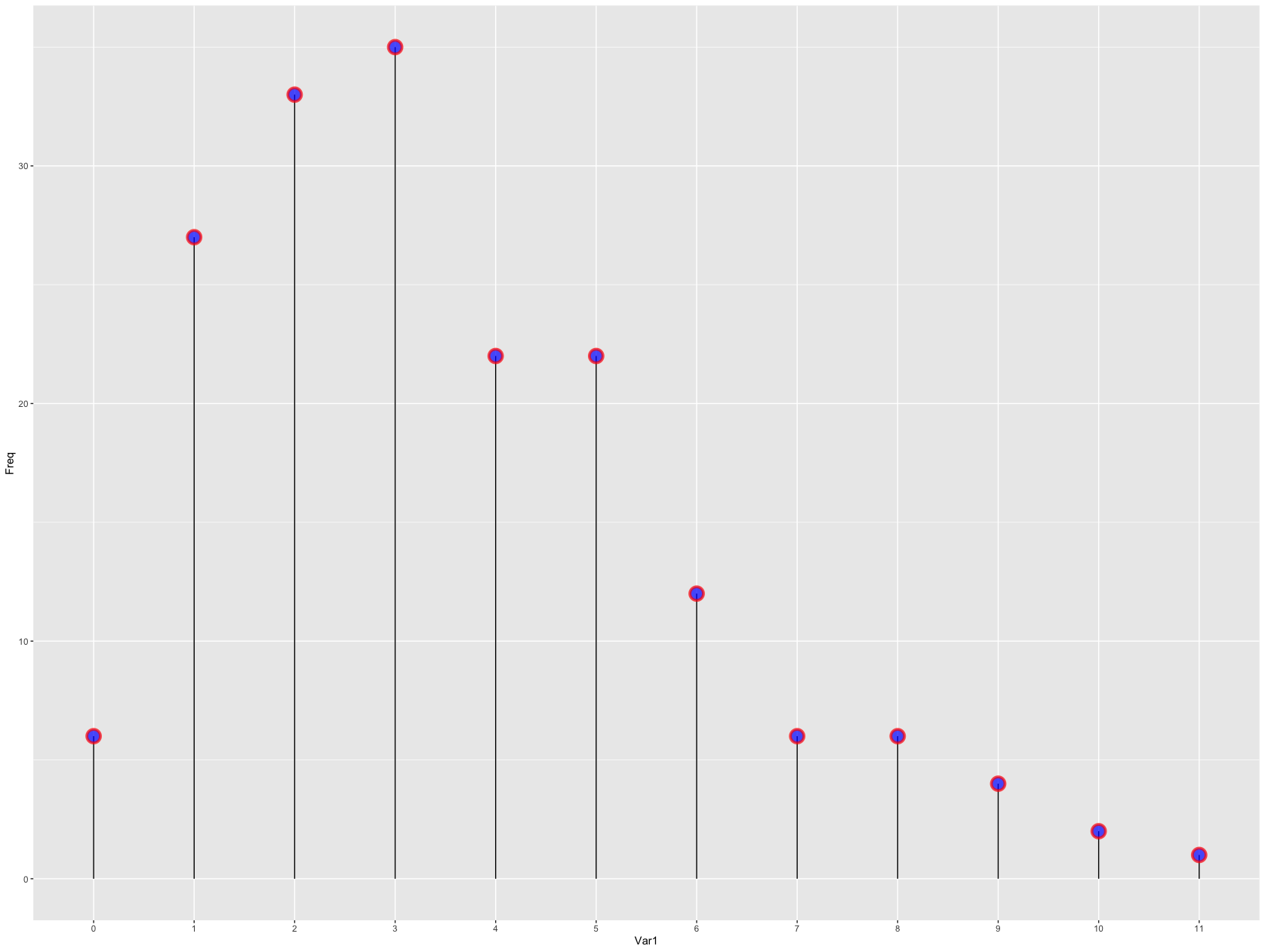}
			\caption{In-Degree distribution}
			\label{fig: indegree}
		\end{subfigure}%
		\begin{subfigure}{.5\textwidth}
			\centering
			\includegraphics[width=75mm]{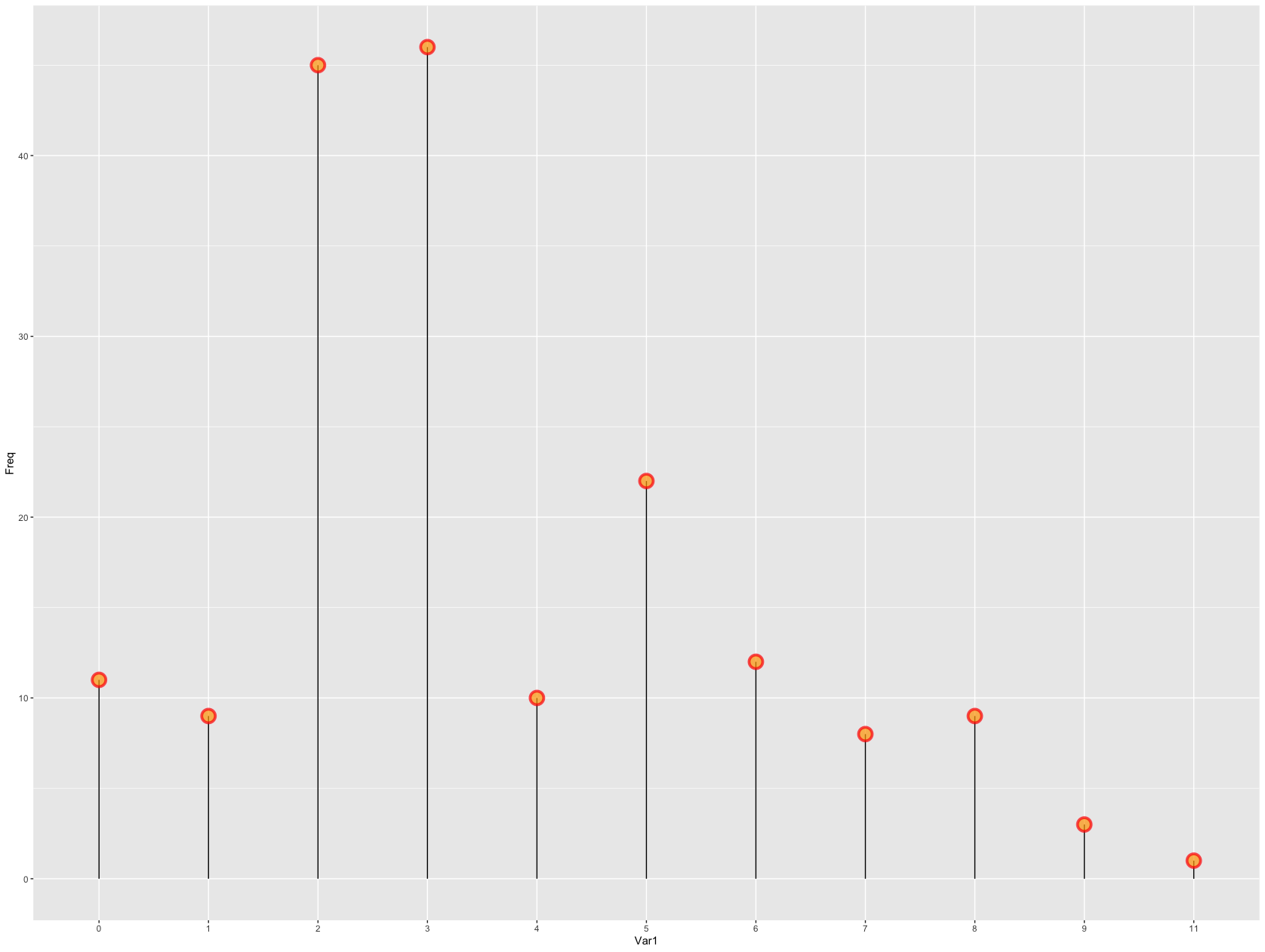}
			\caption{Out-Degree distribution}
			\label{fig: outdegree}
		\end{subfigure}
		\caption[Degree Distribution]{Degree distribution, observed data.}
	\end{figure}

	In our setting, relations between students in a given class $c$ are fully described by the adjacency matrix $\boldsymbol{A}_{c}$, where the generic element $A_{c}(i,j)$ equals 1 if the student $i$, enrolled in the class $c$, has nominated student $j$ in the same class $c$ as one of his friends. Note that friendship may be asymmetric: $i$ may regard $j$ as a friend, but not vice versa. We denote by $\mathcal{N}^{out}_{i}$ the set of students that unit $i$ has nominated as friends and by $\mathcal{N}^{in}_{i}$ the set of students that nominated $i$. $N^{out}_{i}$ and $N^{in}_{i}$ denote the corresponding cardinality, i.e., the \emph{out-degree} and \emph{in-degree}, respectively. Figure \ref{fig: indegree} shows the in-degree distribution, in the entire population, while Figure \ref{fig: outdegree} displays the out-degree distribution . 

	\subsection{Partially Unobserved Social Network: Link Prediction Model  \label{subsubsubsec: ea_pecul_netw}}
	The experiment does not consider the possibility of inter-class links. Although they have not been explicitly reported in the survey, they are likely to be present. Therefore, while intra-class links are observed, inter-class links are missing.
	In fact, students belonging to different classes, but enrolled in the same school, are likely to know students in other classes with whom they
	similar hobbies and activities.
	The missing information about inter-class links is crucial here because these links might have been vectors for the spread of the treatment. Therefore, imputing missing inter-class links is required to be able to reconstruct the diffusion process and, in turn, conduct a sensitivity analysis for the treatment effect.
	
	In a wide variety of empirical settings the network information is incomplete. For this reason, link-prediction has become a growing research topic within the field of network science and within the statistical literature of network data. The key idea of link-prediction models is to use the observed network to predict missing links. 
	
\begin{figure}[H]
		\centering
		\centering
		\includegraphics[width=90mm]{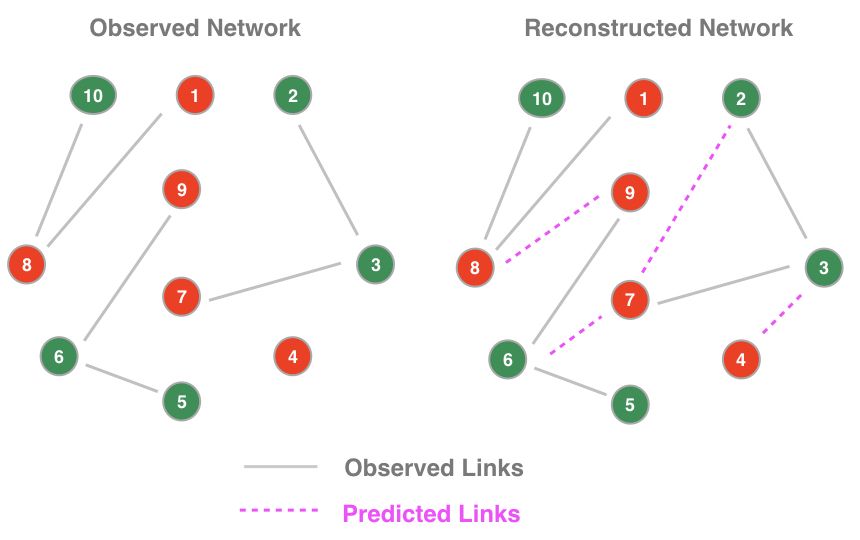}
		\caption[Link prediction: example]{Observed network vs reconstructed network: missing links are imputed by implementing a link prediction algorithm}
		\label{fig: reconsnet}
	\end{figure}

	Figure \ref{fig: reconsnet} provides a graphical intuition concerning the missing-links issue: given an observed network (left side network), link-prediction models use various and heterogeneous statistical techniques to impute missing links, generating a complete network (right side network). Links are predicted (purple-dotted arcs) according to a prediction model, which is determined by the specific setting.
	
	There are a variety of prediction models for link imputation. 
	Prediction models  can be classified into the following broad classes: i) purely statistical models, which use a parametric strategy for imputation, first estimating each link probability, then imputing missing links according to the estimated probability \citep{cranmer2011inferential,fellows2012exponential}; ii) network reconstruction models, developed in the growing literature of statistical physics for complex networks. These approaches are highly flexible and impute missing links while considering either similarity-based methods, likelihood-based criteria or entropy-based strategies \citep{liben2007link,zhou2009predicting,lu2011link}. The former approach assumes independent link formation between dyads, whereas the latter approach imputes missing links by taking into account the characteristics and the link patterns observed in the entire network.
	
	Here, we rely on a machine learning algorithm, which predicts missing inter-class links using a flexible approach, based on recursive partitioning \citep{buuren2010mice}. 
	Specifically, we use \emph{random forests} \citep{breiman1984classification},
	an extension of the classification and regression trees (CART) \citep{friedman1984classification}.
	Random forests are particularly flexible, perform well in managing with possible nonlinearities or interactions, and do not require specific assumptions \citep{shah2014comparison,doove2014recursive}. 
	
	
	The imputation algorithm trains the model on the observed ties, then predicting missing ties. Here, the imputation process takes as inputs a batch of dyadic covariates, which are used by the algorithm to recursively split the data. Splits are chosen to best predict the presence (or absence) of the dyadic tie. These dyadic covariates represent four measures of similarities that have been defined based on unit-level characteristics (the set of unit-level characteristics can be found in \cite{lattarulo2017nudging}). Specifically, we include in the analysis four variables, representing the level of similarity between two given students with respect to hobbies, school attitudes, cultural interests and personal background (details about these similarity indicators can be found in the Appendix \ref{app: similarity}). These similarities measures capture the key mechanisms that might prompt a friendship tie between students belonging to different classes. They are not affected by the treatment variable and they represent the key inputs of multiple imputation algorithm. In addition to the similarity measures, we have included in the imputation model two individual-specific indicators: the number of inter-class friends and the specific school environment where they are particularly inclined to establish friendship ties (mostly within their class or mostly outside their class). 
	
\begin{figure}[H]
		\centering
		\includegraphics[width=70mm]{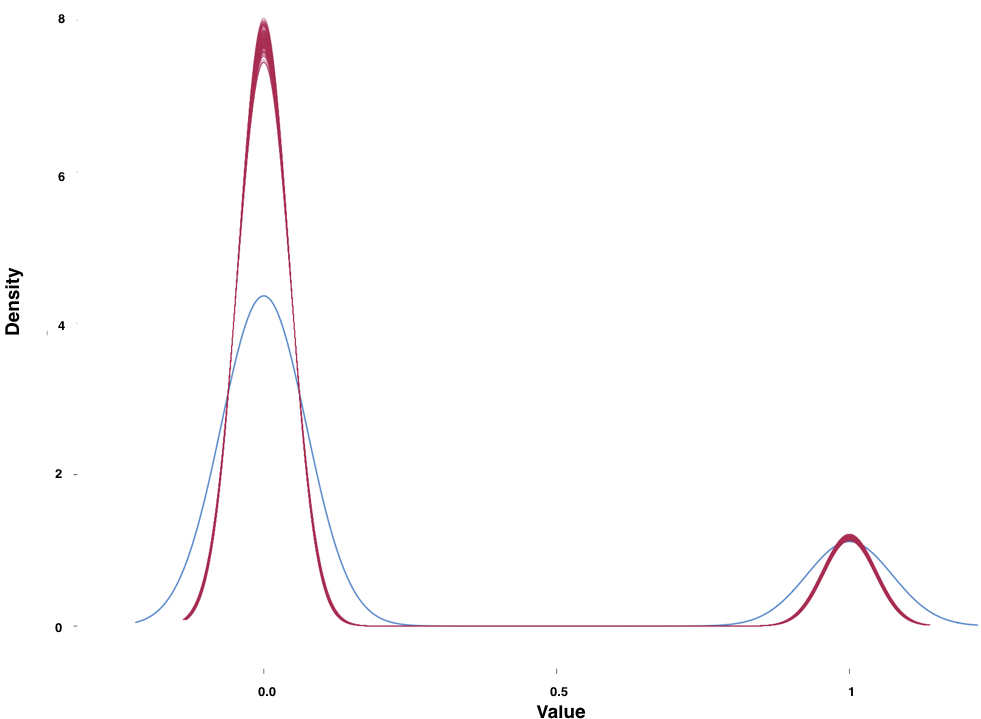}
		\caption[Densities of observed vs imputed links]{Densities of observed vs imputed links indicators. 1 denotes the presence of the link, 0 means the absence of a tie. The blue line shows the density in the original dataset, while the red line depicts densities in the imputed links}
		\label{fig: dens}
	\end{figure}

\begin{figure}[h]
		\centering
		\includegraphics[width=85mm]{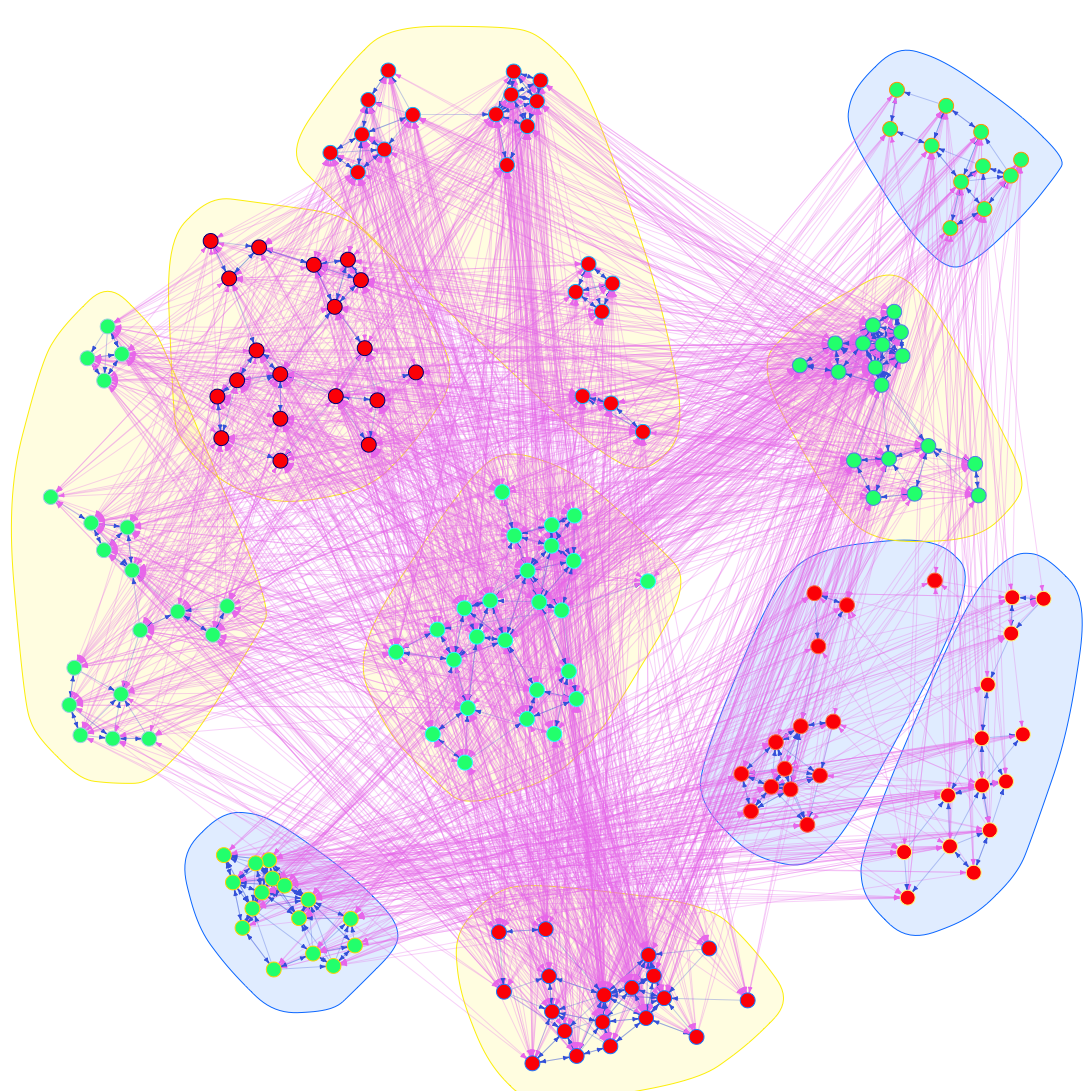}
		\caption[Reconstructed Network]{Reconstructed network: the first of the $M$ generated networks. Blue links represent ties that have been directly observed from data. Conversely, violet links depict relationships that have been imputed by the imputation algorithm. The former type of relationships remains unvaried across the $M$ networks, while the latter one varies.}
		\label{fig: rec}
	\end{figure}

	Through multiple imputation, we generate $M=500$ distinct reconstructed networks. Note that these networks are identical with respect to the known links, but differ in the imputed ties. Figure \ref{fig: dens} shows the densities of the tie indicators in the original (blue line) and imputed (red lines) datasets. As expected, the percentage of present links is less in the imputed datasets. This finding is in line with the general intuitive idea that it is easier to become friends for students who belong to the same class and it demonstrates that the algorithm performs fairly well in predicting the links (the stability of the multiple imputation algorithm can be inspected in Appendix \ref{app: stability}). In fact, the similarity measures that are used to predict the links are intrinsically higher for pairs of students enrolled in the same class and therefore, it makes sense that imputed inter-class links are (in percentage) fewer with respect to the intra-class observed ties. Therefore, even if observed ties and missing ties are intrinsically different, the dyadic covariates we account for allow to catch the diverse nature of links and guarantee the empirical validity of the imputation algorithm.
	Figures \ref{fig: rec} provides a graphical example of how a (complete) reconstructed network looks like, in our setting. The plot refers to the first of the 500 generated networks. Nodes are colored according to their initial assignment status (red nodes are treated units, while green characterizes untreated units). The figure displays two kinds of links: blue links denote observed intra-class links, while violet edges depict inter-class links.

	\subsection{Sensitivity Analysis: Empirical Results \label{subsec: ea_res}}
	The empirical analysis proposed by \cite{lattarulo2017nudging} is likely to be affected by the presence of an hidden treatment diffusion process, as scholars may share the link about the video presentation with their friends. We applied the proposed sensitivity analysis developed in Section \ref{subsec: trd_sens}. Given the cluster randomized experiment and the presence of missing links, we used the Horvitz-Thomson estimator in Equation \ref{eq: HT_diff_cr}
	and the multiple imputation algorithm as in 
	Section \ref{subsec: ea_sensanal}.
	
	We here discuss the key empirical findings of the proposed sensitivity analysis based on the estimator $\widehat{\tau}_2^{\star}$. The multiple imputation procedure generates $M$ separated reconstructed networks, which embrace the observed intra-class links as well as the predicted inter-class ties. The ensemble of the $M$ generated networks encompasses the variation boundary of the entire network structure and it represents one of the inputs of the sensitivity analysis algorithm we have introduced in Subsection \ref{subsec: trd_sens}. Here, we set the grid of the eligible values for the diffusion parameter $\overline{p}$ to $\mathcal{\overline{P}}= \{0.01,0.05,0.10,0.20,0.25\}$. Finally, we fix the number of sampled configurations of the (unknown) treatment vector at time $t'$ to $R=500$. 
	
\begin{figure}[H]
		\centering
		\includegraphics[width=90mm,height=80mm]{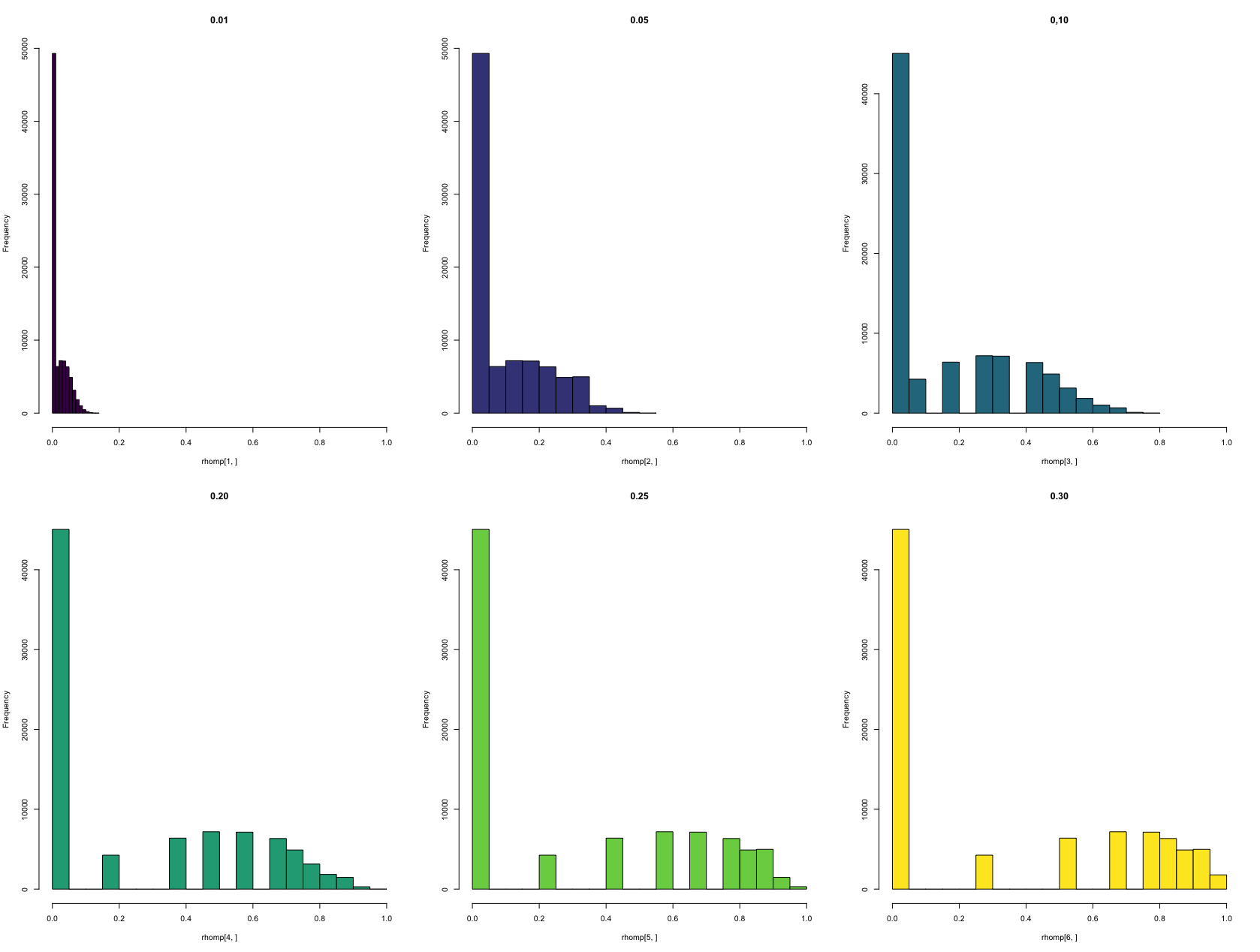}
		\caption[Probability of receiving the treatment due to the diffusion process, under increasing $\overline{p}$]{Probability of receiving the treatment by means of the diffusion process: histograms of the probabilities of receiving the treatment by diffusion.}
		\label{fig: rho}
	\end{figure}

\begin{figure}[H]
		\centering
		\includegraphics[width=100mm]{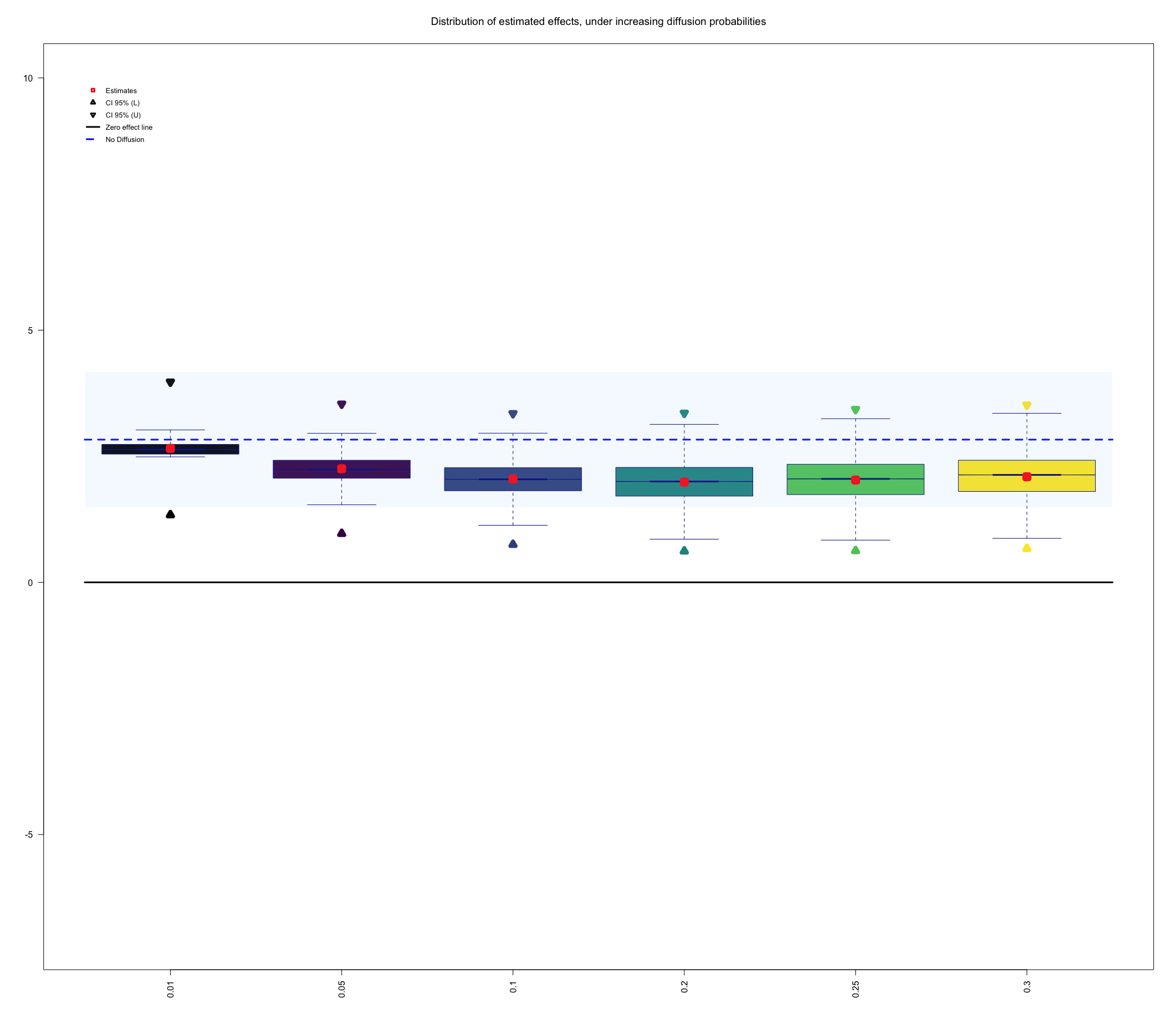}
		\caption[General Results]{Box-plots of estimated treatment effects and their corresponding 95\% Confidence Intervals, under increasing treatment diffusion probabilities. \label{fig: eff}}
	\end{figure}

	Figure \ref{fig: rho} gives an idea about the switching status process that happens in the presence of a plausible treatment spreading. In particular, it depicts the distributions of the probability of receiving the treatment by diffusion, for various configurations of the fixed diffusion parameter. Under small values of $\overline{p}$, very few units are eligible to gain the treatment by diffusion. As the fixed diffusion parameter increases, the number of initially untreated units who receive the treatment gets higher. 
	Figure \ref{fig: eff} graphically summarizes the key empirical finding of the sensitivity analysis. The procedure accounts both for the uncertain network structure (multiply imputing the $M$ reconstructed networks) and for the unknown after-diffusion treatment vector (generating $R$ different  treatment assignment vectors at time $t'$). In particular, Figure  \ref{fig: eff} shows the box-plots of the treatment effect estimates, obtained under the various configurations of $m$ and $r$. In addition, it shows the extremes of the 95\% confidence interval, that, in case of a positive treatment diffusion probability, have been constructed so to incorporate both the between variation and the within variation (as shown in Subsection \ref{subsec: trd_sens}). Box plots refer to various possible characterizations of the diffusion parameter $\overline{p}$. Under the no-diffusion assumption, the intervention has a positive and significant impact on students' museum visits ($\widehat{\tau}_{obs}^{b}=2.8295$ with a 95\% confidence interval which equals $\big[ 1.4863, 4.1727 \big])$ . The graphs suggest that ignoring the treatment diffusion process may lead to an overestimation of the treatment effect, but it also shows that the eventual diffusion spreading does not have a significant impact on results: even under considerable (and probably unplausible) treatment diffusion probabilities the estimates remain in the positive domain. On the other hand, the presence of a strong treatment diffusion process heavily increases the variability of the estimates, so that even a very tiny shift in the specification of the fixed diffusion parameter causes a relevant increase in the estimated total standard errors.
	
	To summarize, we can state that treatment diffusion could have plausibly affected the results of the experiment that we have revisited in this work. However, because the sign of the treatment effect is preserved even under high diffusion probabilities, we can state that the sensitivity analysis is in support of the major finding of the experiment, that is, the video-presentation has a positive impact on students and encourages them to attend museums visits. 
	\newpage
	\section{Concluding Remarks and Future Developments \label{sec: concl}}
	A wide variety of policy evaluation studies in the field of social sciences can be affected by the presence of a treatment diffusion process \citep{an2018causal,an2019opening}. For instance, the empirical setting proposed by \cite{lattarulo2017nudging} studies the effect of different types of school incentives aimed at encouraging students to attend cultural events and the intervention of interest is a promotional video, which can be tangibly spread among scholars through virtual or real social connections. When it is likely to arise, treatment diffusion induces a missclassification in the treatment variable and, if the correct classification cannot be retrieved by observing the exact treatment diffusion network, this process introduces a bias in the estimates and leads to inaccurate conclusions on the real effect of the intervention. It is not possible to explicitly correct this estimation bias, as the treatment diffusion process is usually unknown.
	
	This contribution represents the first methodological attempt of explicitly handling an unknown treatment diffusion process, in evaluating the causal effect of an intervention. Although treatment diffusion represents one of the three mechanisms which sparks interference \citep{hudgens2008toward,tchetgen2012causal}, it has not been extensively discussed yet, as most of the existing contributions analysing spillovers treat spillovers as a whole, without disentangling among the specific mechanisms which rule the interference mechanism \citep{an2019opening}. Here, the proposed approach is based on a sensitivity analysis on the unknown after-diffusion treatment assignment vector: specifically, it intends to compare the na\"ive estimate of the treatment effect obtained while neglecting any diffusion process with the estimates computed under credible diffusion scenarios. The proposed methodology is flexible and can be reworked so to more suitably target a wider ensemble of real world applications: for instance, given the methodological peculiarities of the proposed empirical setting, we rearrange the framework so to account for i) cluster randomized designs and ii) partially unknown baseline networks. Furthermore, it is possible to easily extend the theoretical framework in order to include some covariates inside the diffusion parameter: more specifically, we can replace the fixed single diffusion parameter $\overline{p}$ by a collection of diffusion parameters $\overline{p}_i=\overline{p}_i(\mathbf{X}_i)$, where $\mathbf{X}_i$ is a vector of observed covariates associated to unit $i$.   
	
	The paper formalizes the treatment diffusion bias and it has proven that, as long as the unconfoundedness assumption holds, the direction of the bias depends on the sign of the average treatment effect on those units who have a higher probability of receiving the intervention by diffusion. If they positively respond (in mean) to the treatment, then ignoring diffusion leads to an underestimation of the treatment effect. If vice versa, the intervention has a negative mean impact on their behavior, then ruling out the possibility of any diffusion process taking place in the analysis implies an overestimation of the effect. Inspired by this methodological finding, we observe how the sensitivity analysis procedure works in some simulated scenarios, where the experiment is affected by either an underestimation or overestimation bias, caused by the diffusion process. In the simulations' setting, the effect is heterogeneous with respect of the degree and the sensitivity analysis can also be employed for reducing the treatment diffusion bias, without simply assessing the robustness of results against plausible diffusion scenarios. Simulations' results show that the proposed methodology effectively reduces the treatment diffusion bias while towing the estimates towards the true value. The illustrated procedure can be applied in all those randomized settings where the diffusion process is likely to take place, independently on the characteristics of the initial randomization design.
	
	Our empirical findings suggest that ignoring the treatment diffusion process, when it plausibly arises, paves the way to an inaccurate evaluation of the causal effect of interest. Specifically, in the empirical application,  the effect of the intervention is reasonably smaller when accounting for the treatment diffusion mechanism. However, estimates appear to be robust even under highly relevant diffusion probabilities. This finding suggests that, even if the treatment spreads over the network and the intervention is slightly less effective that what it seemed from the initial estimates (obtained under the no-diffusion assumption), the promotional video still emerges as positive incentive for encouraging scholars in visiting museums. Nevertheless, the possible diffusion process does not compromise the validity of the main finding of the study by \cite{lattarulo2017nudging}, i.e the video positively encourages students towards museums attendance: conversely, the present analysis corroborates this argument further.
	
	The whole methodological framework that we illustrate in this work is highly flexible and it can be implemented in all those randomized settings, where diffusion plausibly arises. The sensitivity analysis procedure is presented under a very general setting so that it can be easily reworked by the user to account for the peculiarities, which characterize the specific empirical evaluation. These peculiarities may either improve or weaken the performance of the sensitivity analysis in catching plausible diffusion scenarios. For instance, the issue of a partially unobserved baseline network which characterize the empirical application of this work adds an uncertain component to the framework and may reduce the performance of the analysis. On the opposite, the whole framework would highly benefit from some eventual a priori information on the specific treatment diffusion process: for instance, if a researcher were able to more precisely characterize the process due to available information about the role played by some baseline covariates in affecting the treatment spreading, the sensitivity analysis would increase its performance. Clearly, the more information is available on the treatment diffusion process, the more accurate is the algorithm in depicting plausible diffusion scenarios. To conclude, the methodology proposed in the present contribution, together with the assumptions on which it relies, are highly versatile and can be revisited so to suitably model a wide variety of empirical scenarios, in all the fields of social sciences. 
	
	Due to its flexibility, the illustrated methodology can be easily improved through a lot of possible theoretical extensions. The whole procedure can accommodate for more complex methodological settings,  and explore specific aspects related to treatment diffusion, as the issues related to the hidden treatment diffusion process in program evaluation studies have not been explicitly addressed in causal inference literature yet \citep{an2018causal,an2019opening}. Future developments might involve a different temporal characterization of the treatment diffusion process, different definitions of the diffusion probabilities (for instance, they could be a function of the proportion of initially treated neighbors or be dependent on specific dyadic covariates) or an extended theoretical identification of the general estimating framework, which could be designed to allow for the possibility of indirect effects \citep{hudgens2008toward,forastiere2020identification}. It would be also interesting to study how to explicitly account for treatment diffusion in designing experiments \citep{angelucci2015program, baird2018optimal, kang2016peer}. The acquired information about treatment spreading in a network may be used in the experimental design, and the randomization strategy may be planned with the aim of maximizing the total number of individuals who can benefit from the intervention, either by the initial design or by diffusion. 
	\newpage
	\noindent {\bf Acknowledgement}\\
	Data regarding the field experiment on the effect of different types of school incentives on students' museum attendance have been collected and organized by Patrizia Lattarulo (IRPET – Tuscany’s Regional Institute for Economic Planning,), Marco Mariani (IRPET – Tuscany’s Regional Institute for Economic Planning) and Laura Razzolini (University of Alabama). An extensive discussion about data can be found in \cite{lattarulo2017nudging}. 
	\\
	Irene Crimaldi, Fabrizia Mealli and Costanza Tort\`u are members of the Italian Group Gruppo
	Nazionale per l'Analisi Matematica, la Probabilit\`a e le loro Applicazioni of the Italian Institute
	"Istituto Nazionale di Alta Matematica".
	\\[10pt]
	{\bf Declaration}\\
	All the authors developed the methodology and contributed to the
	final version of the manuscript. Costanza Tort\`u performed the simulations and the analysis of the empirical data, under the supervision of Irene Crimaldi and Laura Forastiere.
	\\[10pt]
	{\bf Funding Status}\\
	Irene Crimaldi is partially supported by the Italian Programma di Attivit\`a Integrata (PAI), project
	"TOol for Fighting FakEs (TOFFE)", funded by IMT School for Advanced Studies Lucca. Costanza
	Tort\`u has been supported by the Frontier Proposal Fellowship (FPF), funded by IMT School for Advanced Studies Lucca.
	\newpage
	\bibliographystyle{elsarticle-harv}
	\bibliography{document.bib}

\begin{thebibliography}{118}
\expandafter\ifx\csname natexlab\endcsname\relax\def\natexlab#1{#1}\fi
\expandafter\ifx\csname url\endcsname\relax
  \def\url#1{\texttt{#1}}\fi
\expandafter\ifx\csname urlprefix\endcsname\relax\def\urlprefix{URL }\fi

\bibitem[{An(2018)}]{an2018causal}
An, W., 2018. Causal inference with networked treatment diffusion. Sociological
  Methodology 48~(1), 152--181.

\bibitem[{An and VanderWeele(2019)}]{an2019opening}
An, W., VanderWeele, T.~J., 2019. Opening the blackbox of treatment
  interference: Tracing treatment diffusion through network analysis.
  Sociological Methods \& Research, 0049124119852384.

\bibitem[{Angelucci and Di~Maro(2015)}]{angelucci2015program}
Angelucci, M., Di~Maro, V., 2015. Program evaluation and spillover effects. The
  World Bank.

\bibitem[{Aronow and Middleton(2013)}]{aronow2013class}
Aronow, P.~M., Middleton, J.~A., 2013. A class of unbiased estimators of the
  average treatment effect in randomized experiments. Journal of Causal
  Inference 1~(1), 135--154.

\bibitem[{Aronow and Samii(2017)}]{aronow2017estimating}
Aronow, P.~M., Samii, C., 2017. Estimating average causal effects under general
  interference, with application to a social network experiment. The Annals of
  Applied Statistics 11~(4), 1912--1947.

\bibitem[{Aronow et~al.(2019)Aronow, Samii, and Wang}]{aronow2019design}
Aronow, P.~M., Samii, C., Wang, Y., 2019. Design-based inference for spatial
  experiments with interference.

\bibitem[{Arpino et~al.(2016)Arpino, Mattei, et~al.}]{arpino2016assessing}
Arpino, B., Mattei, A., et~al., 2016. Assessing the causal effects of financial
  aids to firms in tuscany allowing for interference. Annals of Applied
  Statistics 10~(3), 1170--1194.

\bibitem[{Athey et~al.(2018)Athey, Eckles, and Imbens}]{AtheyEcklesImbens2018}
Athey, S., Eckles, D., Imbens, G.~W., 2018. Exact p-values for network
  interference. Journal of the American Statistical Association 113~(521),
  230--240.

\bibitem[{Babanezhad et~al.(2010)Babanezhad, Vansteelandt, and
  Goetghebeur}]{babanezhad2010comparison}
Babanezhad, M., Vansteelandt, S., Goetghebeur, E., 2010. Comparison of causal
  effect estimators under exposure misclassification. Journal of Statistical
  Planning and Inference 140~(5), 1306--1319.

\bibitem[{Baird et~al.(2018)Baird, Bohren, McIntosh, and
  {\"O}zler}]{baird2018optimal}
Baird, S., Bohren, J.~A., McIntosh, C., {\"O}zler, B., 2018. Optimal design of
  experiments in the presence of interference. Review of Economics and
  Statistics 100~(5), 844--860.

\bibitem[{Bargagli~Stoffi et~al.(2020)Bargagli~Stoffi, Tort{\'u}, and
  Forastiere}]{bargagli2020heterogeneous}
Bargagli~Stoffi, F., Tort{\'u}, C., Forastiere, L., 2020. Heterogeneous
  treatment and spillover effects under clustered network interference.
  Costanza and Forastiere, Laura, Heterogeneous Treatment and Spillover Effects
  Under Clustered Network Interference (August 3, 2020).

\bibitem[{Basse and Feller(2018)}]{basse2018analyzing}
Basse, G., Feller, A., 2018. Analyzing two-stage experiments in the presence of
  interference. Journal of the American Statistical Association 113~(521),
  41--55.

\bibitem[{Bound et~al.(2001)Bound, Brown, and
  Mathiowetz}]{bound2001measurement}
Bound, J., Brown, C., Mathiowetz, N., 2001. Measurement error in survey data.
  In: Handbook of econometrics. Vol.~5. Elsevier, pp. 3705--3843.

\bibitem[{Bourdieu(2011)}]{bourdieu2011forms}
Bourdieu, P., 2011. The forms of capital.(1986). Cultural theory: An anthology
  1, 81--93.

\bibitem[{Bourigault et~al.(2014)Bourigault, Lagnier, Lamprier, Denoyer, and
  Gallinari}]{bourigault2014learning}
Bourigault, S., Lagnier, C., Lamprier, S., Denoyer, L., Gallinari, P., 2014.
  Learning social network embeddings for predicting information diffusion. In:
  Proceedings of the 7th ACM international conference on Web search and data
  mining. pp. 393--402.

\bibitem[{Braun et~al.(2014)Braun, Gorfine, Zigler, Dominici, and
  Parmigiani}]{braun2014adjustment}
Braun, D., Gorfine, M., Zigler, C., Dominici, F., Parmigiani, G., 2014.
  Adjustment for mismeasured exposure using validation data and propensity
  scores.

\bibitem[{Braun et~al.(2016)Braun, Zigler, Dominici, and
  Gorfine}]{braun2016using}
Braun, D., Zigler, C., Dominici, F., Gorfine, M., 2016. Using validation data
  to adjust the inverse probability weighting estimator for misclassified
  treatment. Using Validation Data to Adjust the Inverse Probability Weighting
  Estimator for Misclassified Treatment.

\bibitem[{Breiman et~al.(1984)Breiman, Friedman, Stone, and
  Olshen}]{breiman1984classification}
Breiman, L., Friedman, J., Stone, C.~J., Olshen, R.~A., 1984. Classification
  and regression trees. CRC press.

\bibitem[{Bridges et~al.(2000)Bridges, Thompson, Meltzer, Reeve, Talamonti,
  Cox, Lilac, Hall, Klimov, and Fukuda}]{bridges2000effectiveness}
Bridges, C.~B., Thompson, W.~W., Meltzer, M.~I., Reeve, G.~R., Talamonti,
  W.~J., Cox, N.~J., Lilac, H.~A., Hall, H., Klimov, A., Fukuda, K., 2000.
  Effectiveness and cost-benefit of influenza vaccination of healthy working
  adults: a randomized controlled trial. Jama 284~(13), 1655--1663.

\bibitem[{Buuren and Groothuis-Oudshoorn(2010)}]{buuren2010mice}
Buuren, S.~v., Groothuis-Oudshoorn, K., 2010. mice: Multivariate imputation by
  chained equations in r. Journal of statistical software, 1--68.

\bibitem[{Carroll et~al.(2006)Carroll, Ruppert, Stefanski, and
  Crainiceanu}]{carroll2006measurement}
Carroll, R.~J., Ruppert, D., Stefanski, L.~A., Crainiceanu, C.~M., 2006.
  Measurement error in nonlinear models: a modern perspective. CRC press.

\bibitem[{Centola(2010)}]{centola2010spread}
Centola, D., 2010. The spread of behavior in an online social network
  experiment. science 329~(5996), 1194--1197.

\bibitem[{Chin et~al.(2013)Chin, Daysal, and Imberman}]{chin2013impact}
Chin, A., Daysal, N.~M., Imberman, S.~A., 2013. Impact of bilingual education
  programs on limited english proficient students and their peers: Regression
  discontinuity evidence from texas. Journal of Public Economics 107, 63--78.

\bibitem[{Chuang and Lin(1999)}]{chuang1999foreign}
Chuang, Y.-C., Lin, C.-M., 1999. Foreign direct investment, r\&d and spillover
  efficiency: Evidence from taiwan's manufacturing firms. The Journal of
  Development Studies 35~(4), 117--137.

\bibitem[{Cohen et~al.(2002)Cohen, Goto, Nagata, Nelson, and
  Walsh}]{cohen2002r}
Cohen, W.~M., Goto, A., Nagata, A., Nelson, R.~R., Walsh, J.~P., 2002. R\&d
  spillovers, patents and the incentives to innovate in japan and the united
  states. Research policy 31~(8-9), 1349--1367.

\bibitem[{Cowan and Jonard(2004)}]{cowan2004network}
Cowan, R., Jonard, N., 2004. Network structure and the diffusion of knowledge.
  Journal of economic Dynamics and Control 28~(8), 1557--1575.

\bibitem[{Cox(1958)}]{cox1958planning}
Cox, D.~R., 1958. Planning of experiments.

\bibitem[{Cranmer and Desmarais(2011)}]{cranmer2011inferential}
Cranmer, S.~J., Desmarais, B.~A., 2011. Inferential network analysis with
  exponential random graph models. Political analysis 19~(1), 66--86.

\bibitem[{Cr{\'e}pon et~al.(2013)Cr{\'e}pon, Duflo, Gurgand, Rathelot, and
  Zamora}]{crepon2013labor}
Cr{\'e}pon, B., Duflo, E., Gurgand, M., Rathelot, R., Zamora, P., 2013. Do
  labor market policies have displacement effects? evidence from a clustered
  randomized experiment. The quarterly journal of economics 128~(2), 531--580.

\bibitem[{Crimaldi et~al.(2020)Crimaldi, Forastiere, Mealli, and
  Tort{\'u}}]{tortu2020immigration}
Crimaldi, I., Forastiere, L., Mealli, F., Tort{\'u}, C., 2020. The causal
  effect of immigration policy on income inequality. Book of short paper - SIS
  2020, Italian Statistical Society, 1549--1556.

\bibitem[{Del~Prete et~al.(2019)Del~Prete, Forastiere, and
  Leone~Sciabolazza}]{del2019causal}
Del~Prete, D., Forastiere, L., Leone~Sciabolazza, V., 2019. Causal inference on
  networks under continuous treatment interference: an application to trade
  distortions in agricultural markets. Available at SSRN 3363173.

\bibitem[{D{\'\i}az and van~der Laan(2013)}]{diaz2013sensitivity}
D{\'\i}az, I., van~der Laan, M.~J., 2013. Sensitivity analysis for causal
  inference under unmeasured confounding and measurement error problems. The
  international journal of biostatistics 9~(2), 149--160.

\bibitem[{DiMaggio(1982)}]{dimaggio1982cultural}
DiMaggio, P., 1982. Cultural capital and school success: The impact of status
  culture participation on the grades of us high school students. American
  sociological review, 189--201.

\bibitem[{DiMaggio and Useem(1978)}]{dimaggio1978origins}
DiMaggio, P., Useem, M., 1978. The origins and consequences of class
  differences in exposure to the arts in america. Theory and Society 5~(2),
  141--161.

\bibitem[{Doove et~al.(2014)Doove, Van~Buuren, and
  Dusseldorp}]{doove2014recursive}
Doove, L.~L., Van~Buuren, S., Dusseldorp, E., 2014. Recursive partitioning for
  missing data imputation in the presence of interaction effects. Computational
  Statistics \& Data Analysis 72, 92--104.

\bibitem[{Duong et~al.(2011)Duong, Wellman, and Singh}]{duong2011modeling}
Duong, Q., Wellman, M.~P., Singh, S., 2011. Modeling information diffusion in
  networks with unobserved links. In: 2011 IEEE Third International Conference
  on Privacy, Security, Risk and Trust and 2011 IEEE Third International
  Conference on Social Computing. IEEE, pp. 362--369.

\bibitem[{Erd{\H{o}}s and R{\'e}nyi(1959)}]{erdos1959random}
Erd{\H{o}}s, P., R{\'e}nyi, A., 1959. On random graphs i. Publicationes
  Mathematicae 6~(290-297), 18.

\bibitem[{Fellows and Handcock(2012)}]{fellows2012exponential}
Fellows, I., Handcock, M.~S., 2012. Exponential-family random network models.
  arXiv preprint arXiv:1208.0121.

\bibitem[{Forastiere et~al.(2020{\natexlab{a}})Forastiere, Airoldi, and
  Mealli}]{forastiere2020identification}
Forastiere, L., Airoldi, E.~M., Mealli, F., 2020{\natexlab{a}}. Identification
  and estimation of treatment and interference effects in observational studies
  on networks. Journal of the American Statistical Association~(just-accepted),
  1--49.

\bibitem[{Forastiere et~al.(2020{\natexlab{b}})Forastiere, Airoldi, and
  Mealli}]{forastiere2016identification}
Forastiere, L., Airoldi, E.~M., Mealli, F., 2020{\natexlab{b}}. Identification
  and estimation of treatment and interference effects in observational studies
  on networks. Forthcoming in the Journal of the American Statistical
  Association. Preprint at arXiv:1609.06245.

\bibitem[{Forastiere et~al.(2019{\natexlab{a}})Forastiere, Lattarulo, Mariani,
  Mealli, and Razzolini}]{forastiere2019museums}
Forastiere, L., Lattarulo, P., Mariani, M., Mealli, F., Razzolini, L.,
  2019{\natexlab{a}}. Exploring encouragement, treatment, and spillover effects
  using principal stratification, with application to a field experiment on
  teens’ museum attendance. Journal of Business \& Economic Statistics 0~(0),
  1--15.

\bibitem[{Forastiere et~al.(2019{\natexlab{b}})Forastiere, Lattarulo, Mariani,
  Mealli, and Razzolini}]{forastiere2019exploring}
Forastiere, L., Lattarulo, P., Mariani, M., Mealli, F., Razzolini, L.,
  2019{\natexlab{b}}. Exploring encouragement, treatment, and spillover effects
  using principal stratification, with application to a field experiment on
  teens’ museum attendance. Journal of Business \& Economic Statistics,
  1--15.

\bibitem[{Forastiere et~al.(2016)Forastiere, Mealli, and
  VanderWeele}]{forastiere2016clusters}
Forastiere, L., Mealli, F., VanderWeele, T.~J., 2016. Identification and
  estimation of causal mechanisms in clustered encouragement designs:
  Disentangling bed nets using bayesian principal stratification. Journal of
  the American Statistical Association 111~(514), 510--525.

\bibitem[{Forastiere et~al.(2018)Forastiere, Mealli, Wu, and
  Airoldi}]{forastiere2018estimating}
Forastiere, L., Mealli, F., Wu, A., Airoldi, E., 2018. Estimating causal
  effects under interference using bayesian generalized propensity scores.
  arXiv preprint arXiv:1807.11038.

\bibitem[{Friedman et~al.(1984)Friedman, Olshen, Stone,
  et~al.}]{friedman1984classification}
Friedman, J.~H., Olshen, R.~A., Stone, C.~J., et~al., 1984. Classification and
  regression trees. Belmont, CA: Wadsworth \& Brooks.

\bibitem[{Fuller(2009)}]{fuller2009measurement}
Fuller, W.~A., 2009. Measurement error models. Vol. 305. John Wiley \& Sons.

\bibitem[{Gai and Kapadia(2010)}]{gai2010contagion}
Gai, P., Kapadia, S., 2010. Contagion in financial networks. Proceedings of the
  Royal Society A: Mathematical, Physical and Engineering Sciences 466~(2120),
  2401--2423.

\bibitem[{Gertler(2004)}]{gertler2004conditional}
Gertler, P., 2004. Do conditional cash transfers improve child health? evidence
  from progresa's control randomized experiment. American economic review
  94~(2), 336--341.

\bibitem[{Goel et~al.(2016)Goel, Anderson, Hofman, and
  Watts}]{goel2016structural}
Goel, S., Anderson, A., Hofman, J., Watts, D.~J., 2016. The structural virality
  of online diffusion. Management Science 62~(1), 180--196.

\bibitem[{Grace(2017)}]{grace2017statistical}
Grace, Y.~Y., 2017. Statistical analysis with measurement error or
  misclassification strategy, method and application.

\bibitem[{Grandjean et~al.(2004)Grandjean, Budtz-J{\o}rgensen, Keiding, and
  Weihe}]{grandjean2004underestimation}
Grandjean, P., Budtz-J{\o}rgensen, E., Keiding, N., Weihe, P., 2004.
  Underestimation of risk due to exposure misclassification. International
  journal of occupational medicine and environmental health.

\bibitem[{Green and Vavreck(2008)}]{green2008analysis}
Green, D.~P., Vavreck, L., 2008. Analysis of cluster-randomized experiments: A
  comparison of alternative estimation approaches. Political Analysis,
  138--152.

\bibitem[{Hamby(1994)}]{hamby1994review}
Hamby, D.~M., 1994. A review of techniques for parameter sensitivity analysis
  of environmental models. Environmental monitoring and assessment 32~(2),
  135--154.

\bibitem[{Horvitz and Thompson(1952)}]{horvitz1952generalization}
Horvitz, D.~G., Thompson, D.~J., 1952. A generalization of sampling without
  replacement from a finite universe. Journal of the American statistical
  Association 47~(260), 663--685.

\bibitem[{Hudgens and Halloran(2008)}]{hudgens2008toward}
Hudgens, M.~G., Halloran, M.~E., 2008. Toward causal inference with
  interference. Journal of the American Statistical Association 103~(482),
  832--842.

\bibitem[{Imai et~al.(2020)Imai, Jiang, and Malani}]{imai2020causal}
Imai, K., Jiang, Z., Malani, A., 2020. Causal inference with interference and
  noncompliance in two-stage randomized experiments. Journal of the American
  Statistical Association~(just-accepted), 1--39.

\bibitem[{Imai and Yamamoto(2010)}]{imai2010causal}
Imai, K., Yamamoto, T., 2010. Causal inference with differential measurement
  error: Nonparametric identification and sensitivity analysis. American
  Journal of Political Science 54~(2), 543--560.

\bibitem[{Imbens and Rubin(2015)}]{imbens2015causal}
Imbens, G.~W., Rubin, D.~B., 2015. Causal inference in statistics, social, and
  biomedical sciences. Cambridge University Press.

\bibitem[{Iooss and Lema{\^\i}tre(2015)}]{iooss2015review}
Iooss, B., Lema{\^\i}tre, P., 2015. A review on global sensitivity analysis
  methods. In: Uncertainty management in simulation-optimization of complex
  systems. Springer, pp. 101--122.

\bibitem[{Kang and Imbens(2016)}]{kang2016peer}
Kang, H., Imbens, G., 2016. Peer encouragement designs in causal inference with
  partial interference and identification of local average network effects.
  arXiv preprint arXiv:1609.04464.

\bibitem[{Katona et~al.(2011)Katona, Zubcsek, and Sarvary}]{katona2011network}
Katona, Z., Zubcsek, P.~P., Sarvary, M., 2011. Network effects and personal
  influences: The diffusion of an online social network. Journal of marketing
  research 48~(3), 425--443.

\bibitem[{Kempe et~al.(2003)Kempe, Kleinberg, and Tardos}]{kempe2003maximizing}
Kempe, D., Kleinberg, J., Tardos, {\'E}., 2003. Maximizing the spread of
  influence through a social network. In: Proceedings of the ninth ACM SIGKDD
  international conference on Knowledge discovery and data mining. pp.
  137--146.

\bibitem[{Kempe et~al.(2005)Kempe, Kleinberg, and
  Tardos}]{kempe2005influential}
Kempe, D., Kleinberg, J., Tardos, {\'E}., 2005. Influential nodes in a
  diffusion model for social networks. In: International Colloquium on
  Automata, Languages, and Programming. Springer, pp. 1127--1138.

\bibitem[{Kisida et~al.(2014)Kisida, Greene, and Bowen}]{kisida2014creating}
Kisida, B., Greene, J.~P., Bowen, D.~H., 2014. Creating cultural consumers: The
  dynamics of cultural capital acquisition. Sociology of Education 87~(4),
  281--295.

\bibitem[{Koskinen et~al.(2013)Koskinen, Robins, Wang, and
  Pattison}]{koskinen2013bayesian}
Koskinen, J.~H., Robins, G.~L., Wang, P., Pattison, P.~E., 2013. Bayesian
  analysis for partially observed network data, missing ties, attributes and
  actors. Social Networks 35~(4), 514--527.

\bibitem[{La~Torre et~al.(2020)La~Torre, D’Egidio, Guastamacchia, Barbagallo,
  and Mannocci}]{la2020diffusion}
La~Torre, G., D’Egidio, V., Guastamacchia, S., Barbagallo, A., Mannocci, A.,
  2020. Diffusion of the italian social media campaign against smoking on a
  social network and youtube. Journal of Preventive Medicine and Hygiene
  61~(2), E200.

\bibitem[{Lamberson(2016)}]{lamberson2016diffusion}
Lamberson, P., 2016. Diffusion in networks. Bramoull{\'e}, Yann, Andrea
  Galeotti, and Brian Rogers, editors.

\bibitem[{Lattarulo et~al.(2017)Lattarulo, Mariani, and
  Razzolini}]{lattarulo2017nudging}
Lattarulo, P., Mariani, M., Razzolini, L., 2017. Nudging museums attendance: a
  field experiment with high school teens. Journal of Cultural Economics
  41~(3), 259--277.

\bibitem[{Leitner(2005)}]{leitner2005financial}
Leitner, Y., 2005. Financial networks: Contagion, commitment, and private
  sector bailouts. The Journal of Finance 60~(6), 2925--2953.

\bibitem[{Leung(2020{\natexlab{a}})}]{leung2020treatment}
Leung, M.~P., 2020{\natexlab{a}}. Treatment and spillover effects under network
  interference. Review of Economics and Statistics 102~(2), 368--380.

\bibitem[{Leung(2020{\natexlab{b}})}]{leung2020}
Leung, M.~P., 2020{\natexlab{b}}. Treatment and spillover effects under network
  interference. The Review of Economics and Statistics 102~(2), 368--380.

\bibitem[{Lewbel(2007)}]{lewbel2007estimation}
Lewbel, A., 2007. Estimation of average treatment effects with
  misclassification. Econometrica 75~(2), 537--551.

\bibitem[{Liben-Nowell and Kleinberg(2007)}]{liben2007link}
Liben-Nowell, D., Kleinberg, J., 2007. The link-prediction problem for social
  networks. Journal of the American society for information science and
  technology 58~(7), 1019--1031.

\bibitem[{Liu et~al.(2016)Liu, Hudgens, and Becker-Dreps}]{Liu2016}
Liu, L., Hudgens, M., Becker-Dreps, S., 2016. On inverse probability-weighted
  estimators in the presence of interference. Biometrika 103~(4), 829--842.
\newline\urlprefix\url{https://doi.org/10.1093/biomet/asw047}

\bibitem[{Liu and Hudgens(2014)}]{liu2014large}
Liu, L., Hudgens, M.~G., 2014. Large sample randomization inference of causal
  effects in the presence of interference. Journal of the american statistical
  association 109~(505), 288--301.

\bibitem[{Liu et~al.(2019)Liu, Chen, Jeon, Chen, and Chen}]{liu2019influence}
Liu, W., Chen, X., Jeon, B., Chen, L., Chen, B., 2019. Influence maximization
  on signed networks under independent cascade model. Applied Intelligence
  49~(3), 912--928.

\bibitem[{Liu et~al.(2017)Liu, Sidhu, Beacom, and Valente}]{liu2017social}
Liu, W., Sidhu, A., Beacom, A.~M., Valente, T.~W., 2017. Social network theory.
  The international encyclopedia of media effects, 1--12.

\bibitem[{Loh et~al.(2020)Loh, Hudgens, Clemens, Ali, and
  Emch}]{loh2020randomization}
Loh, W.~W., Hudgens, M.~G., Clemens, J.~D., Ali, M., Emch, M.~E., 2020.
  Randomization inference with general interference and censoring. Biometrics
  76~(1), 235--245.

\bibitem[{L{\'o}pez-Pintado(2008)}]{lopez2008diffusion}
L{\'o}pez-Pintado, D., 2008. Diffusion in complex social networks. Games and
  Economic Behavior 62~(2), 573--590.

\bibitem[{L{\"u} and Zhou(2011)}]{lu2011link}
L{\"u}, L., Zhou, T., 2011. Link prediction in complex networks: A survey.
  Physica A: statistical mechanics and its applications 390~(6), 1150--1170.

\bibitem[{Mahajan(2010)}]{mahajan2010innovation}
Mahajan, V., 2010. Innovation diffusion. Wiley International Encyclopedia of
  Marketing.

\bibitem[{McCaffrey et~al.(2013)McCaffrey, Lockwood, and
  Setodji}]{mccaffrey2013inverse}
McCaffrey, D.~F., Lockwood, J., Setodji, C.~M., 2013. Inverse probability
  weighting with error-prone covariates. Biometrika 100~(3), 671--680.

\bibitem[{Miles et~al.(2019)Miles, Petersen, and van~der
  Laan}]{miles2019causal}
Miles, C.~H., Petersen, M., van~der Laan, M.~J., 2019. Causal inference when
  counterfactuals depend on the proportion of all subjects exposed. Biometrics
  75~(3), 768--777.

\bibitem[{Nichol et~al.(1995)Nichol, Lind, Margolis, Murdoch, McFadden, Hauge,
  Magnan, and Drake}]{nichol1995effectiveness}
Nichol, K.~L., Lind, A., Margolis, K.~L., Murdoch, M., McFadden, R., Hauge, M.,
  Magnan, S., Drake, M., 1995. The effectiveness of vaccination against
  influenza in healthy, working adults. New England Journal of Medicine
  333~(14), 889--893.

\bibitem[{Nyblom et~al.(2003)Nyblom, Borgatti, Roslakka, and
  Salo}]{nyblom2003statistical}
Nyblom, J., Borgatti, S., Roslakka, J., Salo, M.~A., 2003. Statistical analysis
  of network data—an application to diffusion of innovation. Social Networks
  25~(2), 175--195.

\bibitem[{Ogburn et~al.(2017)Ogburn, Sofrygin, Diaz, and van~der
  Laan}]{Ogburn2017}
Ogburn, E., Sofrygin, O., Diaz, I., van~der Laan, M., 2017. Causal inference
  for social network data. arXiv:1705.08527.

\bibitem[{Onnela and Christakis(2012)}]{onnela2012spreading}
Onnela, J.-P., Christakis, N.~A., 2012. Spreading paths in partially observed
  social networks. Physical Review E 85~(3), 036106.

\bibitem[{Paluck et~al.(2016)Paluck, Shepherd, and Aronow}]{paluck2016changing}
Paluck, E.~L., Shepherd, H., Aronow, P.~M., 2016. Changing climates of
  conflict: A social network experiment in 56 schools. Proceedings of the
  National Academy of Sciences 113~(3), 566--571.

\bibitem[{Pan et~al.(2016)Pan, Zhou, L{\"u}, and Hu}]{pan2016predicting}
Pan, L., Zhou, T., L{\"u}, L., Hu, C.-K., 2016. Predicting missing links and
  identifying spurious links via likelihood analysis. Scientific reports 6~(1),
  1--10.

\bibitem[{Papadogeorgou et~al.(2019)Papadogeorgou, Mealli, and
  Zigler}]{papadogeorgou2019causal}
Papadogeorgou, G., Mealli, F., Zigler, C.~M., 2019. Causal inference with
  interfering units for cluster and population level treatment allocation
  programs. Biometrics 75~(3), 778--787.

\bibitem[{Robins et~al.(2000)Robins, Rotnitzky, and
  Scharfstein}]{robins2000sensitivity}
Robins, J.~M., Rotnitzky, A., Scharfstein, D.~O., 2000. Sensitivity analysis
  for selection bias and unmeasured confounding in missing data and causal
  inference models. In: Statistical models in epidemiology, the environment,
  and clinical trials. Springer, pp. 1--94.

\bibitem[{Rogers et~al.(2012)Rogers, Chapman, and
  Giotsas}]{rogers2012measuring}
Rogers, M., Chapman, C., Giotsas, V., 2012. Measuring the diffusion of
  marketing messages across a social network. Journal of Direct, Data and
  Digital Marketing Practice 14~(2), 97--130.

\bibitem[{Rosenbaum(2014)}]{rosenbaum2014sensitivity}
Rosenbaum, P.~R., 2014. Sensitivity analysis in observational studies. Wiley
  StatsRef: Statistics Reference Online.

\bibitem[{Rosenbaum and Rubin(1983)}]{rosenbaum1983assessing}
Rosenbaum, P.~R., Rubin, D.~B., 1983. Assessing sensitivity to an unobserved
  binary covariate in an observational study with binary outcome. Journal of
  the Royal Statistical Society: Series B (Methodological) 45~(2), 212--218.

\bibitem[{Rubin(1974)}]{rubin1974estimating}
Rubin, D.~B., 1974. Estimating causal effects of treatments in randomized and
  nonrandomized studies. Journal of educational Psychology 66~(5), 688.

\bibitem[{Rubin(1996)}]{rubin1996multiple}
Rubin, D.~B., 1996. Multiple imputation after 18+ years. Journal of the
  American statistical Association 91~(434), 473--489.

\bibitem[{Rubin(2004)}]{rubin2004multiple}
Rubin, D.~B., 2004. Multiple imputation for nonresponse in surveys. Vol.~81.
  John Wiley \& Sons.

\bibitem[{Saito et~al.(2008)Saito, Nakano, and Kimura}]{saito2008prediction}
Saito, K., Nakano, R., Kimura, M., 2008. Prediction of information diffusion
  probabilities for independent cascade model. In: International conference on
  knowledge-based and intelligent information and engineering systems.
  Springer, pp. 67--75.

\bibitem[{S{\"a}vje et~al.(2017)S{\"a}vje, Aronow, and
  Hudgens}]{savje2017average}
S{\"a}vje, F., Aronow, P.~M., Hudgens, M.~G., 2017. Average treatment effects
  in the presence of unknown interference. arXiv preprint arXiv:1711.06399.

\bibitem[{Shah et~al.(2014)Shah, Bartlett, Carpenter, Nicholas, and
  Hemingway}]{shah2014comparison}
Shah, A.~D., Bartlett, J.~W., Carpenter, J., Nicholas, O., Hemingway, H., 2014.
  Comparison of random forest and parametric imputation models for imputing
  missing data using mice: a caliber study. American journal of epidemiology
  179~(6), 764--774.

\bibitem[{Shu and Yi(2019{\natexlab{a}})}]{shu2019causal}
Shu, D., Yi, G.~Y., 2019{\natexlab{a}}. Causal inference with measurement error
  in outcomes: Bias analysis and estimation methods. Statistical methods in
  medical research 28~(7), 2049--2068.

\bibitem[{Shu and Yi(2019{\natexlab{b}})}]{shu2019weighted}
Shu, D., Yi, G.~Y., 2019{\natexlab{b}}. Weighted causal inference methods with
  mismeasured covariates and misclassified outcomes. Statistics in medicine
  38~(10), 1835--1854.

\bibitem[{Sobel(2006)}]{sobel2006}
Sobel, M.~E., 2006. What do randomized studies of housing mobility demonstrate?
  Journal of the American Statistical Association 101~(476), 1398--1407.

\bibitem[{Sofrygin and van~der Laan(2017)}]{Sofrygin2017}
Sofrygin, O., van~der Laan, M., 2017. Semi-parametric estimation and inference
  for the mean outcome of the single time-point intervention in a causally
  connected population. Journal of Causal Inference 5~(1), 20160003.

\bibitem[{Steckler et~al.(1992)Steckler, Goodman, McLeroy, Davis, and
  Koch}]{steckler1992measuring}
Steckler, A., Goodman, R.~M., McLeroy, K.~R., Davis, S., Koch, G., 1992.
  Measuring the diffusion of innovative health promotion programs. American
  journal of health promotion 6~(3), 214--224.

\bibitem[{Tan(2006)}]{tan2006distributional}
Tan, Z., 2006. A distributional approach for causal inference using propensity
  scores. Journal of the American Statistical Association 101~(476),
  1619--1637.

\bibitem[{Tan(2010)}]{tan2010bounded}
Tan, Z., 2010. Bounded, efficient and doubly robust estimation with inverse
  weighting. Biometrika 97~(3), 661--682.

\bibitem[{Tan(2013)}]{tan2013variance}
Tan, Z., 2013. Variance estimation under misspecified models.

\bibitem[{Tchetgen and VanderWeele(2012)}]{tchetgen2012causal}
Tchetgen, E. J.~T., VanderWeele, T.~J., 2012. On causal inference in the
  presence of interference. Statistical methods in medical research 21~(1),
  55--75.

\bibitem[{Tort{\`u} et~al.(2020)Tort{\`u}, Forastiere, Crimaldi, and
  Mealli}]{tortu2020modelling}
Tort{\`u}, C., Forastiere, L., Crimaldi, I., Mealli, F., 2020. Modelling
  network interference with multi-valued treatments: the causal effect of
  immigration policy on crime rates. arXiv preprint arXiv:2003.10525.

\bibitem[{Valente(1993)}]{valente1993diffusion}
Valente, T.~W., 1993. Diffusion of innovations and policy decision-making.
  Journal of Communication 43~(1), 30--45.

\bibitem[{Valente(2005)}]{valente2005network}
Valente, T.~W., 2005. Network models and methods for studying the diffusion of
  innovations. Models and methods in social network analysis 28, 98--116.

\bibitem[{Vanderweele(2012)}]{vanderweele2012inference}
Vanderweele, T.~J., 2012. Inference for additive interaction under exposure
  misclassification. Biometrika 99~(2), 502--508.

\bibitem[{Wang et~al.(2012)Wang, Chen, and Wang}]{wang2012scalable}
Wang, C., Chen, W., Wang, Y., 2012. Scalable influence maximization for
  independent cascade model in large-scale social networks. Data Mining and
  Knowledge Discovery 25~(3), 545--576.

\bibitem[{Yanagi(2018)}]{yanagi2018inference}
Yanagi, T., 2018. Inference on local average treatment effects for
  misclassified treatment. Available at SSRN 3065923.

\bibitem[{Yang and Zhou(2013)}]{yang2013credit}
Yang, J., Zhou, Y., 2013. Credit risk spillovers among financial institutions
  around the global credit crisis: Firm-level evidence. Management Science
  59~(10), 2343--2359.

\bibitem[{Zheng et~al.(2013)Zheng, L{\"u}, Zhao, et~al.}]{zheng2013spreading}
Zheng, M., L{\"u}, L., Zhao, M., et~al., 2013. Spreading in online social
  networks: The role of social reinforcement. Physical Review E 88~(1), 012818.

\bibitem[{Zhou et~al.(2009)Zhou, L{\"u}, and Zhang}]{zhou2009predicting}
Zhou, T., L{\"u}, L., Zhang, Y.-C., 2009. Predicting missing links via local
  information. The European Physical Journal B 71~(4), 623--630.

\end{thebibliography}
	
	\newpage
	\appendix
	\begin{center}
		{\bf \huge Online Appendix}
	\end{center}
	
	\pagenumbering{arabic}
	\setcounter{page}{1}

	\section{Diffusion Bias}
	\label{app: bias}
	Here, we derive the bias due to the hidden treatment diffusion process. 
	If the policy maker neglected the possibility of any diffusion process playing a role in the analysis, she would estimate the quantity $\tau^{no-diff}= \mathbb{E} \; \Big [ Y_{it''}|Z_{it}=1 \Big ] -  \mathbb{E} \; \Big [ Y_{it''}|Z_{it}=0 \Big ]$, that is 
	\begin{align*}
		\tau^{no-diff} & =  \mathbb{E} \; \Big [ Y_{it''}|Z_{it}=1 \Big ] - \mathbb{E} \; \Big [ Y_{it''}|Z_{it}=0 \Big ] \\
		& = \Bigg\{ \; \mathbb{E} \; \Big [ Y_{it''}|Z_{it'}=1, Z_{it}=1 \Big ] P(Z_{it'}=1|Z_{it}=1) +  \mathbb{E} \; \Big [ Y_{it''}|Z_{it'}=0, Z_{it}=1 \Big ] P(Z_{it'}=0|Z_{it}=1) \; \Bigg\} - \\
		& \;\;\;\;\; \Bigg\{ \; \mathbb{E} \; \Big [ Y_{it''}|Z_{it'}=1, Z_{it}=0 \Big ] P(Z_{it'}=1|Z_{it}=0) + \mathbb{E} \; \Big [ Y_{it''}|Z_{it'}=0, Z_{it}=0 \Big ] P(Z_{it'}=0|Z_{it}=0)\; \Bigg\} \\
		&=\; \mathbb{E} \; \Big [ Y_{it''}(1)|Z_{it'}=1, Z_{it}=1 \Big ] - 
		\mathbb{E} \; \Big [ Y_{it''}(1)|Z_{it'}=1, Z_{it}=0 \Big ] \mathbb{E}[\rho_i|Z_{it}=0] - \\
		& \;\;\;\;\; 
		\mathbb{E} \; \Big [ Y_{it''}(0)|Z_{it'}=0, Z_{it}=0 \Big ] (1-\mathbb{E}[\rho_i|Z_{it}=0])\
		\\
		&=\; \mathbb{E} \; \Big [ Y_{it''}(1)|Z_{it}=1 \Big ] - 
		\mathbb{E} \; \Big [ Y_{it''}(1)|Z_{it'}=1, Z_{it}=0 \Big ] \mathbb{E}[\rho_i|Z_{it}=0 ]-\\
		&\;\;\;\;\;  
		\mathbb{E} \; \Big [ Y_{it''}(0)|Z_{it'}=0, Z_{it}=0 \Big ] (1-\mathbb{E}[\rho_i|Z_{it}=0 ])\,,
	\end{align*}
	where we have used that
	\begin{align*}
		P(Z_{it'}=1|Z_{it}=1)=&1\quad \mbox{(if a node was treated at time $t$, she is surely treated at time $t'$)},\\
		P(Z_{it'}=0|Z_{it}=1)=&0\quad\mbox{(if a node was treated at time $t$, there is no possibility that she is untreated at time $t'$ )}
	\end{align*}
	and $\rho_i=\rho_i(\mathbf{Z}_{-it},\mathbf{G})=P(Z_{it'}=1|Z_{it}=0,\boldsymbol{Z}_{-it},\mathbf{G})$. (The quantity $\mathbb{E}[\rho_i|Z_{it}=0 ]$ denotes the averaged value of $\rho_i$ on all the possible configurations of $\mathbf{Z}_{-it}$ and $\mathbf{G}$ given $Z_{it}=0$.)
	%
	%
	
	The treatment diffusion bias $b$ can be obtained by computing the difference between $\tau^{no-diff}$ and $\tau^\star$, that is $b=\tau^{no-diff}-\tau^{\star}$, where $\tau^\star=\mathbb{E}[Y_{it''}(1)]-\mathbb{E}[Y_{it''}(0)]$. Under the first assumption included in Assumption \ref{ass:unc} we have $\mathbb{E} \; \Big [ Y_{it''}(1)|Z_{it}=1 \Big ]=\mathbb{E}[Y_{it''}(1)] $ and so 
	\begin{align*}
		\small
		b&=  \mathbb{E} \; \Big [ Y_{it''}(1) \Big ] - 
		\mathbb{E} \; \Big [ Y_{it''}(1)|Z_{it'}=1, Z_{it}=0 \Big ] \mathbb{E}[\rho_i |Z_{it}=0 ]- \\
		&\;\;\;\;\;\mathbb{E} \; \Big [ Y_{it''}(0)|Z_{it'}=0, Z_{it}=0 \Big ] (1-\mathbb{E}[\rho_i|Z_{it}=0 ])-
		\mathbb{E}[Y_{it''}(1)]+\mathbb{E}[Y_{it''}(0)]\\
		&=  \mathbb{E} \; \Big [ Y_{it''}(0) \Big ] -\mathbb{E} \; \Big [ Y_{it''}(0)|Z_{it'}=0, Z_{it}=0 \Big ] (1-\mathbb{E}[\rho_i |Z_{it}=0]) -\\&\;\;\;\;\; 
		\mathbb{E} \; \Big [ Y_{it''}(1)|Z_{it'}=1, Z_{it}=0 \Big ] \mathbb{E}[\rho_i|Z_{it}=0 ]. 
	\end{align*}
	
	{\bf Remark:} If the Assumption \ref{ass:unc} is not satisfied, the quantity $b$ is given by the above quantity plus $\mathbb{E}[Y_{it''}(1)|Z_{it}=1]-\mathbb{E}[Y_{it''}(1)]$.
	
	{\bf Remark:} If the entire Assumption \ref{ass:unc}  (first and second part) is satisfied, then 
	we have 
	\begin{equation*}
		\begin{split}
			\mathbb{E}\Big [ Y_{it''}(0)|Z_{it'}=0, Z_{it}=0 \Big ]&=
			\frac{\mathbb{E}\Big[Y_{it''}(0)I(Z_{it'}=0)| Z_{it}=0 \Big]}{P(Z_{it'}=0|Z_{it}=0)}\\
			&=\frac{\mathbb{E}\Big[\mathbb{E}[Y_{it''}(0)I(Z_{it'}=0)| Z_{it}=0, \mathbf{Z}_{-it},\mathbf{G}] | Z_{it}=0\Big]}{P(Z_{it'}=0|Z_{it}=0)}\\
			&=\frac{\mathbb{E}\Big[\mathbb{E}[Y_{it''}(0)| Z_{it}=0, \mathbf{Z}_{-it},\mathbf{G}] \mathbb{E}[I(Z_{it'}=0)| Z_{it}=0, \mathbf{Z}_{-it},\mathbf{G}]| Z_{it}=0\Big]}{P(Z_{it'}=0|Z_{it}=0)}\\
			&=\frac{\mathbb{E}\Big[\mathbb{E}[Y_{it''}(0)] P(Z_{it'}=0| Z_{it}=0, \mathbf{Z}_{-it},\mathbf{G})| Z_{it}=0\Big]}{P(Z_{it'}=0|Z_{it}=0)}\\
			&=\mathbb{E}[Y_{it''}(0)].
		\end{split}
	\end{equation*}
	Similarly, we have $\mathbb{E}\Big [ Y_{it''}(1)|Z_{it'}=1, Z_{it}=0 \Big ]=\mathbb{E}[Y_{it''}(1)]$.
	Hence the expression of the bias $b$ results simpler.
	
	\section{Proofs for the proposed estimators}
	\label{app:estimators}
	\subsection{Estimator $\tau_1^{\star}$ for Bernoulli randomized experiments }
	\label{app:estimators_indep}
	\begin{proposition}
		Under the independence between $Z_{it}$ and $[\mathbf{Z}_{-it},\mathbf{G}]$, for all $i\in\mathcal{N}$,  and under the 
		Assumption \ref{ass:unc},  the estimator 
		\begin{align*}
			\widehat{\tau}_1^{\star} = \frac{1}{N} \:\Big[ \sum_{i=1}^{N}Z_{it'}\frac{Y_{it''}}{\pi_{it'}(1;\mathbf{Z}_{-it},\mathbf{G})}-
			\sum_{i=1}^{N}(1-Z_{it'})\frac{Y_{it''}}{1-\pi_{it'}(1;\mathbf{Z}_{-it},\mathbf{G})}\:\Big]
		\end{align*}
		is unbiased.
	\end{proposition}
	
	\begin{proof}
		We recall that, under the independence between $Z_{it}$ and $[\mathbf{Z}_{-it},\mathbf{G}]$, we have 
		$$
		\pi_{it'}(1;\mathbf{Z}_{-it},\mathbf{G})=
		\pi_{it}(1;\mathbf{Z}_{-it},\mathbf{G})+
		\rho_i(\mathbf{Z}_{-it},\mathbf{G})
		(1-\pi_{it}(1;\mathbf{Z}_{-it},\mathbf{G}))=
		\pi_{it}(1)+\rho_i(\mathbf{Z}_{-it},\mathbf{G})(1-\pi_{it}(1))\in (0,1)
		$$
		(since $\pi_{it}(1)\in (0,1)$ and $\rho_i<1$). Moreover, we have 
		\begin{align*}
			&\mathbb{E}\Bigg[\frac{Y_{it''}Z_{it'}}{\pi_{it'}(1;\mathbf{Z}_{-it},\mathbf{G})}\Bigg]
			=\mathbb{E}\Bigg[\frac{Y_{it''}I(Z_{it'}=1)}{\pi_{it'}(1; \mathbf{Z}_{-it},\mathbf{G})}\Bigg]
			=\mathbb{E}\Bigg[\frac{Y_{it''}(1)I(Z_{it'}=1)}{\pi_{it'}(1; \mathbf{Z}_{-it},\mathbf{G})}\Bigg]
			=\mathbb{E}\Bigg[\mathbb{E}\Bigg[\frac{Y_{it''}(1)I(Z_{it'}=1)}{\pi_{it'}(1; \mathbf{Z}_{-it},\mathbf{G})}| \mathbf{Z}_{-it},\mathbf{G}\Bigg]\Bigg]\\
			&=\mathbb{E}\Bigg[\frac{1}{\pi_{it'}(1;\mathbf{Z}_{-it},\mathbf{G})}
			\mathbb{E}\Bigg[Y_{it''}(1)I(Z_{it'}=1)| \mathbf{Z}_{-it},\mathbf{G}\Bigg]\Bigg].
		\end{align*}
		Now, it is enough to observe that 
		\begin{align*}
			&\mathbb{E}\Bigg[Y_{it''}(1)I(Z_{it'}=1)| \mathbf{Z}_{-it},\mathbf{G}\Bigg]=
			\mathbb{E}\Bigg[Y_{it''}(1)I(Z_{it'}=1)| Z_{it}=1,\mathbf{Z}_{-it},\mathbf{G}\Bigg]
			P(Z_{it}=1|\mathbf{Z}_{-it},\mathbf{G})+\\
			&\mathbb{E}\Bigg[Y_{it''}(1)I(Z_{it'}=1)| Z_{it}=0,\mathbf{Z}_{-it},\mathbf{G}\Bigg]
			P(Z_{it}=0|\mathbf{Z}_{-it},\mathbf{G})=\\
			&\mathbb{E}\Bigg[Y_{it''}(1)| Z_{it}=1,\mathbf{Z}_{-it},\mathbf{G}\Bigg]
			\pi_{it}(1;\mathbf{Z}_{-it},\mathbf{G})+\\
			&\mathbb{E}\Bigg[Y_{it''}(1)|Z_{it}=0,\mathbf{Z}_{-it},\mathbf{G}\Bigg]P(Z_{it'}=1| Z_{it}=0,\mathbf{Z}_{-it},\mathbf{G})
			(1-\pi_{it}(1;\mathbf{Z}_{-it},\mathbf{G}))=\\
			&\mathbb{E}[Y_{it''}(1)]
			\pi_{it}(1;\mathbf{Z}_{-it},\mathbf{G})+
			\mathbb{E}[Y_{it''}(1)]\rho_i(\mathbf{Z}_{-it},\mathbf{G})
			(1-\pi_{it}(1;\mathbf{Z}_{-it},\mathbf{G}))=\\
			&\mathbb{E}[Y_{it''}(1)]\pi_{it'}(1;\mathbf{Z}_{-it},\mathbf{G}).
		\end{align*}
		(For the second equality we have used that $Z_{it}=1$ implies $Z_{it'}=1$, the second part of Assumption \ref{ass:unc} and the definition of $\pi_{it}(1; \mathbf{Z}_{-it},\mathbf{G})$. For the third equality we have used the first part of Assumption \ref{ass:unc} and the definition of $\rho_i$. Finally, for the last equality, we have used the definition of $\pi_{it'}(1; \mathbf{Z}_{-it},\mathbf{G})$.)
		\\
		\indent Similarly, we have 
		\begin{align*}
			&\mathbb{E}\Bigg[\frac{Y_{it''}(1-Z_{it'})}{(1-\pi_{it'}(1;\mathbf{Z}_{-it},\mathbf{G}))}\Bigg]
			=\mathbb{E}\Bigg[\frac{Y_{it''}I(Z_{it'}=0)}{(1-\pi_{it'}(1; \mathbf{Z}_{-it},\mathbf{G}))}\Bigg]
			=\mathbb{E}\Bigg[\frac{Y_{it''}(0)I(Z_{it'}=0)}{(1-\pi_{it'}(1; \mathbf{Z}_{-it},\mathbf{G}))}\Bigg]
			=\\
			&\mathbb{E}\Bigg[\mathbb{E}\Bigg[\frac{Y_{it''}(0)I(Z_{it'}=0)}{(1-\pi_{it'}(1; \mathbf{Z}_{-it},\mathbf{G}))}| \mathbf{Z}_{-it},\mathbf{G}\Bigg]\Bigg]
			=\mathbb{E}\Bigg[\frac{1}{(1-\pi_{it'}(1;\mathbf{Z}_{-it},\mathbf{G}))}
			\mathbb{E}\Bigg[Y_{it''}(0)I(Z_{it'}=0)| \mathbf{Z}_{-it},\mathbf{G}\Bigg]\Bigg],
		\end{align*}
		where
		\begin{align*}
			&\mathbb{E}\Bigg[Y_{it''}(0)I(Z_{it'}=0)| \mathbf{Z}_{-it},\mathbf{G}\Bigg]=
			\mathbb{E}\Bigg[Y_{it''}(0)I(Z_{it'}=0)| Z_{it}=1,\mathbf{Z}_{-it},\mathbf{G}\Bigg]
			P(Z_{it}=1|\mathbf{Z}_{-it},\mathbf{G})+\\
			&\mathbb{E}\Bigg[Y_{it''}(0)I(Z_{it'}=0)| Z_{it}=0,\mathbf{Z}_{-it},\mathbf{G}\Bigg]
			P(Z_{it}=0|\mathbf{Z}_{-it},\mathbf{G})=\\
			&\mathbb{E}\Bigg[Y_{it''}(0)|Z_{it}=0,\mathbf{Z}_{-it},\mathbf{G}\Bigg]P(Z_{it'}=0| Z_{it}=0,\mathbf{Z}_{-it},\mathbf{G})
			(1-\pi_{it}(1;\mathbf{Z}_{-it},\mathbf{G}))=\\
			&\mathbb{E}[Y_{it''}(0)](1-\rho_i(\mathbf{Z}_{-it},\mathbf{G}))
			(1-\pi_{it}(1;\mathbf{Z}_{-it},\mathbf{G}))=
			\mathbb{E}[Y_{it''}(0)](1-\pi_{it'}(1;\mathbf{Z}_{-it},\mathbf{G})).
		\end{align*}
	\end{proof}
	
	\subsection{Estimator $\tau_2^{\star}$ for cluster randomized experiments }
	\label{app:estimators_clust}
	
	\begin{proposition}
		Under the independence between $Z_{it}$ and $\mathbf{G}$, for all $i\in\mathcal{N}$, and under Assumption \ref{ass:unc}, the estimator 
		$$
		\widehat{\tau}_2^{\star} = 
		\frac{1}{N} \:\Big[ \sum_{i=1}^{N}Z_{it'}\frac{Y_{it''}}{\pi_{it'}(1;\mathbf{G})}-
		\sum_{i=1}^{N}(1-Z_{it'})\frac{Y_{it''}}{1-\pi_{it'}(1;\mathbf{G})}\:\Big]\,.
		$$
		is unbiased.
	\end{proposition}
	
	\begin{proof} We recall that, under the independence between $Z_{i,t}$ and $\mathbf{G}$, we have 
		$$
		\pi_{it'}(1;\mathbf{G})=\pi_{it}(1;\mathbf{G})+\mathbb{E}[\rho_i|Z_{it}=0,\mathbf{G}](1-\pi_{it}(1;\mathbf{G}))=\pi_{it}(1)+\mathbb{E}[\rho_i|Z_{it}=0,\mathbf{G}](1-\pi_{it}(1))
		\in (0,1)
		$$
		(since $\pi_{it}(1)\in (0,1)$ and $\rho_i<1$). Moreover, we have 
		\begin{align*}
			&\mathbb{E}\Bigg[\frac{Y_{it''}Z_{it'}}{\pi_{it'}(1;\mathbf{G})}\Bigg]
			=\mathbb{E}\Bigg[\frac{Y_{it''}I(Z_{it'}=1)}{\pi_{it'}(1;\mathbf{G})}\Bigg]
			=\mathbb{E}\Bigg[\frac{Y_{it''}(1)I(Z_{it'}=1)}{\pi_{it'}(1;\mathbf{G})}\Bigg]
			=\mathbb{E}\Bigg[\mathbb{E}\Bigg[\frac{Y_{it''}(1)I(Z_{it'}=1)}{\pi_{it'}(1;\mathbf{G})}| \mathbf{G}\Bigg]\Bigg]\\
			&=\mathbb{E}\Bigg[\frac{1}{\pi_{it'}(1;\mathbf{G})}
			\mathbb{E}\Bigg[Y_{it''}(1)I(Z_{it'}=1)|\mathbf{G}\Bigg]\Bigg]
		\end{align*}
		and 
		\begin{align*}
			\mathbb{E}[Y_{it''}(1)I(Z_{it'}=1)|\mathbf{G}]=
			\mathbb{E}[Y_{it''}(1)I(Z_{it'}=1)| Z_{it}=1,\mathbf{G}]
			P(Z_{it}=1|\mathbf{G})+
			\mathbb{E}[Y_{it''}(1)I(Z_{it'}=1)| Z_{it}=0,\mathbf{G}]
			P(Z_{it}=0|\mathbf{G})\,,
		\end{align*}
		where 
		the first term in the sum is 
		$$
		\mathbb{E}[Y_{it''}(1)I(Z_{it'}=1)| Z_{it}=1,\mathbf{G}]P(Z_{it}=1|\mathbf{G})=
		\mathbb{E}[Y_{it''}(1)| Z_{it}=1,\mathbf{G}]\pi_{it}(1;\mathbf{G})=
		\mathbb{E}[Y_{it''}(1)]\pi_{it}(1;\mathbf{G})
		$$
		(because $Z_{it}=1$ implies $Z_{it'}=1$ and by the first part of Assumption \ref{ass:unc}) and 
		the second one is 
		\begin{align*}
			&\mathbb{E}[Y_{it''}(1)I(Z_{it'}=1)| Z_{it}=0,\mathbf{G}]P(Z_{it}=0|\mathbf{G})=
			\\
			&\mathbb{E}\left[\mathbb{E}[Y_{it''}(1)I(Z_{it'}=1)| Z_{it}=0, \mathbf{Z}_{-it},\mathbf{G}] | Z_{it}=0,\mathbf{G}\right]
			(1-\pi_{it}(1;\mathbf{G}))=\\
			&\mathbb{E}\left[
			\mathbb{E}[Y_{it''}(1)| Z_{it}=0, \mathbf{Z}_{-it},\mathbf{G}] 
			\mathbb{E}[I(Z_{it'}=1)| Z_{it}=0, \mathbf{Z}_{-it},\mathbf{G}]
			| Z_{it}=0,\mathbf{G}\right]
			(1-\pi_{it}(1;\mathbf{G}))=\\
			&\mathbb{E}[Y_{it''}(1)] 
			\mathbb{E}\left[ P(Z_{it'}=1| Z_{it}=0, \mathbf{Z}_{-it},\mathbf{G})| Z_{it}=0,\mathbf{G}\right]
			(1-\pi_{it}(1;\mathbf{G}))=\\
			&\mathbb{E}[Y_{it''}(1)] \mathbb{E}[ \rho_i | Z_{it}=0,\mathbf{G}](1-\pi_{it}(1;\mathbf{G}))
		\end{align*}
		(because of Assumption \ref{ass:unc} and the definition of $\rho_i$). 
		Therefore, combining together the two terms and using the expression of $\pi_{it'}(1;\mathbf{G})$, 
		we get  
		$\mathbb{E}[Y_{it''}(1)I(Z_{it'}=1)|\mathbf{G}]=
		\mathbb{E}[Y_{it''}(1)]\pi_{it'}(1;\mathbf{G})\,.
		$
		Similarly, we have 
		\begin{align*}
			&\mathbb{E}\Bigg[\frac{Y_{it''}(1-Z_{it'})}{(1-\pi_{it'}(1;\mathbf{G}))}\Bigg]
			=\mathbb{E}\Bigg[\frac{Y_{it''}I(Z_{it'}=0)}{(1-\pi_{it'}(1;\mathbf{G}))}\Bigg]
			=\mathbb{E}\Bigg[\frac{Y_{it''}(0)I(Z_{it'}=0)}{(1-\pi_{it'}(1;\mathbf{G}))}\Bigg]
			=\\
			&\mathbb{E}\Bigg[\mathbb{E}\Bigg[\frac{Y_{it''}(0)I(Z_{it'}=0)}{(1-\pi_{it'}(1;\mathbf{G}))}| \mathbf{G}\Bigg]\Bigg]
			=\mathbb{E}\Bigg[\frac{1}{(1-\pi_{it'}(1;\mathbf{G}))}
			\mathbb{E}\Bigg[Y_{it''}(0)I(Z_{it'}=0)| \mathbf{G}\Bigg]\Bigg]
		\end{align*}
		and
		\begin{align*}
			\mathbb{E}[Y_{it''}(0)I(Z_{it'}=0)|\mathbf{G}]&=
			\mathbb{E}[Y_{it''}(0)I(Z_{it'}=0)| Z_{it}=1,\mathbf{G}]
			P(Z_{it}=1|\mathbf{G})+
			\mathbb{E}[Y_{it''}(0)I(Z_{it'}=0)| Z_{it}=0,\mathbf{G}]
			P(Z_{it}=0|\mathbf{G})\\
			&=\mathbb{E}[Y_{it''}(0)I(Z_{it'}=0)| Z_{it}=0,\mathbf{G}](1-\pi_{it}(1;\mathbf{G}))\,,
		\end{align*}
		where
		\begin{align*}
			&\mathbb{E}[Y_{it''}(0)I(Z_{it'}=0)| Z_{it}=0,\mathbf{G}]=
			\mathbb{E}\left[\mathbb{E}[Y_{it''}(0)I(Z_{it'}=0)| Z_{it}=0, \mathbf{Z}_{-it},\mathbf{G}] 
			| Z_{it}=0,\mathbf{G}\right]=\\
			&\mathbb{E}\left[
			\mathbb{E}[Y_{it''}(0)| Z_{it}=0, \mathbf{Z}_{-it},\mathbf{G}] 
			\mathbb{E}[I(Z_{it'}=0)| Z_{it}=0, \mathbf{Z}_{-it},\mathbf{G}] 
			| Z_{it}=0,\mathbf{G}\right]=\\
			&\mathbb{E}[Y_{it''}(0)] 
			\mathbb{E}\left[P(Z_{it'}=0| Z_{it}=0, \mathbf{Z}_{-it},\mathbf{G}) | Z_{it}=0,\mathbf{G}\right]
			=\mathbb{E}[Y_{it''}(0)] 
			\left(1-\mathbb{E}[\rho_i| Z_{it}=0,\mathbf{G}]\right)\,.
		\end{align*}
		Taking into account that $\left(1-\mathbb{E}[\rho_i| Z_{it}=0,\mathbf{G}]\right)(1-\pi_{it}(1;\mathbf{G}))=(1-\pi_{it'}(1;\mathbf{G}))$,
		we get
		$$
		\mathbb{E}[Y_{it''}(0)I(Z_{it'}=0)|\mathbf{G}]=
		\mathbb{E}[Y_{it''}(0)](1-\pi_{it'}(1;\mathbf{G}))\,.
		$$
	\end{proof}
	
	
	\section{Technical details for computing the post diffusion ATE in cluster randomized settings}
	\label{app: techstarstar}
	
	In the empirical framework presented in Section \ref{sec: ea}, the initial assignment of the treatments does not depend on neither the individual covariates or the baseline social network (randomly generated starting from the partial information and from the covariates), but the random variables  $Z_{it}$ e $\mathbf{Z}_{-it}$ are not independent (since unit $i$'s classmates are treated as $i$). Therefore we have
	$$
	\pi_{it}(1;\mathbf{Z}_{-it},\mathbf{G})=
	\pi_{it}(1;\mathbf{Z}_{-it})=\begin{cases}
		&1\quad\mbox{if } Z_j=1\;\forall j\in C(i)\setminus\{i\}\\
		&0\quad\mbox{otherwise}\,,
	\end{cases}
	$$
	where  $C(i)$ denotes unit $i$'s class \footnote{Note that we have not independence also among classes, because exactly $5$ classes are drawn and assigned to the treatment.}. It follows that the estimator $\widehat{\tau}^\star_1$ is not suitable for the considered application and so we employ the alternative estimator $\widehat{\tau}_2^{\star}$.  
	For the computation of this last estimator, taking into account  that the initial treatment $\mathbf{Z}_t$ does not depend on the covariates $\mathbf{X}$ and it is independent of $\mathbf{G}$, we need to compute the quantity
	$$
	\pi_{it'}(1; \mathbf{G})=\pi_{it}(1) + \mathbb{E}[\rho_i|Z_{it}=0,\mathbf{G}] ( 1-\pi_{it}(1) )=
	\frac{1}{2}+\mathbb{E}[\rho_i|Z_{it}=0,\mathbf{G}] \frac{1}{2}\,,
	$$
	where 
	\begin{align*}
		E[\rho_i|Z_{it}=0,\mathbf{G}]&=\sum_{\mathbf{z}_{-i}} \rho_i(\mathbf{z}_{-i},\mathbf{G}) P(\mathbf{Z}_{-it}=\mathbf{z}_{-i}|Z_{it}=0,\mathbf{G})\\
		&=\sum_{\mathbf{z}_{-i}} \rho_i(\mathbf{z}_{-i},\mathbf{G}) P(\mathbf{Z}_{-it}=\mathbf{z}_{-i}|Z_{it}=0)\\
		&=\sum_{i_1\neq i_2\neq\dots\neq i_5\in \mathcal{C}\setminus C(i)}
		[1-(1-\overline{p})^{T_{it}}]P(\mathcal{C}_T=\{C_{i_1},\dots,C_{i_5}\}\, | Z_{it}=0)\\
		&=
		\frac{1}{\binom{9}{5}}\sum_{i_1\neq i_2\neq \dots \neq i_5\in \mathcal{C}\setminus C(i)}
		[1-(1-\overline{p})^{T_{it}}]
	\end{align*}
	with $\mathcal{C}=\{1,\dots,10\}=$ possible classes, $\mathcal{C}_T=$ treated classes at time $t$
	and  $T_{it}=T_{it}(i_1,\dots,i_5,\mathbf{G})$=number of unit $i$'s friends, with respect to $\mathbf{G}$, in the treated classes $C_{i_1},\dots,C_{i_5}$.
	
	Instead of computing the quantity $E[\rho_i|Z_{it}=0,\mathbf{G}]$, we can observe that 
	$$
	\mathbb{E}[\rho_i|Z_{it}=0,\mathbf{G}] \frac{1}{2}=
	E[\rho_i|Z_{it}=0,\mathbf{G}]\, P(Z_{it}=0 | \mathbf{G})=
	E[\rho_i\,I(Z_{it}=0) |\mathbf{G}].
	$$
	Therefore, we can replace the above quantity by its empirical mean (given $\mathbf{G}$). More precisely, given $\mathbf{G}$, we simulate $\Omega$ possible scenarios of initial treatments and then we compute
	the empirical mean of the obtained random variables  $\rho_i(\mathbf{Z}_{-it},\mathbf{G}) I(Z_{it}=0)$, that is  
	$\frac{1}{\Omega}\sum_{\omega=1}^\Omega  \rho_i(\mathbf{Z}_{-it}^{(\omega)}, \mathbf{G}) I(Z^{(\omega)}_{it}=0)\,.
	$
	\section{Illustrative Simulations in Cluster Randomized Setting}
	\label{app: simul_clust}
	In Section \ref{sec: simul} we have observed the performance of the sensitivity analysis in a Bernoulli randomized setting, where the treatment assignment of one units is independent on that of any other unit in the sample. However, this could not be the case. For instance, in our motivating application the treatment of interest has been assigned at a cluster level, so that students belonging to the same class have been all assigned to the same treatment status. In Section \ref{subsec: trd_estim}, we present the unbiased estimator $\widehat{\tau}^{\star}_2$, which can be fairly adopted for estimating the average treatment effect in the cluster randomized scenario. Here, we show how this estimator performs in a simulated setting, which is particularly adherent to our proposed application for what concerns the initial treatment assignment. The idea behind the simulated study is identical to the one presented in the Bernoulli randomized scenario. However, here the data generating process accounts for the clustered structure of data and for the cluster randomized design. Here, we consider a sample made up by $N=200$ units, belonging to $C=10$ exogenous clusters, of which $C_{T}=5$ are assigned to the active intervention. Thus the individual probability of being treated at time $t$ is $C_{T}/C=0.5$. Individuals are linked according to a fully observable network $\mathbf{G}$, where within-clusters ties are more likely to occur than between-clusters ties (they respectively show up with probability 0.2 and 0.02, respectively). All the quantities related to the diffusion process have been defined exactly as in the Bernoulli randomized setting. The only difference is that the effect $\tau_{i}^{*}$ has here a slightly different characterization: the dummy indicator $N^{high}_{i}$ signaling those controls who have an high probability of receiving the treatment by diffusion is here replaced by a similar indicator which is based on the distribution of the number of links established with nodes who belong to a different cluster (\emph{extra} cluster links), instead of being related to the simple in-degree distribution. In detail, the indicator that we use here, $N^{high,extra}_i$ equals 1 if the number of extra-cluster links of $i$ is above a given cutoff $\nu$,  ($N^{in,extra}_{i} > \nu$). This cutoff is identified by the distribution of the number of extra cluster ties: specifically, we take $\nu$ as the median of the number of the extra cluster ties. We condition on the distribution of the number of extra cluster ties instead of simply conditioning on the in-degree because in a cluster randomized study, extra cluster ties are exactly the ones that drive the treatment diffusion process: indeed, no diffusion spreading is possible within clusters.    
	
\begin{figure}[H]
		\centering
		\includegraphics[width=150mm]{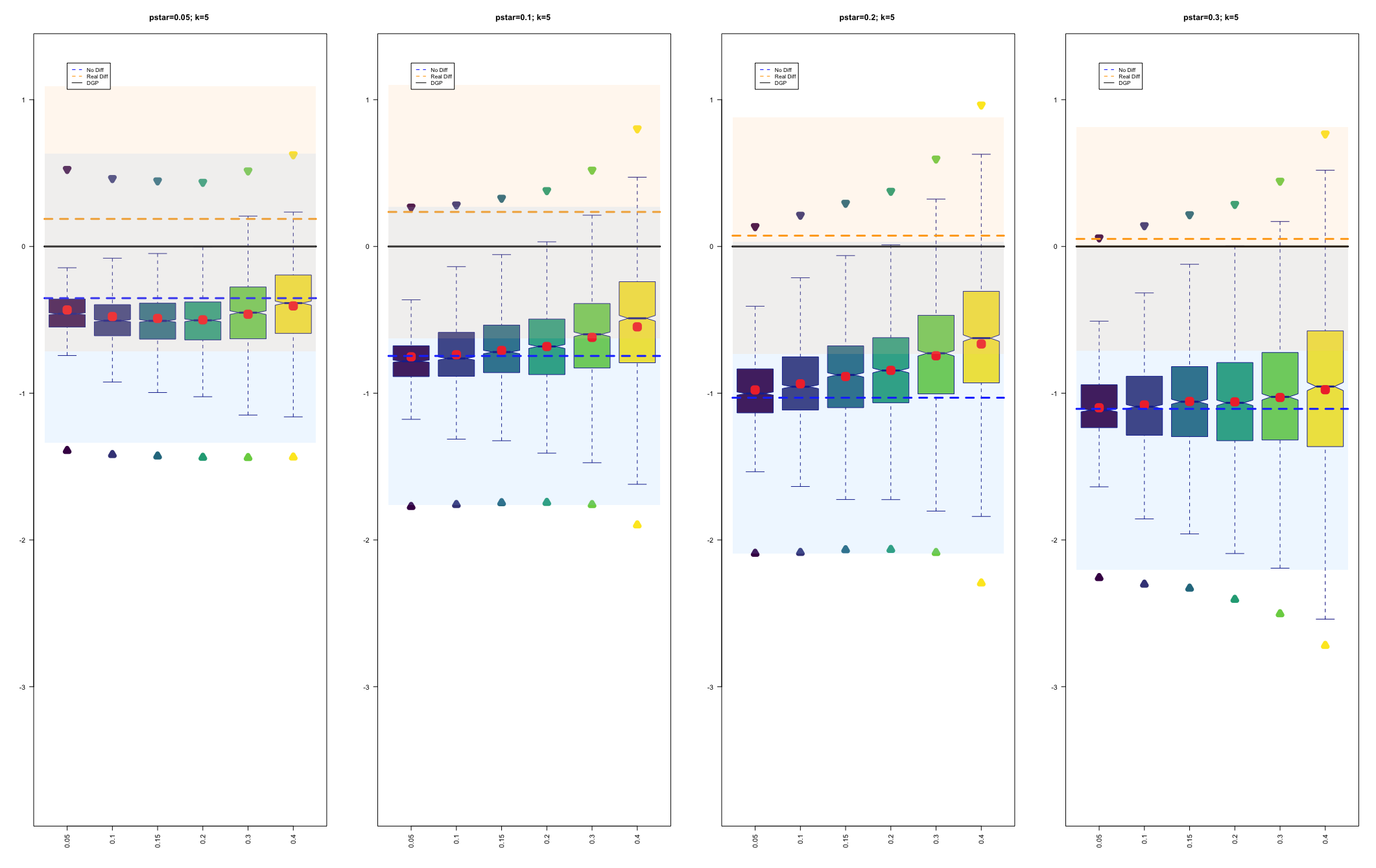}
		\caption[Underestimation, $k=1$: cluster randomized setting]{Underestimation, $k=1$: cluster randomized setting}
		\label{fig: undest_low_clust}
	\end{figure}

\begin{figure}[H]
		\centering
		\includegraphics[width=150mm]{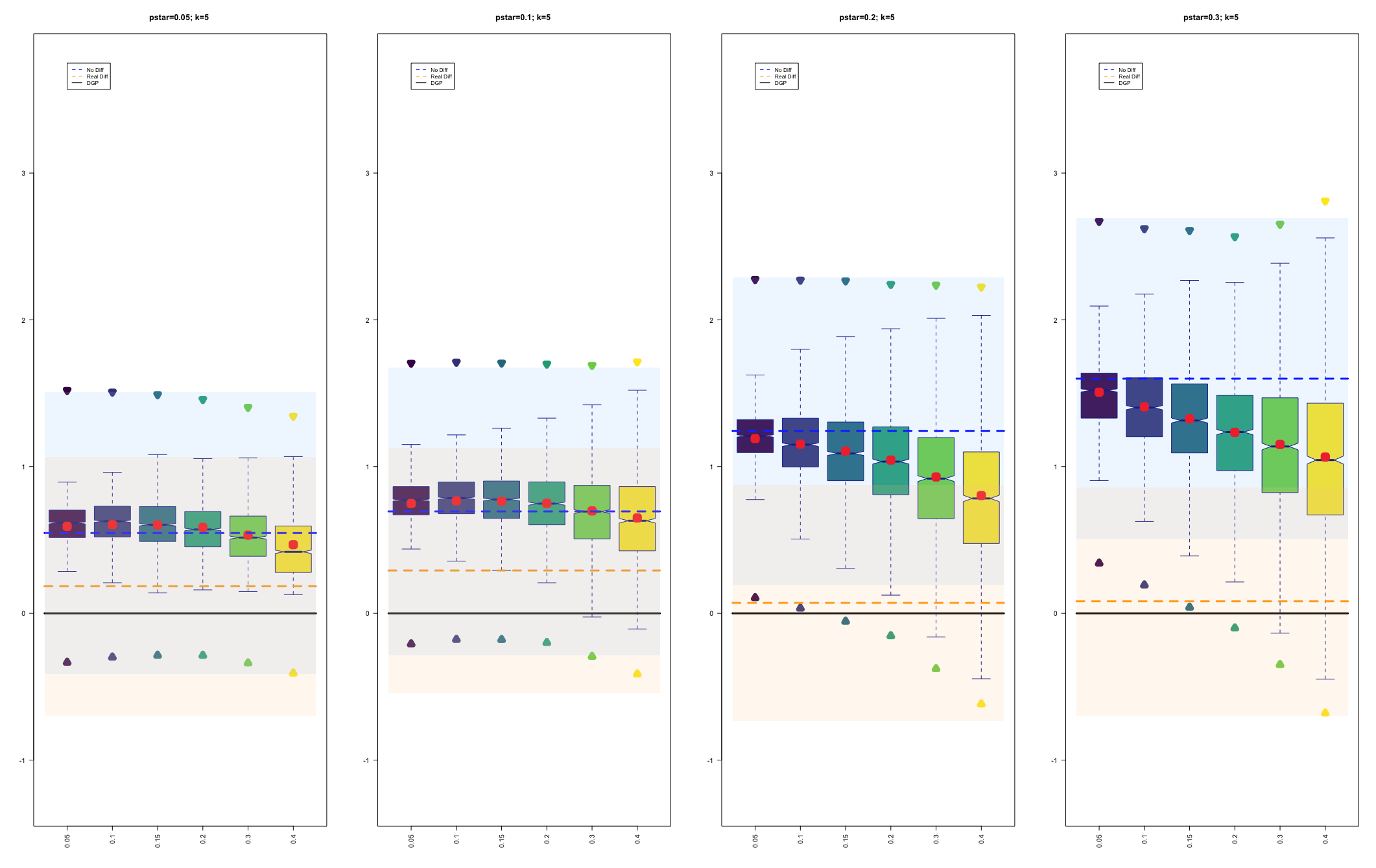}
		\caption[Overestimation, $k=1$: cluster randomized setting]{Overestimation, $k=1$: cluster randomized setting}
		\label{fig: overest_low_clust}
	\end{figure}

	Figure \ref{fig: undest_low_clust} shows results in the underestimation scenario, with $k=5$. It is immediate to note that our proposed methodology with the estimator $\widehat{\tau}_2^{\star}$ performs well in reducing the estimation bias due to the diffusion process.
	Figure \ref{fig: overest_low_clust} shows the corresponding results in the overestimation scenario, with $k=5$. Here, ignoring the treatment diffusion process leads to an overestimation of the real effect of the intervention. The proposed sensitivity analysis produces a downward shifting in the estimates, towing them towards 0. 
	Summing up, we can state that the methodology that we are proposing in this work performs well in reducing the estimation bias due to the diffusion process, both in the overestimation and in the underestimation scenario. The procedure works well in both Bernoulli randomized experimental designs and cluster randomized designs.
	
	
	\section{Similarity Measures}
	\label{app: similarity}
	In this section, we detail the four dyadic similarity measures that we employed for imputing missing links among dyads. They catch the baseline degree of similarity between each pair of students with respect to hobbies, school attitudes, cultural interests and personal background. They all have been derived starting from individual-level covariates and they have been computed as follows:
	\begin{enumerate}
		\item Hobbies Similarity $x1_{ij}$ (Jaccard similarity between hobbies of units $i$ and $j$): specifically, students were asked whether they are interested in politics, whether they practice sports, gymnastics or volunteering and whether they love painting, listening music, chatting on social networks and watching TV. 
		\item School Attitudes Similarity $x2_{ij}$: this variable represents a measure of similarity with respect to the school attitudes. This quantity involves both an evaluation of the individual academic performance, expressed in terms of grade point average (gpa), and the attendance of specific extracurricular activities offered by the school (music lessons, language lessons, humanities lessons). The dyadic similarity in school performance between units $i$ and $j$, that we call $x2^{a}_{ij}$, is  measured according to the formula $x2^{a}_{ij}=1-\frac{|gpa_{i}-gpa_{j}|}{\max(gpa)-\min(gpa)}$; while, the variable $x2^{b}_{ij}$ quantifies the extent of similarity between $i$ and $j$ in terms of school activities through a Jaccard measure.  The final $x2_{ij}$ value results from the mean of these two measures, that is $x2_{ij}=\frac{x2^{a}_{ij}+x2^{b}_{ij}}{2}$.
		\item Cultural Interests Similarity $x3_{ij}$: this variable indicates the level of similarity with respect to the individual baseline attitude towards culture. Students have been asked to grade the frequency of how they practice the following interests: book reading, symphony listening, theatrical shows watching, cinema going. Higher values correspond to a more frequent accomplishment of that specific activity. The $x3_{ij}$ variable is obtained by computing the Euclidean distance among these measures (then subtracting the resulting value from 1, so to get a similarity measure, instead that a measure of distance). This value has been in turn standardized, so to get a measure which varies between 0 and 1.   
		\item Personal Background Similarity $x4_{ij}$: this variable measures the level of similarity among individual personal characteristics. Formally, given some students $i$ and $j$, $x4_{ij}$ is defined as a Jaccard similarity of their respective personal features: in particular, the personal characteristics that are included in this evaluation are related to the gender, to the seniority, to the geographical origin of the individual (the survey asks the student to declare if she/he is born abroad or not) and to the current living area (the survey asks the student to declare if she/he is living in suburban areas or not). 
	\end{enumerate}
	
	
	\section{Multiple Imputation Algorithm: Stability}
	\label{app: stability}
\begin{figure}[H]
		\centering
		\includegraphics[width=120mm,height=40mm]{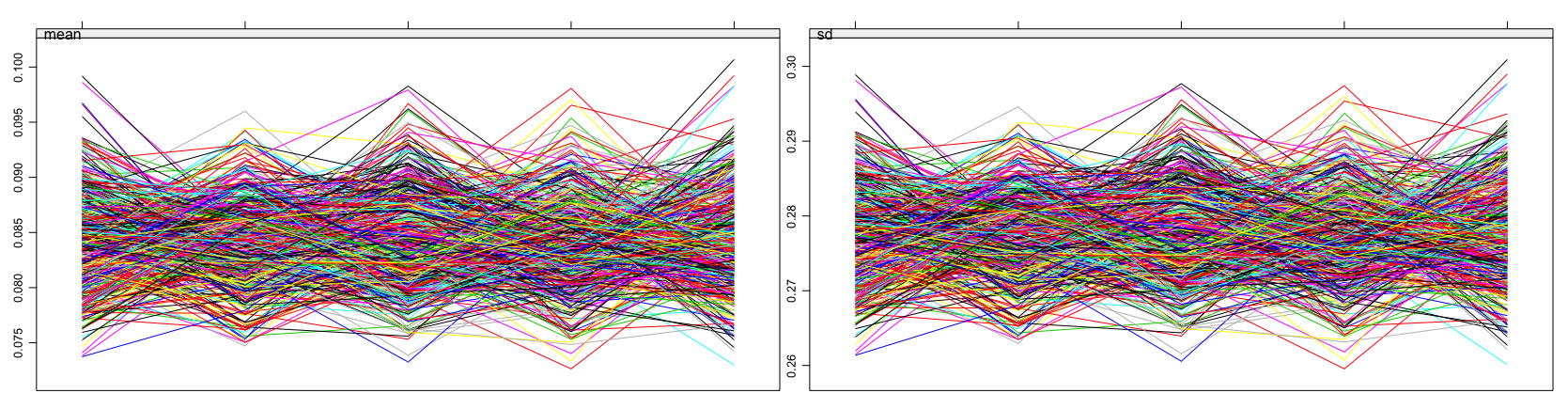}
		\caption{Mean and standard deviation of the link indicator variable: trace plot monitoring the trend of this values over the 5 iterations performed by the algorithm, in the $M=500$ distinct data-imputations.}
		\label{fig: rfor}
	\end{figure}

	Figure \ref{fig: rfor} shows the trace plot concerning the mean and the standard deviation of the link indicator variable. The algorithm multiply imputes missing links after a given number of iterations (we have set this number at 5, which is the default value) for making the prediction more stable. In each iteration, the multiple imputation algorithm is based on a random forest composed by five trees and recursively splits data to predict the dyadic outcome (i.e the presence / absence of the link). The graph suggests that the prediction is highly stable also after a very few number of iterations.
	
\end{document}